\documentclass[11pt]{article}
\pdfoutput=1

\usepackage{jheppub} 

\usepackage{xcolor}
\usepackage{xspace}
\usepackage{amsmath}
\usepackage{graphicx}
\usepackage{multirow}
\usepackage{url}
\usepackage{subfigure}
\usepackage[normalem]{ulem}

\newcommand{\Pe}{\ensuremath{\mathrm{e}}}

\newcommand{\Pgm}{\ensuremath{\mathrm{\mu}}}

\newcommand{\jet}{\ensuremath{\mathrm{j}}}

\newcommand{\PQd}{\ensuremath{\mathrm{d}}}
\newcommand{\PQu}{\ensuremath{\mathrm{u}}}
\newcommand{\PQs}{\ensuremath{\mathrm{s}}}
\newcommand{\PQc}{\ensuremath{\mathrm{c}}}
\newcommand{\PQb}{\ensuremath{\mathrm{b}}}
\newcommand{\PQt}{\ensuremath{\mathrm{t}}}
\newcommand{\PAQt}{\ensuremath{\bar{\PQt}}}

\newcommand{\Ph}{\ensuremath{\mathrm{h}}} 
\newcommand{\PW}{\ensuremath{\mathrm{W}}}
\newcommand{\PZ}{\ensuremath{\mathrm{Z}}}
\newcommand{\PX}{\ensuremath{\mathrm{X}}}

\newcommand{\ttbar}{\ensuremath{\mathrm{\PQt\PAQt}}\xspace}
\newcommand{\tth}{\ensuremath{\mathrm{\PQt\PAQt\Ph}}\xspace}
\newcommand{\ttZ}{\ensuremath{\mathrm{\PQt\PAQt\PZ}}\xspace}
\newcommand{\ttW}{\ensuremath{\mathrm{\PQt\PAQt\PW}}\xspace}
\newcommand{\ttX}{\ensuremath{\mathrm{\PQt\PAQt\PX}}\xspace}
\newcommand{\tthj}{\ensuremath{\tth{+}\jet}\xspace}
\newcommand{\ttWj}{\ensuremath{\ttW{+}\jet}\xspace}
\newcommand{\ttZj}{\ensuremath{\ttZ{+}\jet}\xspace}
\newcommand{\ttXj}{\ensuremath{\ttX{+}\jet}\xspace}

\newcommand{\TeV}{\ensuremath{\,\text{Te\hspace{-.08em}V}}\xspace}

\newcommand{\PYTHIA} {{\tt\textsc{Pythia}}\xspace}
\newcommand{\MADGRAPH} {{\tt\textsc{MadGraph}}\xspace}

\newcommand*{\eftOp}[4]{\ensuremath{%
    {#4%
    \ifx\empty#3\empty\ifx\empty#1\empty\else^{#1}\fi\else^{#1(#3)}\fi%
    \ifx\empty#2\empty\else_{#2}\fi}%
}}
\newcommand{\Oup}{\eftOp{}{t\varphi}{}{\mathcal{O}}\xspace}
\newcommand{\Opqa}{\eftOp{1}{\varphi q}{}{\mathcal{O}}\xspace}
\newcommand{\Opqb}{\eftOp{3}{\varphi q}{}{\mathcal{O}}\xspace}
\newcommand{\Opu}{\eftOp{}{\varphi t}{}{\mathcal{O}}\xspace}
\newcommand{\Opud}{\eftOp{}{\varphi tb}{}{\mathcal{O}}\xspace}
\newcommand{\OuW}{\eftOp{}{tW}{}{\mathcal{O}}\xspace}
\newcommand{\OdW}{\eftOp{}{bW}{}{\mathcal{O}}\xspace}
\newcommand{\OuB}{\eftOp{}{tB}{}{\mathcal{O}}\xspace}
\newcommand{\OuG}{\eftOp{}{tG}{}{\mathcal{O}}\xspace}
\newcommand{\OtG}{\eftOp{}{tG}{}{\mathcal{O}}\xspace} 

\newcommand*{\olra}{\protect\overleftrightarrow}
\newcommand{\FDF} {(\varphi^{\dagger}i\!\olra{D}_{\!\mu}\varphi)}
\newcommand{\FDFI}{(\varphi^{\dagger}i\!\olra{D}_{\!\mu}^{\!I}\varphi)}

\newcommand{\ctp}  {\eftOp{}{t\varphi}{}{c}\xspace}

\newcommand{\cpQM} {\eftOp{-}{\varphi Q}{}{c}\xspace}
\newcommand{\cpQa} {\eftOp{3}{\varphi Q}{}{c}\xspace}
\newcommand{\cpt}  {\eftOp{}{\varphi\PQt}{}{c}\xspace}
\newcommand{\cptb} {\eftOp{}{\varphi\PQt\PQb}{}{c}\xspace}

\newcommand{\ctW}  {\eftOp{}{\PQt\PW}{}{c}\xspace}

\newcommand{\ctZ}  {\eftOp{}{\PQt\PZ}{}{c}\xspace}

\newcommand{\cbW}  {\eftOp{}{\PQb\PW}{}{c}\xspace}

\newcommand{\ctG}  {\eftOp{}{\PQt G}{}{c}\xspace}

\newcommand{\OupDef}{\ensuremath{\bar{q} t\tilde\varphi(\varphi^{\dagger}\varphi)}\xspace}
\newcommand{\OpqaDef}{\ensuremath{\FDF (\bar{q}\gamma^\mu q)}\xspace}
\newcommand{\OpqbDef}{\ensuremath{\FDFI (\bar{q}\gamma^\mu\tau^I q)}\xspace}
\newcommand{\OpuDef}{\ensuremath{\FDF (\bar{t}\gamma^\mu t)}\xspace}
\newcommand{\OpudDef}{\ensuremath{(\tilde\varphi^\dagger iD_\mu\varphi)(\bar{t}\gamma^\mu b)}\xspace}
\newcommand{\OuWDef}{\ensuremath{(\bar{q}\sigma^{\mu\nu}\tau^It)\tilde{\varphi}W_{\mu\nu}^I}\xspace}
\newcommand{\OdWDef}{\ensuremath{(\bar{q}\sigma^{\mu\nu}\tau^Ib){\varphi} W_{\mu\nu}^I}\xspace}
\newcommand{\OuBDef}{\ensuremath{(\bar{q}\sigma^{\mu\nu} t)\tilde{\varphi}B_{\mu\nu}}\xspace}    
\newcommand{\OuGDef}{\ensuremath{(\bar{q}\sigma^{\mu\nu}T^At)\tilde{\varphi}G_{\mu\nu}^A}\xspace}

\newcommand{\dimSixTop}{dim6TopEFT\xspace}

\newcommand{\sW}{\ensuremath{s_{\mathrm{W}}}\xspace}
\newcommand{\cW}{\ensuremath{c_{\mathrm{W}}}\xspace}

\newcommand{\xqcut}{{\tt xqcut}\xspace}
\newcommand{\qcut}{{\tt qCut}\xspace}

\newcommand{\muZeroP}{\ensuremath{\mu_{\ttX}}\xspace}
\newcommand{\muPlusOneP}{\ensuremath{\mu_{\ttXj}}\xspace}
\newcommand{\mutth}{\ensuremath{\mu_{\tth}}\xspace}
\newcommand{\mutthj}{\ensuremath{\mu_{\tth{+}\jet}}\xspace}
\newcommand{\muttW}{\ensuremath{\mu_{\ttW}}\xspace}
\newcommand{\muttWj}{\ensuremath{\mu_{\ttW{+}\jet}}\xspace}
\newcommand{\muttZ}{\ensuremath{\mu_{\ttZ}}\xspace}
\newcommand{\muttZj}{\ensuremath{\mu_{\ttZ{+}\jet}}\xspace}

\newcommand{\kfactor}{K-factor\xspace}
\newcommand{\kfactors}{K-factors\xspace}

\newcommand{\ttZvertex}{\ensuremath{t\text{-}t\text{-}Z}\xspace}
\newcommand{\ttZhvertex}{\ensuremath{t\text{-}t\text{-}Z\text{-}h}\xspace}

\newcommand{\tthvertex}{\ensuremath{t\text{-}t\text{-}h}\xspace}
\newcommand{\tbWvertex}{\ensuremath{t\text{-}b\text{-}W}\xspace} 
 
\newcommand{\gluglu}{\ensuremath{\mathrm{gg}}\xspace} 
\newcommand{\qglu}{\ensuremath{\mathrm{gq}}\xspace} 
\newcommand{\qqbar}{\ensuremath{\mathrm{q\bar{q}}}\xspace} 
\newcommand{\qqbarprime}{\ensuremath{\mathrm{q\bar{q}'}}\xspace}

\newcommand{\shat}{\ensuremath{\hat{s}}\xspace}
\newcommand{\sqrtshat}{\ensuremath{\sqrt{\shat}}\xspace}
\newcommand{\vev}{\ensuremath{v}\xspace}

\title{Matching in $pp \to  t \bar{t} W/Z/h + \text{jet}$ SMEFT studies }

\author[a]{Reza Goldouzian,}
\author[a,b]{Jeong Han Kim,}
\author[a]{Kevin Lannon,}
\author[a]{Adam Martin,}
\author[a]{Kelci Mohrman,}
\author[a]{and Andrew Wightman}

\emailAdd{reza.goldouzian@cern.ch}
\emailAdd{jeonghan.kim@cbu.ac.kr}
\emailAdd{klannon@nd.edu}
\emailAdd{amarti41@nd.edu}
\emailAdd{kmohrman@nd.edu}
\emailAdd{andrew.wightman.2@nd.edu}

\affiliation[a]{Department of Physics, 225 Nieuwland Science Hall, University of Notre Dame, Notre Dame, IN\\
46556, USA} 
\affiliation[b]{Department of Physics, Chungbuk National University, Cheongju, Chungbuk 28644, Republic of Korea}

\abstract{
In this paper, we explore the impact of extra radiation on predictions of $pp \to \ttX, \PX = \Ph/\PW^{\pm}/\PZ$ processes within the dimension-6 SMEFT framework.  While full next-to-leading order calculations are of course preferred, they are not always practical, and so it is useful to be able to capture the impacts of extra radiation using leading-order matrix elements matched to the parton shower and merged.  While a matched/merged leading-order calculation for \ttX is not expected to reproduce the next-to-leading order inclusive cross section precisely, we show that it does capture the relative impact of the EFT effects by considering the ratio of matched SMEFT inclusive cross sections to Standard Model values, $\sigma_{\rm SMEFT}(\ttXj)/\sigma_{\rm SM}(\ttXj) \equiv \mu$. Furthermore, we compare leading order calculations with and without extra radiation and find several cases, such as the effect of the operator \OpuDef on \tth and \ttW, for which the relative cross section prediction increases by more than $10\%$---significantly larger than the uncertainty derived by varying the input scales in the calculation, including the additional scales required for matching and merging. Being leading order at heart, matching and merging can be applied to all operators and processes relevant to  $pp \to \ttX, \PX = \Ph/\PW^{\pm}/\PZ + \text{jet}$, is computationally fast and not susceptible to negative weights. Therefore, it is a useful approach in $\ttX{+}\text{jet}$ studies where complete next-to-leading order results are currently unavailable or unwieldy.
}

\begin{document}
\maketitle

\setcounter{page}{2}
\flushbottom

\section{Motivation}
\label{sec:motivation}
The Standard Model effective field theory~\cite{Weinberg:1979sa, Buchmuller:1985jz, Grzadkowski:2010es, Brivio:2017vri} (SMEFT) provides a useful, bottom-up framework for new physics searches at the LHC. The SMEFT assumes that all new particles are too heavy to produce on-shell, and encompasses their effects into the coefficients of higher dimensional operators formed from SM fields and their derivatives. The higher dimensional operators generate new interactions, modify existing interactions, and alter the relationship between Lagrangian parameters and observables.

Recently, the SMEFT approach has been applied to search for new physics in the production of top quarks produced in association with bosons, such as \tth, \ttW, and \ttZ~\cite{Dror:2015nkp, Buckley:2015lku, Buckley:2015nca, Buckley:2016cfg, Hartland:2019bjb,Maltoni_2019}; see Refs.~\cite{Sirunyan:2017uzs,CMS:2019too,Aaboud:2019njj} for experimental analyses. Such analyses rely on predictions for the size and nature of SMEFT effects usually obtained from matrix element calculations interfaced to the parton shower, such as \MADGRAPH~\cite{Alwall:2014hca} interfaced with \PYTHIA~\cite{Sjostrand:2006za, Sjostrand:2014zea}.  In such predictions, the minimal production of \ttX---where ``X'' refers to one W, Z, or h---is seen as the leading contribution, while production with one additional jet, e.g. \ttXj is viewed as a higher-order correction.  

However, there are many examples of processes for which production with one additional jet turns out to be not just a small correction.  An obvious case would be single top production, for which t-channel production dominates over s-channel production.  There are also cases for which the EFT contribution arising from the process with one extra jet is significant, or even dominant over the process with no additional jets. One would like to incorporate these radiation effects into SMEFT analyses.

To study the impact of extra jets, the obvious solution is to calculate an observable such as the inclusive $\ttX$ cross section to next-to-leading order (NLO) accuracy. In fact, an NLO-capable implementation containing all bosonic (dimension-6) operators, all operators with two fermions, and four fermion operators with at least one top quark was recently released~\cite{Degrande:2020evl}. However, SMEFT predictions at NLO require additional counterterms and are a significant undertaking; see Refs.~\cite{Degrande:2016dqg, Mimasu:2015nqa, Franzosi:2015osa, Bylund:2016phk, Maltoni:2016yxb} for \ttX specific NLO details\footnote{NLO is an observable-dependent qualifier. In this paper we will only consider the total inclusive cross section when talking about NLO.}.  The practical consequences of the counterterms are that the NLO Monte Carlo takes significantly more CPU-time to generate events and tends to produce a substantial fraction of events with negative weights, at least within the \MADGRAPH framework.\footnote{For a quantitative example of these differences, see Appendix~\ref{sec:appendix}.}  These considerations can make it challenging to make use of NLO Monte Carlo.

An additional technical complication arises when generating NLO samples involving dimension-6 operators and processes with electroweak vertices in the current \MADGRAPH implementation. Calculations in \MADGRAPH can be done to a fixed order in QCD coupling, QED coupling, new physics coupling, etc.  At NLO, fixing the coupling order is a must, as \MADGRAPH is currently only capable of accounting for NLO QCD effects in SMEFT analyses. Since electroweak loops cannot be included in current \MADGRAPH SMEFT NLO calculation, tree-level diagrams with QED order greater than or equal to two plus the lowest QED order tree-level diagrams are not permitted by the code (as they would have the same coupling order as a QED loop added to the lowest order diagrams). SMEFT Wilson coefficients introduce a further ambiguity, as the effective QED order assigned to the new couplings can impact whether or not diagrams involving these couplings can be included in the NLO calculation. Furthermore, the QED orders assigned to SMEFT couplings is somewhat arbitrary and varies among UFO models in the literature. For example, let us consider the dimension-6 SMEFT operator \OuB = \OuBDef. In some models, the coupling associated with \OuB has a QED order of 0 or 1, while in others it can be 1 or 2 ~\cite{AguilarSaavedra:2018nen,Degrande:2020evl}. Given the role QED coupling counting plays in \MADGRAPH, this subjective assignment controls how \OuB contributes to a process. These coupling technicalities could potentially be overcome, but this would likely require carefully and manually adjusting the order of each operator. Thus, while a full NLO-capable SMEFT treatment is ideal and available for numerous dedicated processes and operators (inclusive cross sections), its automated use within \MADGRAPH may not always be practical, especially for processes with electroweak gauge bosons.

However, studies of SMEFT in processes with additional radiation can also be carried out at tree-level using matching and merging~\cite{Alwall:2007st, Alwall:2007fs, Alwall:2008qv, Englert:2018byk}. Tree level calculations, unlike calculations desiring NLO accuracy, may be run without any {\tt \MADGRAPH} restriction on the coupling order, regardless how new physics couplings are classified (i.e. their QED or QCD order). In the matching/merging procedure, the phase space of the additional radiation is divided up into two regions. Harder  emissions are handled by matrix element (\MADGRAPH, in our case) calculators with the additional parton explicitly listed in the final state, i.e. $pp \to \ttXj$, while soft emissions are handled by the parton shower. Events in the `wrong' region, e.g. a partonic event with $p_T$ below the division, are removed to avoid double counting. Finally, samples with  distinct multiplicities ($+0j, +1j, +2j$, etc.) according to this breakdown are combined to form inclusive $+ \text{jets}$ events. Since matching/merging uses tree-level matrix elements only, it is less expensive than the full NLO treatment---both in terms of CPU-time of the computation and also in terms of the fraction of events generated with negative weights---and it avoids the technical \MADGRAPH complications mentioned above.  Certainly, matched tree-level calculations of the total cross section will lack NLO precision, but for many cases, the loss of precision is a reasonable trade-off in the face of the practical difficulties of NLO Monte Carlo generation.\footnote{For brevity, we will refer to the combined matching plus merging procedure as `matching' going forward, unless otherwise mentioned.}

The matching portion of the procedure does involve some complications when applied to SMEFT. Specifically, in current Monte Carlo generators SMEFT operators can be included in the matrix element portion of the calculation, but they cannot be added (easily) into the parton shower. Thus, it would appear that matching with SMEFT always causes a mismatch, as the events deleted from the matrix element are not replaced by anything. While a reasonable worry, we will show that the impact of this mismatch is small and, at least when focusing on \ttXj processes where X is on-shell, confined to a single operator. Merging with SMEFT samples is no different than with SM.

The goal of this paper is to show that matched/merged calculations capture the relevant physics of extra partons in the inclusive cross section of \ttX final states without introducing uncertainties beyond the usual ones associated with LO compared to NLO QCD, such as the uncertainty from the choice of scales for the matching, factorization/renormalization, and initial/final state radiation.  Because we know that, even in the SM case, the predicted \ttX inclusive cross section receives non-negligible corrections at NLO compared to LO, we will examine the ratio of the SMEFT cross section to the SM value $\sigma_{\rm SMEFT}(\ttXj)/\sigma_{\rm SM}(\ttXj)$ -- the `relative K-factor', which we will denote as $\mu$. We demonstrate the validity of this approach by showing agreement between the cross section ratios calculated for NLO and matched LO in scenarios where both calculations are possible. Once we have established the validity of using matched calculations, we compare the sensitivity of \tth, \ttW, \ttZ matched inclusive cross sections to SMEFT coefficients with and without extra radiation. The extra radiation can come from a SM vertex, or from an additional quark/gluon emitted from a higher dimensional operator.\footnote{ There are two important aspects of the UFO model implementation in \MADGRAPH that need to be accounted for correctly, as outlined in section~\ref{sec:jet_matching_merging}.}

The layout of the rest of this paper is as follows. In section~\ref{sec:framework} we describe the SMEFT framework and the subset of operators that can have a potential impact on \ttX and \ttXj processes.  Section~\ref{sec:jet_matching_merging} is devoted to justifying the use of matching with SMEFT for theory calculations and Monte Carlo studies. The following section, section~\ref{sec:results} contains our results---plots and tables displaying the impact of SMEFT operators on inclusive \ttX and \ttXj processes. We focus on operators whose impact changes dramatically from $+0$ jets to $+1$ jet.  In several cases, the NLO calculation for the operators we examine is challenging to obtain because too many electroweak vertices are needed. In section~\ref{sec:systemaics}, we discuss uncertainties and compare their size to the size of the impact from adding additional jets into the calculation. Finally, in section~\ref{sec:conclusions}, we conclude.

\section{The SMEFT framework}
\label{sec:framework}
The most common basis for characterizing SMEFT effects is the so-called Warsaw basis~\cite{Grzadkowski:2010es}.  At dimension-6, there are 59 operators if one assumes complete fermion flavor universality, or 2499 for flavor anarchy~\cite{Alonso:2013hga, Henning:2015alf}\footnote{At dimension-5, only one operator structure is allowed and it violates lepton number. Therefore, most SMEFT analyses assume lepton number is a good symmetry, making dimension-6 the lowest mass dimension beyond the SM.}.  A flavor symmetry assumption that interpolates between these two extremes is to assume the SMEFT operators are invariant under flavor $U(2)_Q \times U(2)_u \times U(2)_d$.  Under this assumption, first and second generation quarks have universal SMEFT deviations, while operators involving third generation quarks may be different.  This pattern allows the top-quark (or, potentially, the whole third generation) to have sizable SMEFT effects without running afoul of strong flavor constraints from e.g. kaon physics. Motivated by the possibility of larger third generation effects, the set of operators containing two or more third generation fermions was compiled in Ref. \cite{AguilarSaavedra:2018nen}. In this paper, we continue the mindset of Ref.~\cite{AguilarSaavedra:2018nen}, adopting the reduced set of operators and using the UFO model \dimSixTop for all \MADGRAPH purposes. In the \dimSixTop setup, dimension-6 operators are suppressed by a scale $\Lambda$ that is conventionally fixed to $1\,\TeV$. This choice renders the Wilson coefficients dimensionless.  Additionally, the CKM matrix is assumed to be a unit matrix, the masses of $\PQu,\PQd,\PQs,\PQc,\Pe,\Pgm$ fermions are set to zero, and the unitary gauge is used by default.

While the \dimSixTop model contains 33 operators\footnote{This counting refers to the `baseline' model in Ref.~\cite{AguilarSaavedra:2018nen}. Several stricter flavor symmetry assumptions are possible, such as a `top-philic' scenario. See Ref.~\cite{AguilarSaavedra:2018nen} for more details.} we will further narrow our focus to operators that can have potential tree-level impact on \ttX and \ttXj processes with $\Ph/\PW^{\pm}/\PZ$ on-shell. Our goal, after all, is to explore how well matching captures the effects of additional radiation in \ttX process, rather than perform a complete SMEFT analysis, and a smaller set of operators makes this more tractable. This choice limits the number of operators\footnote{In particular, the requirement of an on-shell $\Ph/\PW^{\pm}/\PZ$ removes all four fermion operators.} to 9, which are listed below in Table~\ref{tab:eft-ops}. As some of the operators in Table~\ref{tab:eft-ops} are not self-Hermitian, they may have complex Wilson coefficients. However as imaginary Wilson coefficients inevitably lead to CP violation and are therefore strongly constrained, we consider real coefficients only. Importantly, we include operators that impact \ttX, \ttXj via both QCD and electroweak interactions. 

\begin{table}[hbt!]
\caption{The operators and corresponding Wilson coefficients considered in this paper.  Left-handed fermion doublets are denoted by $q$, and right-handed fermion singlets by $t$ and $b$; the Higgs doublet is denoted by $\varphi$.  The antisymmetric $SU(2)$ tensor is denoted by $\varepsilon\equiv i\tau^2$.  Furthermore, $\tilde{\varphi}=\varepsilon\varphi^*$, $\FDF \equiv \varphi^\dagger(iD_\mu \varphi) - (iD_\mu\varphi^\dagger) \varphi$, and $\FDFI \equiv \varphi^\dagger\tau^I(iD_\mu \varphi) - (iD_\mu\varphi^\dagger) \tau^I\varphi$, where $\tau^I$ are the Pauli matrices.  Finally, $T^A\equiv \lambda^A/2$, where $\lambda^A$ are Gell-Mann matrices. 
The covariant derivative is $D_{\mu}=\partial_{\mu}-ig_{s}\frac{1}{2}\lambda^{A}G_{\mu}^A-ig\frac{1}{2}\tau^{I}W_{\mu}^{I}-ig'YB_{\mu}$, and $G_{\mu\nu}^{A}=\partial_{\mu} G_{\nu}^{A}-\partial_{\nu} G^{A}_{\mu}+g_{s}f^{ABC}G^{B}_{\mu} G^{C}_{\nu}$, $W_{\mu\nu}^{I}=\partial_{\mu} W_{\nu}^{I}-\partial_{\nu} W^{I}_{\mu}+g\epsilon_{IJK}W^{J}_{\mu} W^{K}_{\nu}$, and $B_{\mu\nu}=\partial_{\mu} B_{\nu}-\partial_{\nu}B_{\mu}$ are the gauge field strength tensors.  The abbreviations \sW and \cW denote the sine and cosine of the weak mixing angle (in the unitary gauge).  More details about the operators can be found in Ref.~\cite{AguilarSaavedra:2018nen}.}
\begin{center}
\begin{tabular}{ccc}
Operator & Definition & Wilson Coefficient \\ \hline
\Oup  & \OupDef  & $\ctp$    \\
\Opqa & \OpqaDef & $\cpQM + \cpQa$   \\
\Opqb & \OpqbDef & $\cpQa$           \\
\Opu  & \OpuDef  & $\cpt$            \\
\Opud & \OpudDef & $\cptb$ \\
\OuW  & \OuWDef  & $\ctW$   \\
\OdW  & \OdWDef  & $\cbW$   \\
\OuB  & \OuBDef  & $ (\cW \ctW - \ctZ)/\sW$ \\
\OuG  & \OuGDef  & $\ctG$   \\  
\end{tabular}
\label{tab:eft-ops}
\end{center}
\end{table}

We would like to draw attention to several notational choices regarding the operators and Wilson coefficients outlined in Table~\ref{tab:eft-ops}. First, we note that following the convention in Ref.~\cite{AguilarSaavedra:2018nen}, the \cpQa and \ctW parameters appear in multiple Wilson coefficients. This choice simplifies the form of \ttZvertex (and \ttZhvertex) interactions, so that contributions to $\bar t_L \gamma_\mu t_L Z^{\mu}$ are set by \cpQM alone, contributions to $\bar t_L \gamma_\mu t_R Z^{\mu\nu}$  are set by \ctZ alone, etc. For the rest of this paper, we will refer to each individual parameter as a Wilson coefficient; for example, rather than denoting \cpQa{+}\cpQM as the Wilson coefficient of \Opqa, we will refer to \cpQa and \cpQM as two distinct Wilson coefficients, where the Wilson coefficient \cpQa is associated with both \Opqa and \Opqb. Because multiple operators can involve the same Wilson coefficient, we will describe EFT vertices in terms of the Wilson coefficient associated with the vertex (instead of the operator associated with the Wilson coefficient) to avoid potential ambiguities. Finally, to maintain consistency, we will also use the language of Wilson coefficients when describing the effects of EFT operators. For example, to discuss how the Lagrangian term $\frac{\cpt}{\Lambda^2} \Opu$ affects a particular process, we will refer to the effect of \cpt on the given process. 

\section{Jet matching/merging for EFT samples}
\label{sec:jet_matching_merging}
When considering processes containing an extra parton in the final state, a matching procedure is required to avoid double counting between matrix element (ME) and parton shower (PS) descriptions of phase space. When the ME is matched to a PS, only widely-separated radiation above a certain matching scale $Q$ will be modeled by the ME and therefore include the EFT effects, while radiation below the scale $Q$ is cut off, and not included in the ME description. Radiation below the scale $Q$, namely a soft and colinear regime, are generated by the PS which does not include the EFT vertices. Thus, one may worry that a contribution, namely the soft and colinear contribution from EFT vertices, may be missed. If the soft and colinear contributions from EFT vertices are small, they can safely be neglected.  However, if they are large, then the matching procedure cannot be used without adding EFT effects to the parton shower.

Although there are multiple different ways to implement matching, for these studies we use the $k_T$-jet version of the MLM matching scheme, an event-rejection based approach that matches partons generated by \MADGRAPH to jets clustered by \PYTHIA~\cite{Alwall:2007fs}.  First, the final-state partons in the event produced by \MADGRAPH are clustered according to the $k_T$ algorithm, where the $k_T$ value is required to be above a specified cutoff scale referred to as the \xqcut. The event  is then passed from \MADGRAPH to \PYTHIA for parton showering. After showering (but before hadronization), \PYTHIA clusters the final-state objects using the $k_T$ algorithm with a cutoff scale called \qcut; using \qcut as the maximal $k_T$ distance between jets and partons, the clustered jets are matched to the ME partons. The event is saved if all jets are successfully matched to partons. The event is otherwise discarded, except in the highest jet multiplicity sample, where extra jets (with $k_T$ less than the softest ME parton) are permitted since there is no danger of double counting in these events. To avoid missing a region of phase space, \xqcut is  chosen to be less than \qcut.

In what follows, we explore the validity of matching applied to generation of EFT samples from two perspectives.  First, we explore the theoretical considerations that come into play as applied to the processes and operators of interest in this paper.  Then, we check the results of applying matching to specific samples generated via \MADGRAPH to see if any signs of missing contributions are visible.

\subsection{Theoretical justification}
\label{sec:matching_theory_justification}

As mentioned above, a major concern with matching as applied to EFT samples is that SMEFT operators are included in the matrix element, but they are not implemented in the parton shower. In this section, we will demonstrate that the impact of this mismatch is small, as EFT contributions for the subset of Warsaw basis operators we are interested in (see section~\ref{sec:framework}) to soft and colinear regime are generically suppressed.

Extra QCD radiation requires gluons, so can only arise in operators with gluon field strengths or derivatives of quark fields. Working within the Warsaw basis, where derivatives have been systematically removed in favor of operators with more fields, and focusing on CP-even operators\footnote{CP-odd operators will not interfere with the SM and therefore have no effect at $1/\Lambda^2$, so we ignore them.}, our options are limited to $\mathcal O_G = f_{ABC}G_\mu^{A,\nu}G_\nu^{B,\rho}G_{\rho}^{C,\mu}$, $O_{\phi G} = \phi^\dag \phi \,G^{A,\mu\nu}G^A_{\mu\nu}$,  $\mathcal O_{dG}$ and $O_{uG}$. If we limit ourselves to the operators that are particularly important for $\ttX$ observables, as laid out by Ref.~\cite{AguilarSaavedra:2018nen, Hartland:2019bjb}, the only operator that remains is $\OtG$\footnote{The impact of $\mathcal O_G$ in matching has been addressed in the literature in Ref.~\cite{Englert:2018byk}. The other operators, $\mathcal O_{\phi G}$ and $\mathcal O_{dG}$ play little role in \ttX and are better constrained by Higgs and $pp \to \bar b b$ observables respectively~\cite{Hartland:2019bjb,Bramante:2014hua}. Additionally, due to the chirality structure of $\mathcal O_{dG}$, it is reasonable to assume $c_{dG} \propto y_b \ll 1$, further suppressing any effects.}. 
\begin{eqnarray}
\frac{\ctG}{\Lambda^2} \OtG  = \frac{g_s \ctG}{\Lambda^2}  ( \overline{q}\, \sigma^{\mu \nu}\, T^a\, t) \tilde{\phi}\, G^a_{\mu \nu}  \;,
\label{eq:Otg}
\end{eqnarray}
Notice the explicit factor of $g_s$ in $\OtG$. This factor is necessary in order for \MADGRAPH to identify $\OtG$ as a `QCD-type' operator and include it correctly.

To address the concern of omitting $\OtG$ from the parton shower, we need to understand the circumstances under which $\OtG$ leads to a soft or colinear gluon. If we set the Higgs to its vacuum expectation value (\vev) and pull one gluon out of the field strength, we are left with a new (non-SM) three-point interaction between two top quarks and a gluon. Including this vertex on any top quark line, we have another potential source of soft/colinear gluon radiation. If we extract two gluons, we have new $g$-$g$-$t$-$t$-$h$ and $g$-$g$-$t$-$t$ QCD vertices. Connecting one of the gluons to the initial partons, the second gluon can land in the final state of $pp \to t\bar t +j$, $pp \to t\bar t Z/h + j$. However, four (and higher) point vertices cannot contribute at tree level to $1$ particle $\to 2$ particle branchings -- the source of soft/collinear issues within QCD -- so we expect their effects to be subdominant. Therefore, in the next section we will focus on the anomalous $t$-$t$-$g$ vertex, as any ill soft/collinear effects from $\ctG$ should be largest there, then comment on how out the conclusions of that study apply more broadly.

\subsubsection{An anomalous \texorpdfstring{$t$-$t$-$g$}{t-t-g} coupling}
\label{sec:2}
 
\begin{figure}[ht]
\begin{center}
\subfigure[]{\includegraphics[width=0.27\textwidth,clip]{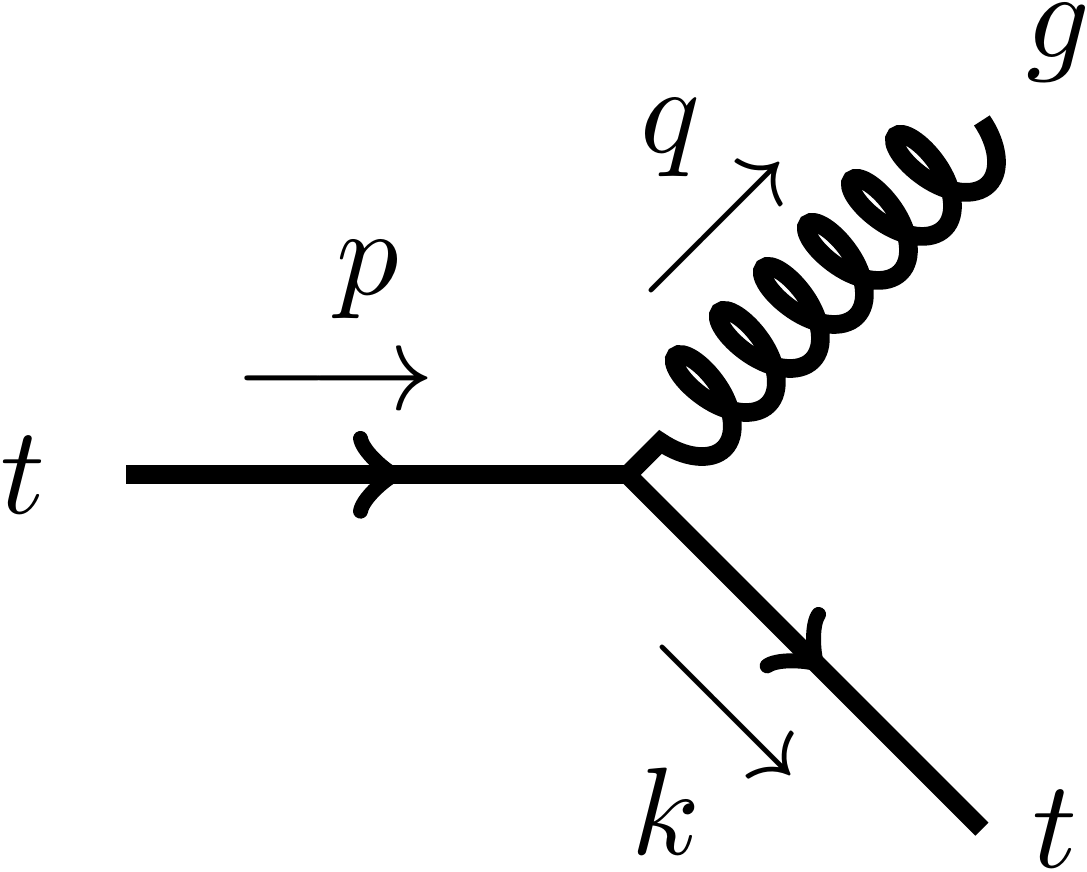} \label{ttg}}\hspace{3.5em}
\subfigure[]{\includegraphics[width=0.27\textwidth,clip]{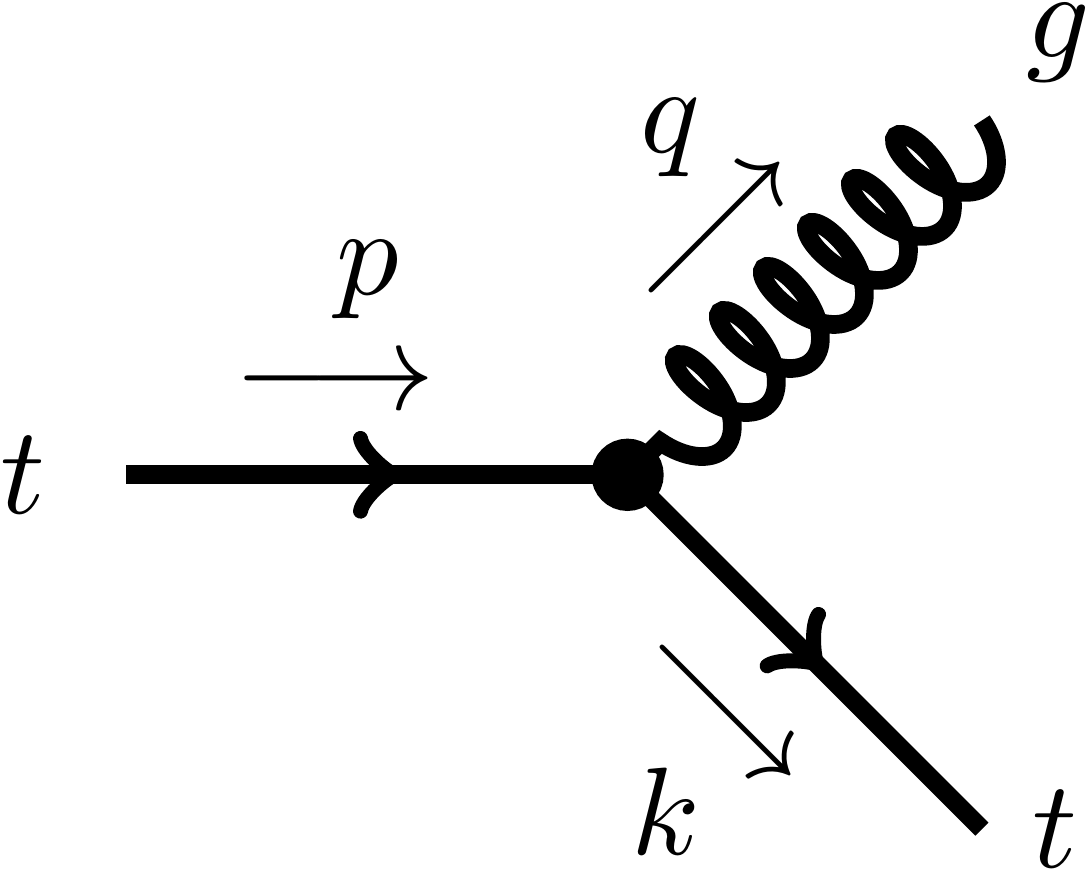} \label{ttg_EFT}}
\caption{The splitting of gluon from the (a) SM $t$-$t$-$g$ vertex, and (b) anomalous $t$-$t$-$g$ vertex generated by the $\OtG$ operator.}\label{ttg_full}
\end{center}
\end{figure}

The anomalous $t$-$t$-$g$ coupling induced by the $\OtG$ operator as shown in Figure \ref{ttg_EFT} is a potential source of soft/collinear radiation. Unlike the SM coupling (Figure \ref{ttg}), this vertex flips the chirality of the top quark, and can contribute to any $pp \to \ttX$, $\PX = \Ph/\PW^{\pm}/\PZ$ processes involving an extra parton. Since the parton shower does not include the anomalous $t$-$t$-$g$ vertex, the information about the EFT below the matching scale $Q$ is missing, resulting in a mismatch.

\begin{figure}[ht!]
\begin{center}
\subfigure[]{\includegraphics[width=0.37\textwidth,clip]{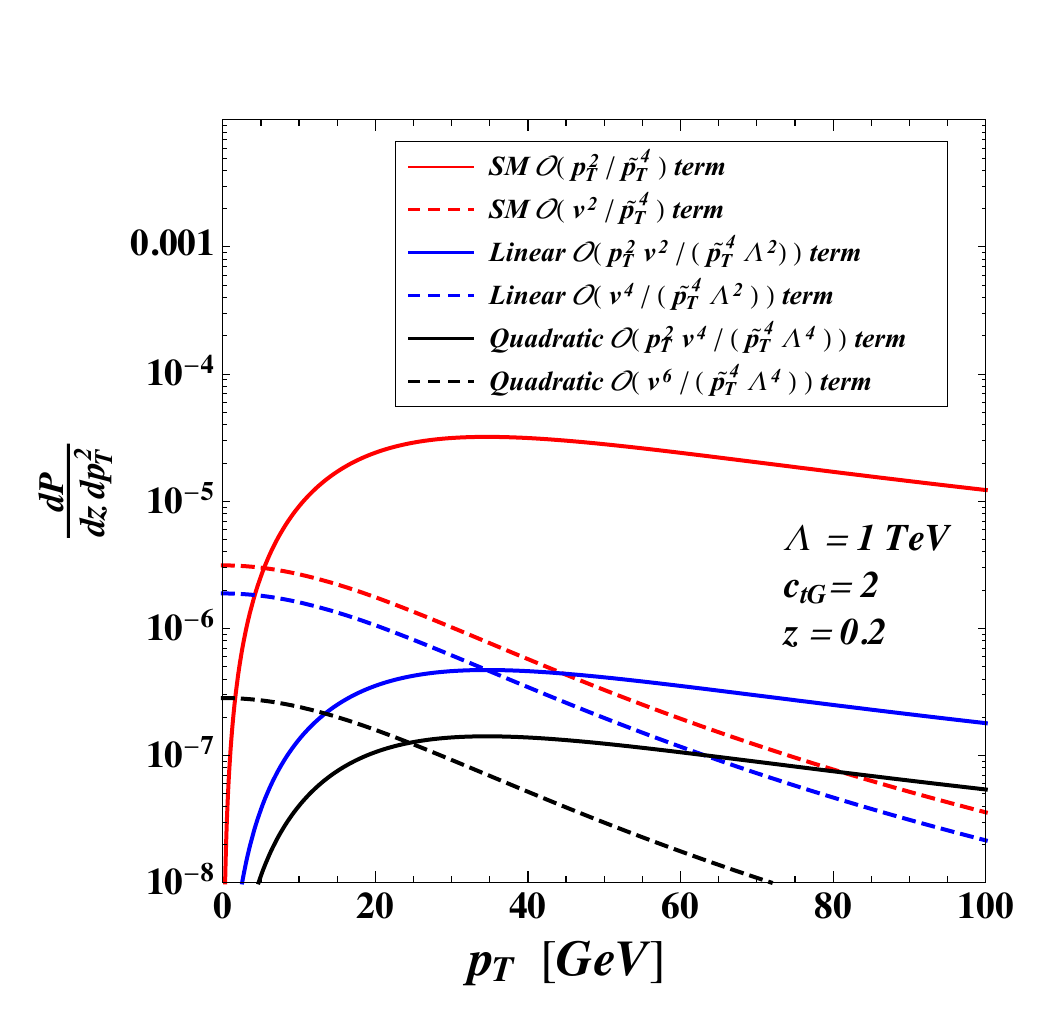} \label{Sp_pT_comp}}\hspace{0.em}
\subfigure[]{\includegraphics[width=0.37\textwidth,clip]{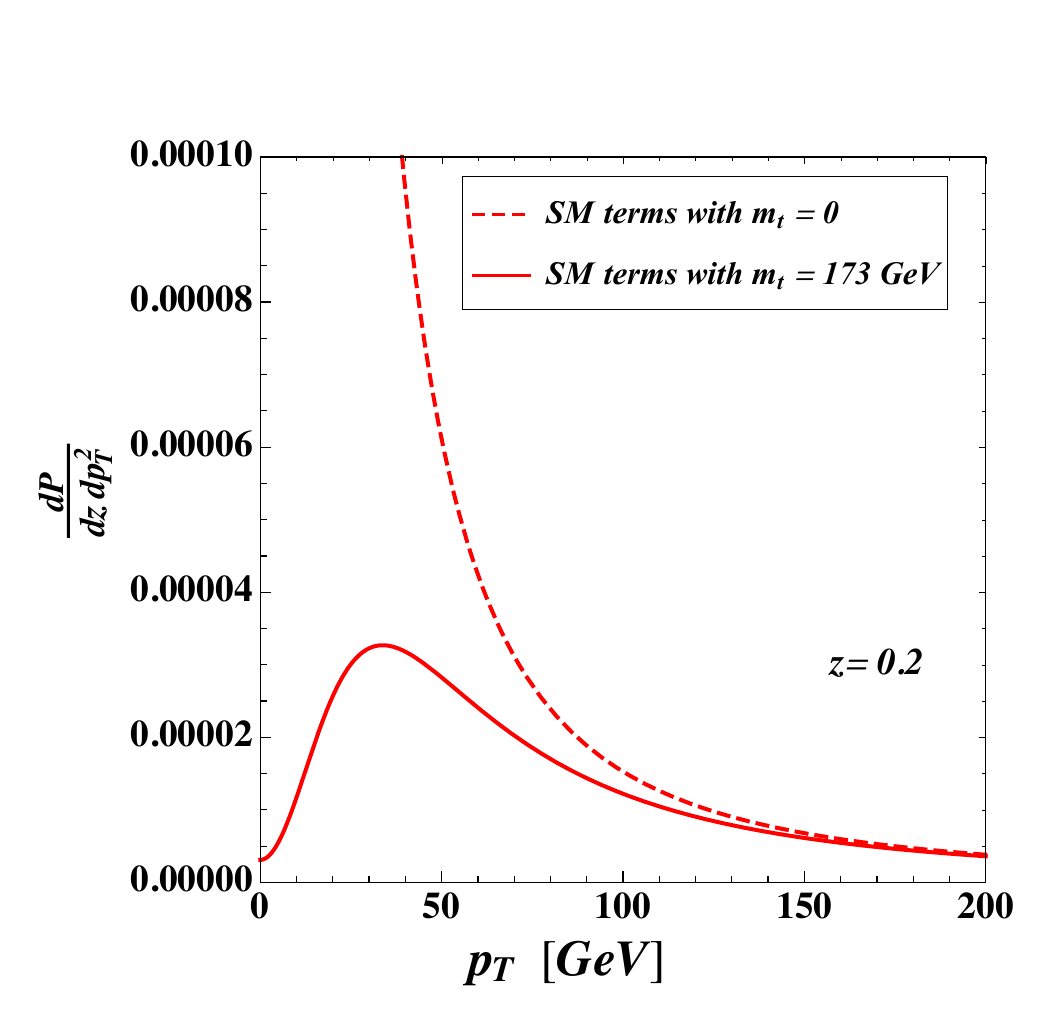} \label{Sp_pT_SM}} \\ \vspace{-2.5em}
\subfigure[]{\includegraphics[width=0.37\textwidth,clip]{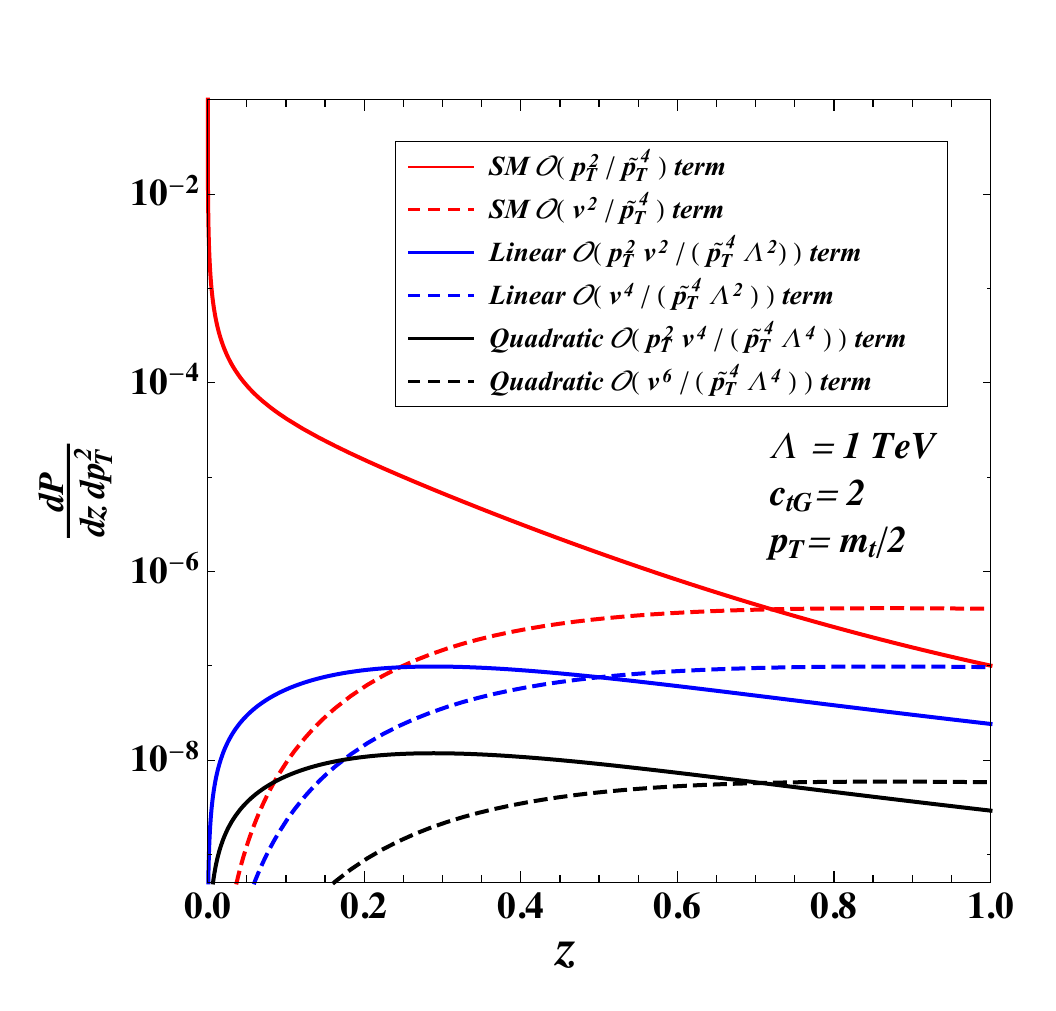} \label{Sp_z_comp}}\hspace{0.em}
\subfigure[]{\includegraphics[width=0.37\textwidth,clip]{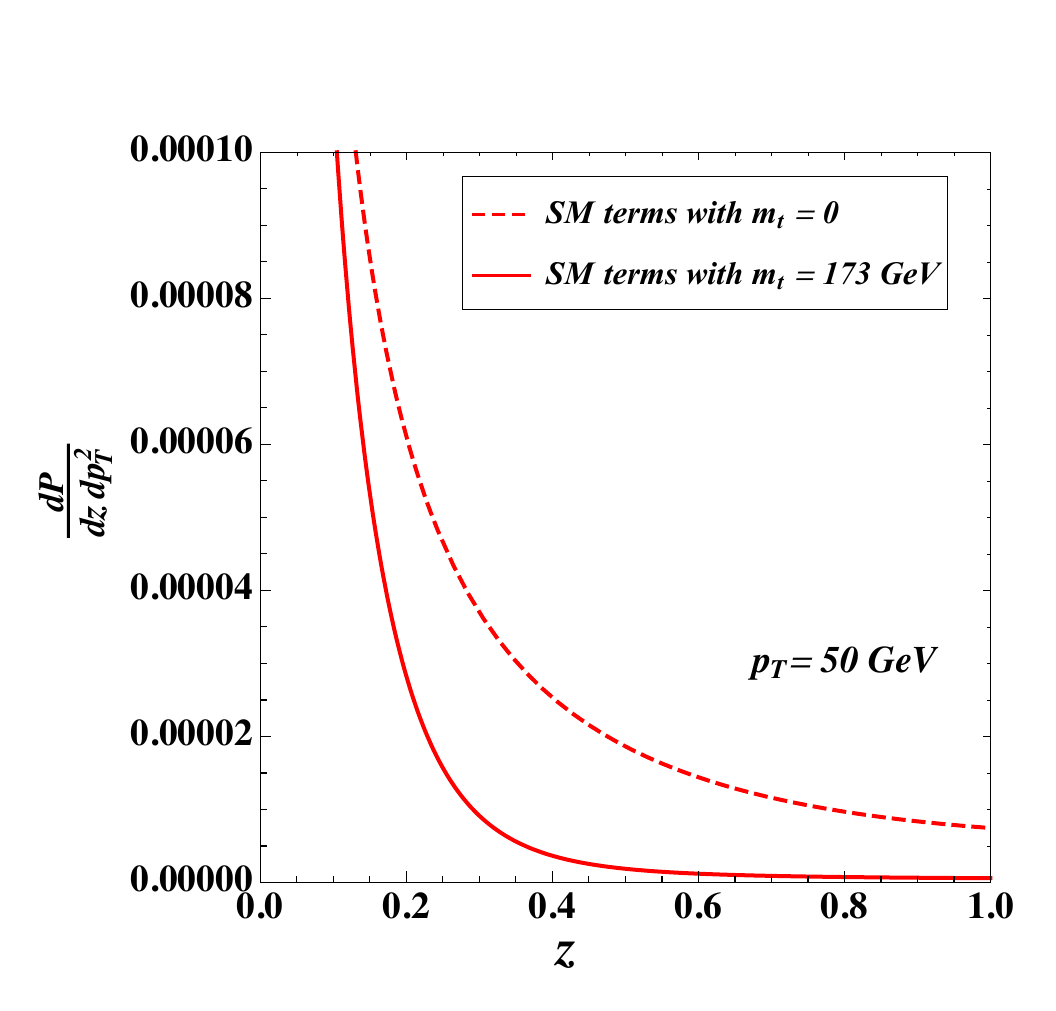} \label{Sp_z_SM}}
\vspace*{-0.2cm}
\caption{Splitting functions for the SM, the interference, and the quadratic EFT pieces in Eq.(\ref{eq:Sp_SM},\ref{eq:Sp_linear},\ref{eq:Sp_quadratic}) respectively. We show the components of each splitting function in terms of (a) $p_T$ with fixed $z = 0.2$ and (c) $z$ with fixed $p_T = m_t/2$. To see the mass effect, the SM splitting function is shown with and without including the mass, in terms of (b) $p_T$ with fixed $z = 0.2$ and (d) $z$ with fixed $p_T = 50$ GeV.
}\label{Sp_appendix}
\end{center}
\end{figure}

To scrutinize the properties of gluons from the EFT (Figure \ref{ttg_EFT}) vertex, we can analytically compute the $q \to q + g$ splitting function \cite{Chen:2016wkt,Larkoski:2011fd,Plehn:2009nd,Catani:2000ef}; Feynman rules for the $t$-$t$-$g$ couplings are summarized in Appendix~\ref{sec:appendix2}. We consider the case where two final particles are almost collinear, with a small relative transverse momentum $p_T$. We choose the coordinate such that the initial top quark is aligned to the $z$-axis, and all particles are on-shell. The four-momenta of three particles can be written as:
\begin{eqnarray} \nonumber
p &=& \big(~ E, 0, 0, p ~\big), \\ 
q &=& \big(~ z E, p_T, 0, \sqrt{z^2 E^2 - p^2_T} ~\big), \\
k &=& \big(~ (1-z) E, -p_T, 0, \sqrt{(1-z)^2 E^2 - m_t^2 - p^2_T} ~\big) ,\nonumber
\end{eqnarray}
where $z = E_q/E_p$ is a dimensionless energy-sharing variable, and $m_t = 173$ GeV is the top quark mass. The splitting function is defined as
\begin{eqnarray}
\frac{d\mathcal{P}}{d z d p^2_T} = \frac{z \bar{z}}{16 \pi^2 \tilde{p}^4_T} \Big( \frac{1}{2}\frac{1}{3} \sum | \mathcal{M}|^2 \Big),
\end{eqnarray}
where $\tilde{p}^2_T \equiv p^2_T + z^2 m^2$ and $\bar{z} \equiv 1-z$, and $\mathcal{M}$ denotes a splitting amplitude shown in Figure \ref{ttg_full}. The sum inside the bracket runs over all polarizations, spins, and colors. The squared amplitude is averaged over initial helicities and colors\footnote{For the detailed expressions of spinors and polarization vectors, see the Refs.\cite{Murayama:1992gi,Chen:2016wkt}.}. Summing the effect of the SM and \OtG vertices and taking the limit $E \gg p_T, m_t$, the splitting functions are:
\begin{eqnarray}
\frac{d\mathcal{P}_{SM}}{d z d p^2_T} &=& \frac{2 \alpha_s}{3 \pi  } \Big( \frac{p^2_T}{\tilde{p}^4_T} \Big)  \frac{ (1+\bar{z}^2)}{z} + \frac{2 \alpha_s}{3 \pi } \Big( \frac{m^2_t}{ \tilde{p}^4_T} \Big) z^3 , \label{eq:Sp_SM} \\
\frac{d\mathcal{P}_{linear}}{d z d p^2_T} &=&  \frac{4 \sqrt{2} \alpha_s \ctG}{3 \pi } \Big( \frac{ p^2_T v m_t  }{  \tilde{p}^4_T \Lambda^2} \Big) z + \frac{4 \sqrt{2} \alpha_s \ctG }{3 \pi  } \Big( \frac{v m^3_t }{ \tilde{p}^4_T \Lambda^2} \Big) z^3 , \label{eq:Sp_linear} \\
\frac{d\mathcal{P}_{quadratic}}{d z d p^2_T} &=&   \frac{8 \alpha_s c^2_{tG}}{3 \pi } \Big( \frac{ p^2_T v^2 m^2_t  }{ \tilde{p}^4_T \Lambda^4} \Big) z + \frac{4 \alpha_s c^2_{tG}}{3 \pi }  \Big( \frac{ v^2 m^4_t}{ \tilde{p}^4_T \Lambda^4} \Big) z^3 , \label{eq:Sp_quadratic}
\label{eq:Sp}
\end{eqnarray}
where we have separated the result into the SM piece, the interference term, and the term quadratic in $c_{tG}$.  

The first term of the SM splitting function in Eq.(\ref{eq:Sp_SM}) scales like $\mathcal{O}( p^2_T/ \tilde{p}^4_T)$ and is relatively enhanced at high $p_T$ with respect to the second term which scales like $\mathcal{O}( v^2/ \tilde{p}^4_T)$. The second term, however, is more enhanced at small $p_T$, and hence named as the ultra-collinear splitting \cite{Chen:2016wkt,Chen:2018uii}. This can be seen in Figure \ref{Sp_pT_comp} that at small $p_T$ the ultra-collinear term (dashed red) is enhanced, whereas the first term (solid red) is relatively suppressed. At high $p_T$, on the other hand, the first term dominates over a wide range of $p_T$. Note that the mass-dependent factor $\tilde{p}^4_T$ in the denominator regulates a collinear divergence at small $p_T$. As shown in Figure \ref{Sp_pT_SM}, in the massless limit ($m_t = 0$), the SM splitting function (dashed red) diverges as $p_T \rightarrow 0$ with fixed value of $z = 0.2$. On the other hand, if we include the mass term, the splitting function (solid red) is regulated at the infrared region.

\begin{figure}[ht]
\begin{center}
\subfigure[]{\includegraphics[width=0.37\textwidth,clip]{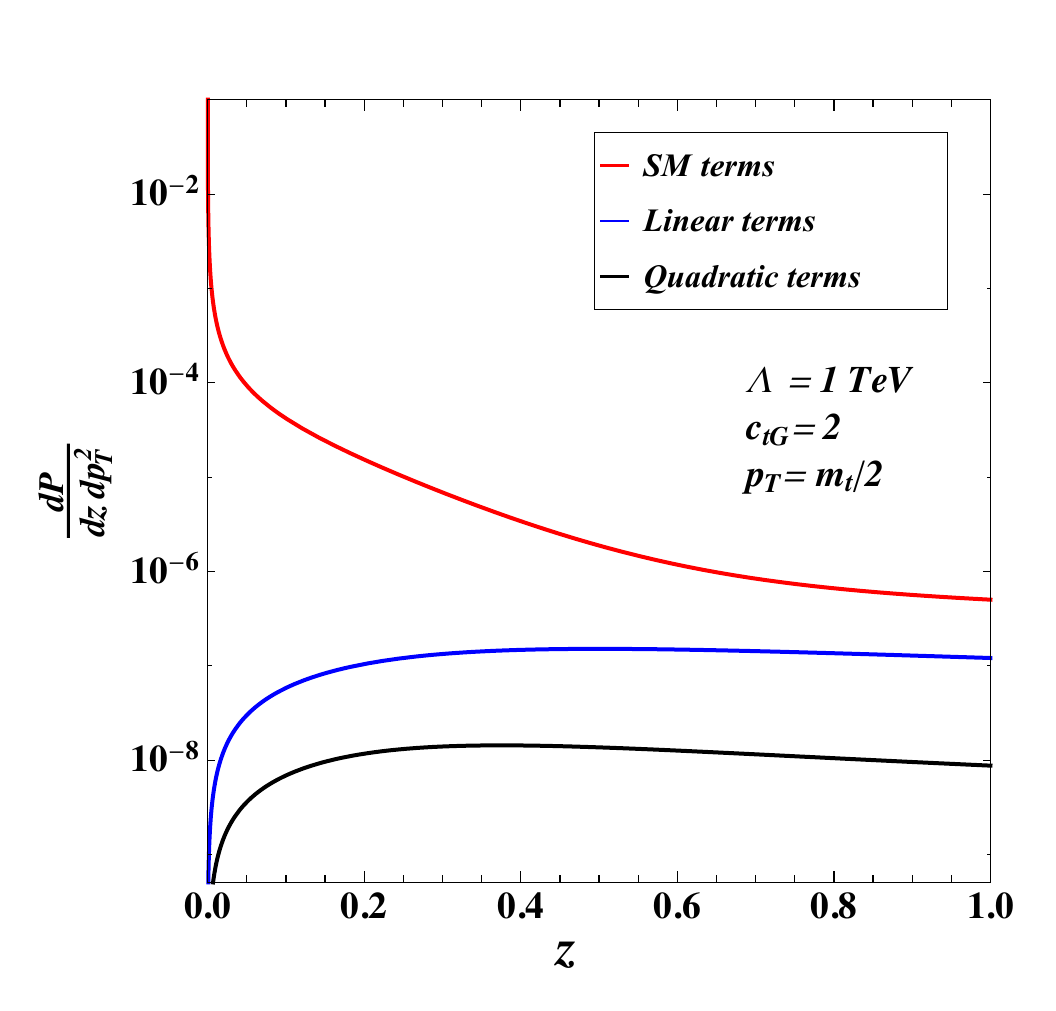} \label{Sp_z_mtd2}}\hspace{0.em}
\subfigure[]{\includegraphics[width=0.37\textwidth,clip]{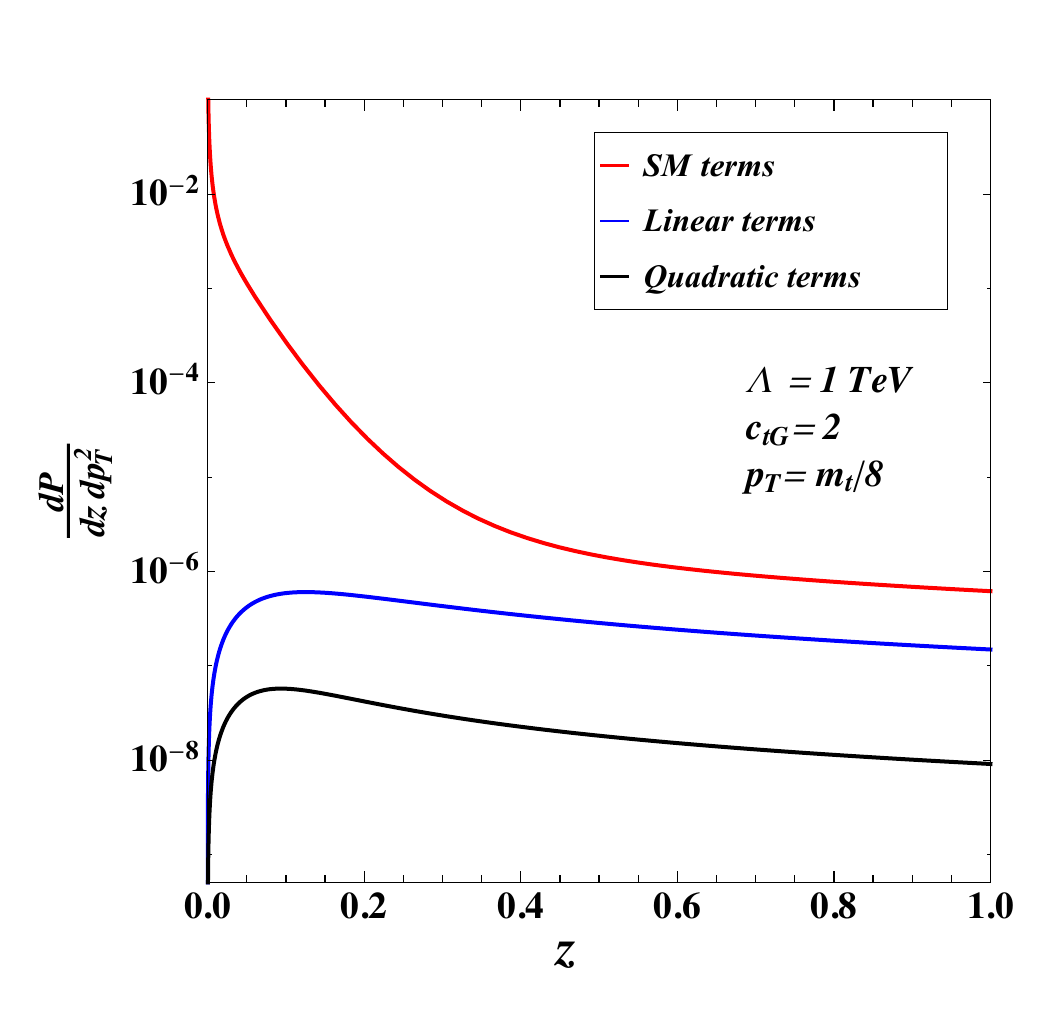} \label{Sp_z_mtd8}} \\ \vspace{-2.em}
\subfigure[]{\includegraphics[width=0.37\textwidth,clip]{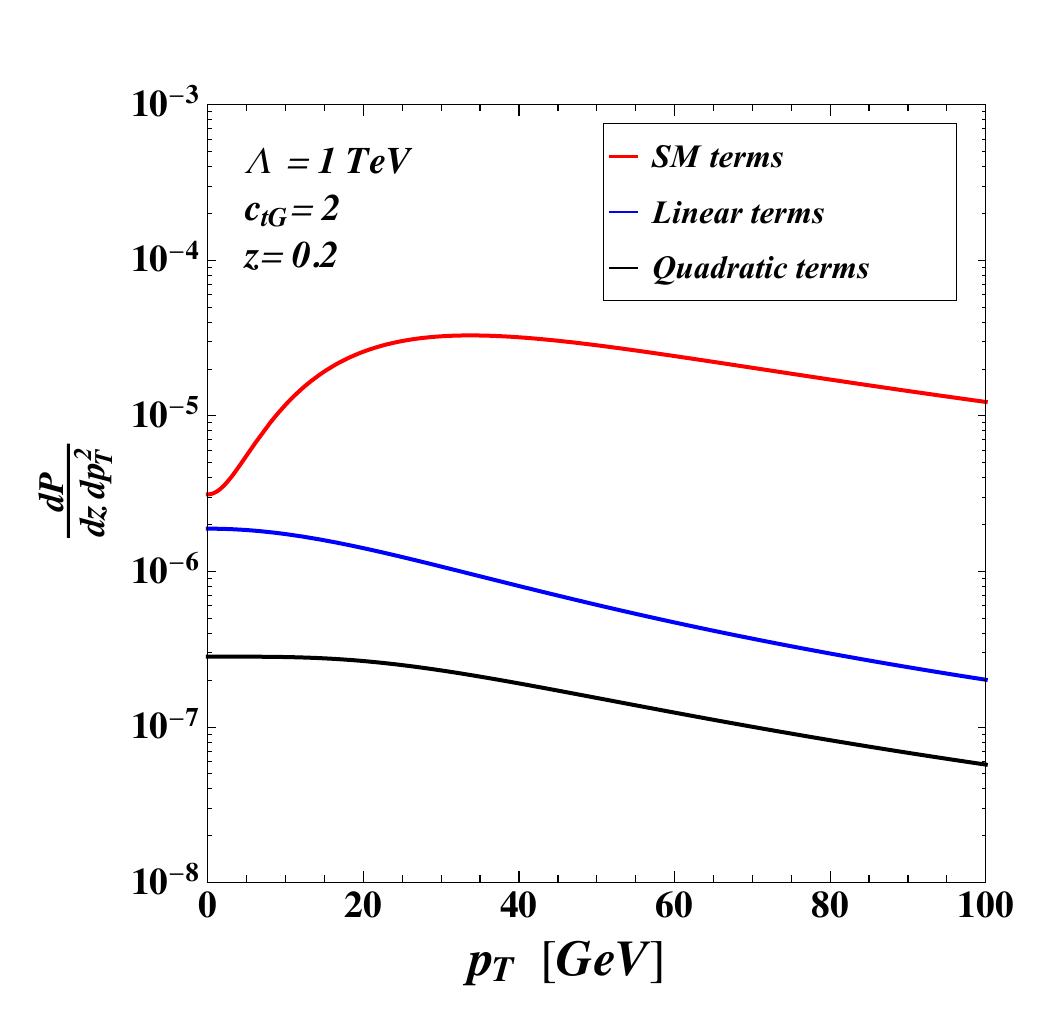} \label{Sp_pT_z0d2}}\hspace{0.em}
\subfigure[]{\includegraphics[width=0.37\textwidth,clip]{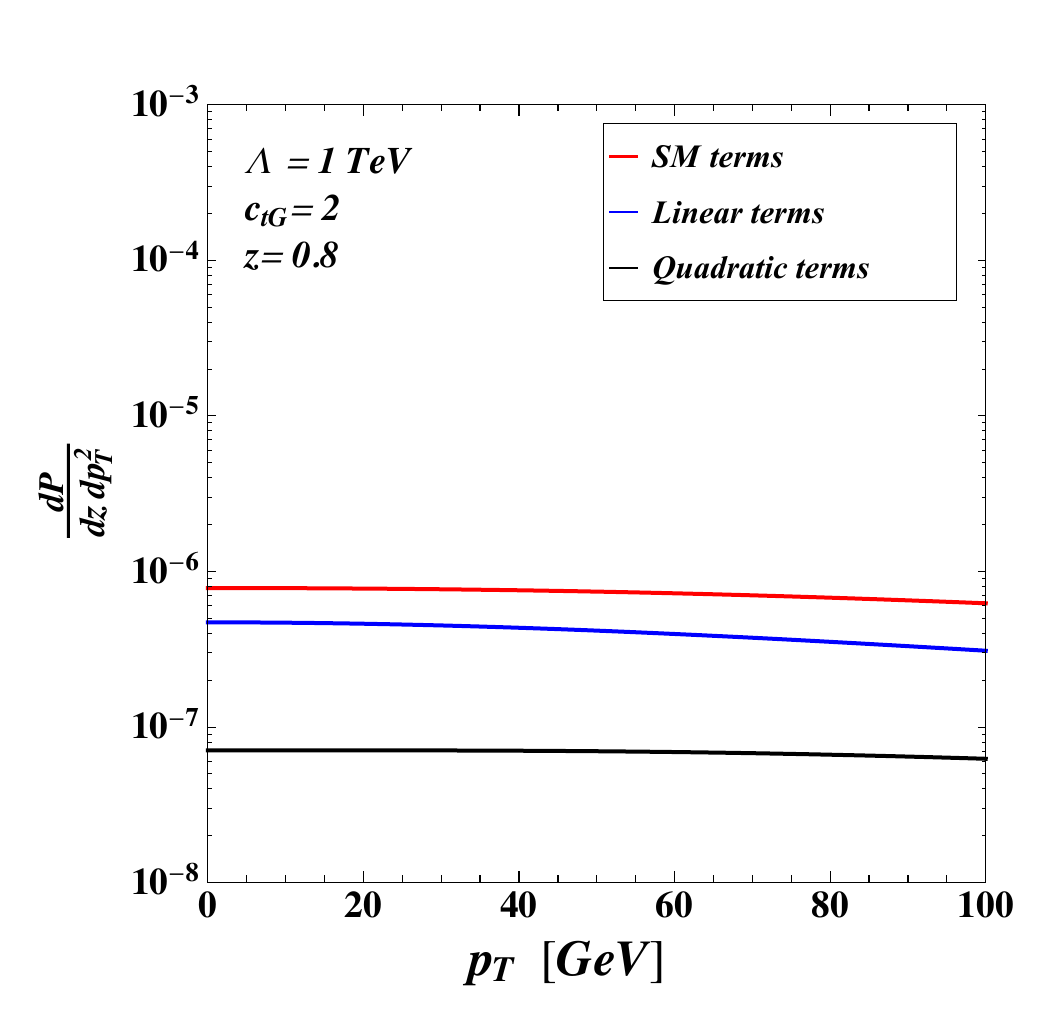} \label{Sp_pT_z0d8}}
\vspace*{-0.2cm}
\caption{Splitting functions for the SM, the interference, and the quadratic EFT pieces in Figure \ref{ttg_full}. The splitting functions in terms of $z$ variable for fixed (a) $p_T = m_t/2$ and (b) $p_T = m_t/8$. The splitting functions in terms of $p_T$ for fixed (c) $z = 0.2$ and (d) $z = 0.8$. As benchmark values, we choose $\ctG = 2$ and $\Lambda = 1$ TeV.
}\label{Sp_final}
\end{center}
\end{figure}

The splitting function for the interference piece in Eq.(\ref{eq:Sp_linear}) receives an overall suppression factor of $\mathcal{O}( v^2/ \Lambda^2)$. The first term (solid blue) scales as $\mathcal{O}( p^2_T v^2/  (\tilde{p}^4_T \Lambda^2))$, and dominates over the second term at high $p_T$ as shown in Figure \ref{Sp_pT_comp}. The second term (dashed blue) is of an order of $\mathcal{O}( v^4/  (\tilde{p}^4_T \Lambda^2))$ that shows the similar ultra-collinear behavior as it is enhanced at small $p_T$.  Although the SM and interference splitting functions behave similarly in terms of $p_T$ up to the overall suppression factor, there is a clear disparity between them in terms of $z$. Note that the first terms of Eq.(\ref{eq:Sp_SM}) and Eq.(\ref{eq:Sp_linear}) have a different $z$-dependence. Their distinctive $z$-dependences are plotted in Figure \ref{Sp_z_comp} where the interference piece (solid blue) is highly suppressed as $z \rightarrow 0$ while the SM term (solid red) is dramatically enhanced. The splitting function for the pure EFT piece in Eq.(\ref{eq:Sp_quadratic}) is qualitatively similar to the interference one, up to the overall suppression factor of $\mathcal{O}( v^4/ \Lambda^4)$, and hence highly suppressed.

Figure \ref{Sp_final} shows total splitting functions for the SM, the interference, and the quadratic EFT terms. As can be seen in the upper panels, the amount of the interference and the pure EFT contribution is highly suppressed near the $z = 0$ region, while the SM contribution is much more enhanced. For $z=0.05$, the size of the SM splitting function is ${\sim}1200~(4200)$ times larger than the size of the interference (pure EFT) splitting function. This means that it is less likely that the soft gluon is emitted due to the interference and the pure EFT portions, while a majority of soft gluon is generated from the SM vertex. As we move away from the $z = 0$ region, the both interference and the pure EFT splitting functions rapidly grow, closing a wide gap with the SM. For $z=0.8$, the SM splitting function is only ${\sim}2~(10)$ times larger than the interference (pure EFT) one. 

The suppression of the anomalous three point vertex in the soft/colliner regime compared to the SM can be understood on dimensional grounds. The anomalous vertex has a $1/\Lambda^2$, which must be compensated by {\em positive} powers of momenta/energy to match the dimensionality of the SM three point vertex\footnote{For example, the anomalous vertex involving a negative helicity quark, negative helicity antiquark and a negative helicity gluon is $\sim \langle 1 3\rangle \langle 2 3 \rangle\, \frac{v}{\Lambda^2}$~\cite{Ma:2019gtx}, where $1,2$ are the quark momenta and $3$ is the gluon momentum (and taking all particles massless and all momenta outgoing). The SM three point vertex -- for negative helicity quark and gluon but a positive helicity antiquark -- $\sim \frac{\langle 1 3\rangle^2}{\langle 1 2 \rangle}$  momentum~\cite{Elvang:2013cua}. The swap in the antiquark helicity reflects the structural difference between a current operator (SM) and magnetic moment type operator (anomalous vertex). Incorporating masses for the fermions makes the expressions more complicated but does not alter the conclusion. Furthermore, by gauge invariance, the forms for the anomalous three point \ctG vertex apply to the four and five point vertices as well.}. Positive powers of momenta may go $\to 0$ in the soft/collinear limit, but never as $\to 1/0$. Therefore, the anomalous vertices do not blow up in the soft/collinear regime and are therefore suppressed (in that limit) compared to SM interactions. This dimensional argument applies to the four and five particle vertices within \ctG as well as to the three particle vertex. Thus, we conclude that all \ctG effects are suppressed in the phase space regime controlled by the parton shower, and expect their absence in \PYTHIA to have minimal impact on matching studies.


\subsection{Monte Carlo validation of matching}
\label{sec:validation_matching}

The prior section demonstrated that the impact of neglecting EFT contributions in the parton shower, at least for the processes of interest in this study, is small. To quantify the impact of applying matching, we look at the concrete case of specific physics processes generated with non-zero values for the Wilson coefficients.  We will use the differential jet rate (DJR) to characterize whether the matching has introduced any discontinuities~\cite{Lenzi:2009fi,Alwall:2008qv}.  For the $k_T$ algorithm, the DJR histogram represents the distribution of $k_T$ values for which  an $n$ jet event transitions to an $n+1$ jet event. A smooth transition between the $n$ and $n+1$ curves indicates that the chosen matching scales have allowed \MADGRAPH and \PYTHIA to work together to smoothly populate the overlapping region of phase space. 

The matched samples used in this section are generated with a model based on the one described in Ref.~\cite{AguilarSaavedra:2018nen}, with an extra factor of $g_s$ applied to the \ctG vertices, as described in section~\ref{sec:matching_theory_justification}. The five-flavor scheme is used, so the mass of the b quark is set to zero. We use the PDF set NNPDF3.1 \cite{Ball:2014uwa} and the default dynamical scale choice for the renormalization and factorization scales; we do not make any parton level cuts. The samples are generated with an \xqcut of 10 and a \qcut of 19.

\begin{figure} [ht!]
\centering
\includegraphics[width=.32\textwidth]{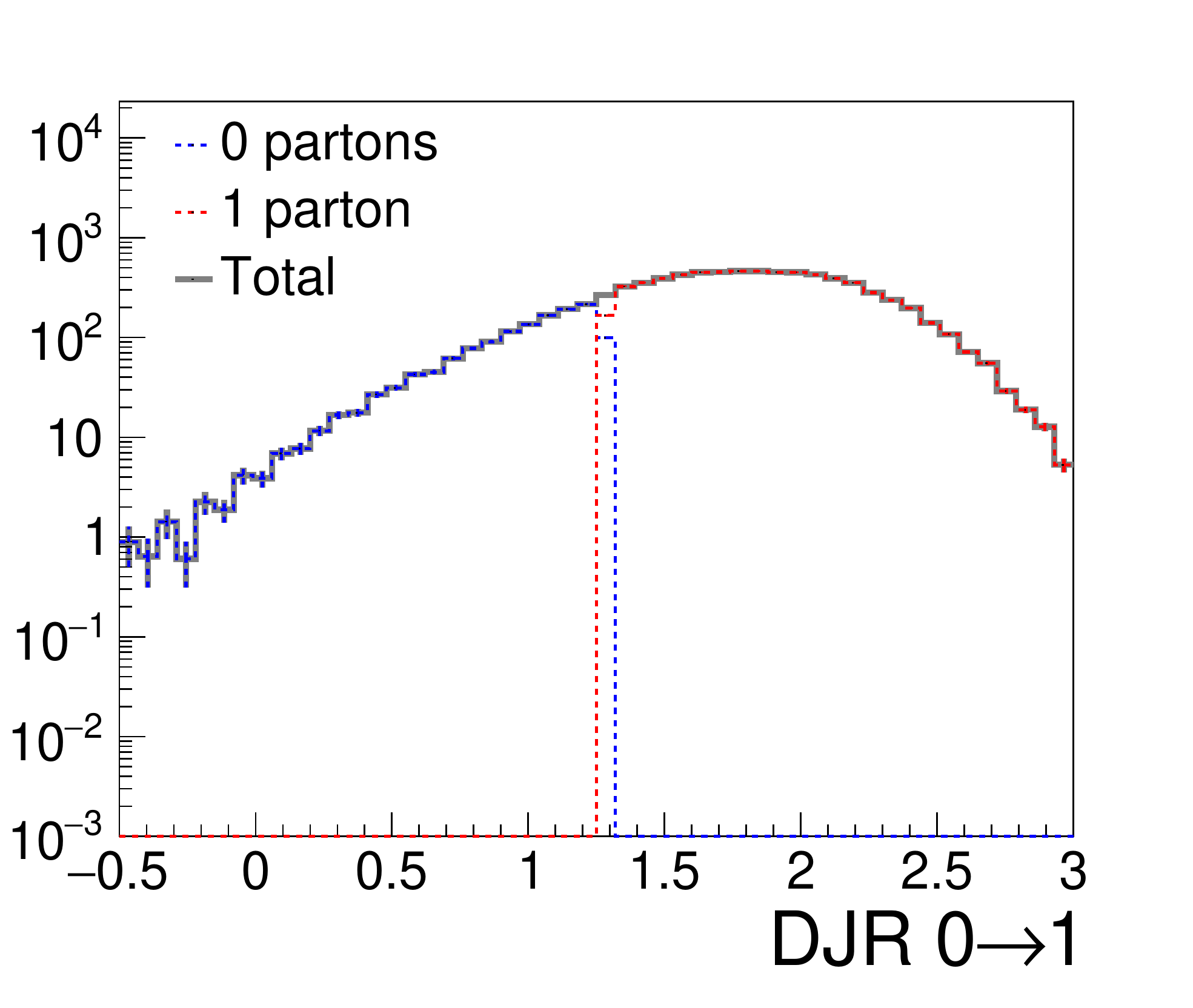}
\includegraphics[width=.32\textwidth]{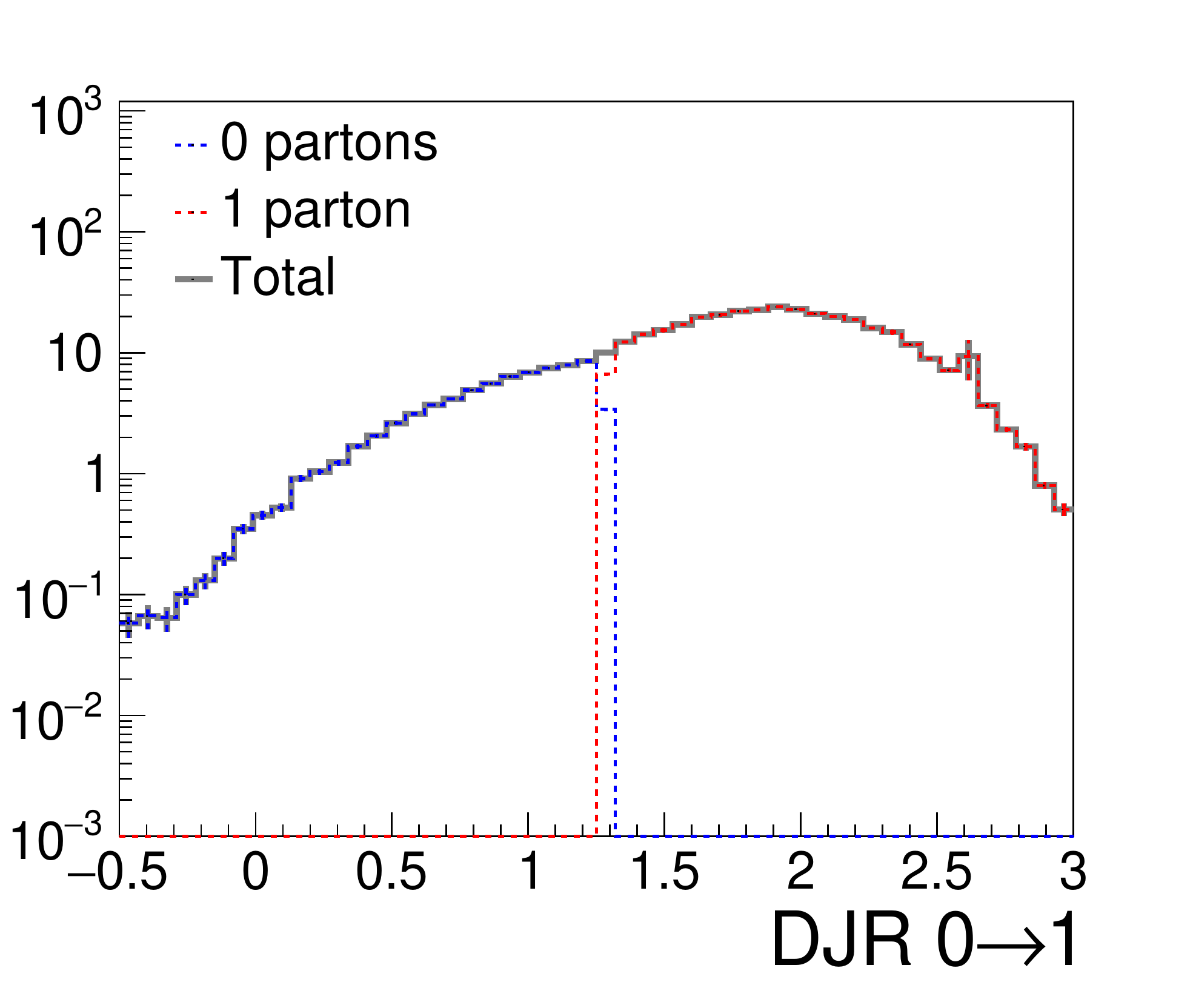}
\includegraphics[width=.32\textwidth]{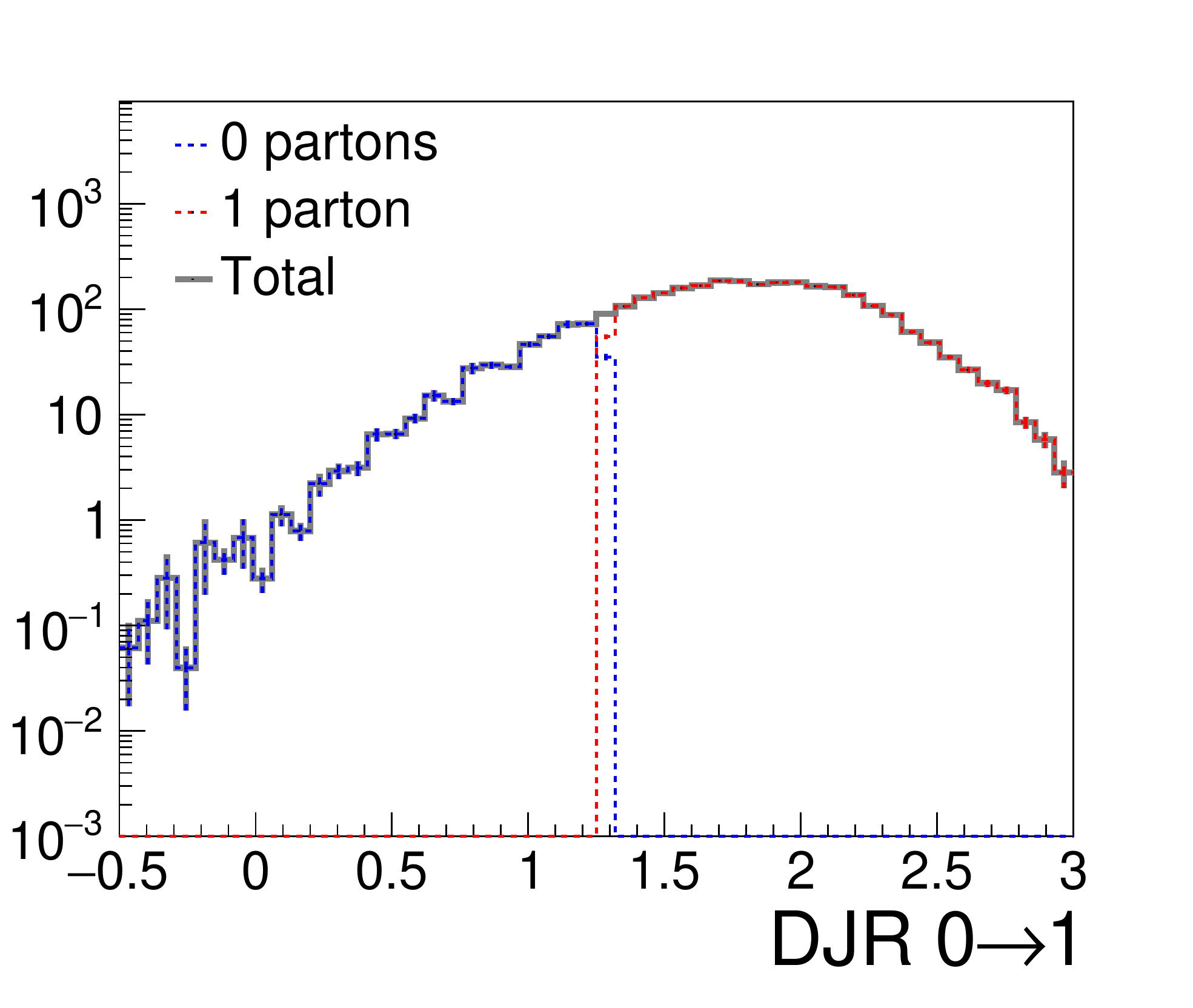}
\caption{DJR plots for $\xqcut=10$, $\qcut=19$ for \tth (left), \ttW (middle) and \ttZ (right). Here, all 9 Wilson coefficients considered in this study are set to non-SM values. The x axis shows the log base 10 of the scale at which an $n$ jet event transitions into an $n{+}1$ jet event. The line labeled ``0 partons" refers to the contribution from the parton shower, while the line labeled ``1 parton" refers to the contribution from the matrix element. The line labeled ``Total" is the sum of the two contributions.}
\label{fig:djr_ttH-ttW-ttZ_NOT_ttll-ttlnu}
\end{figure}

\begin{figure} [ht!]
\centering
\includegraphics[width=.32\textwidth]{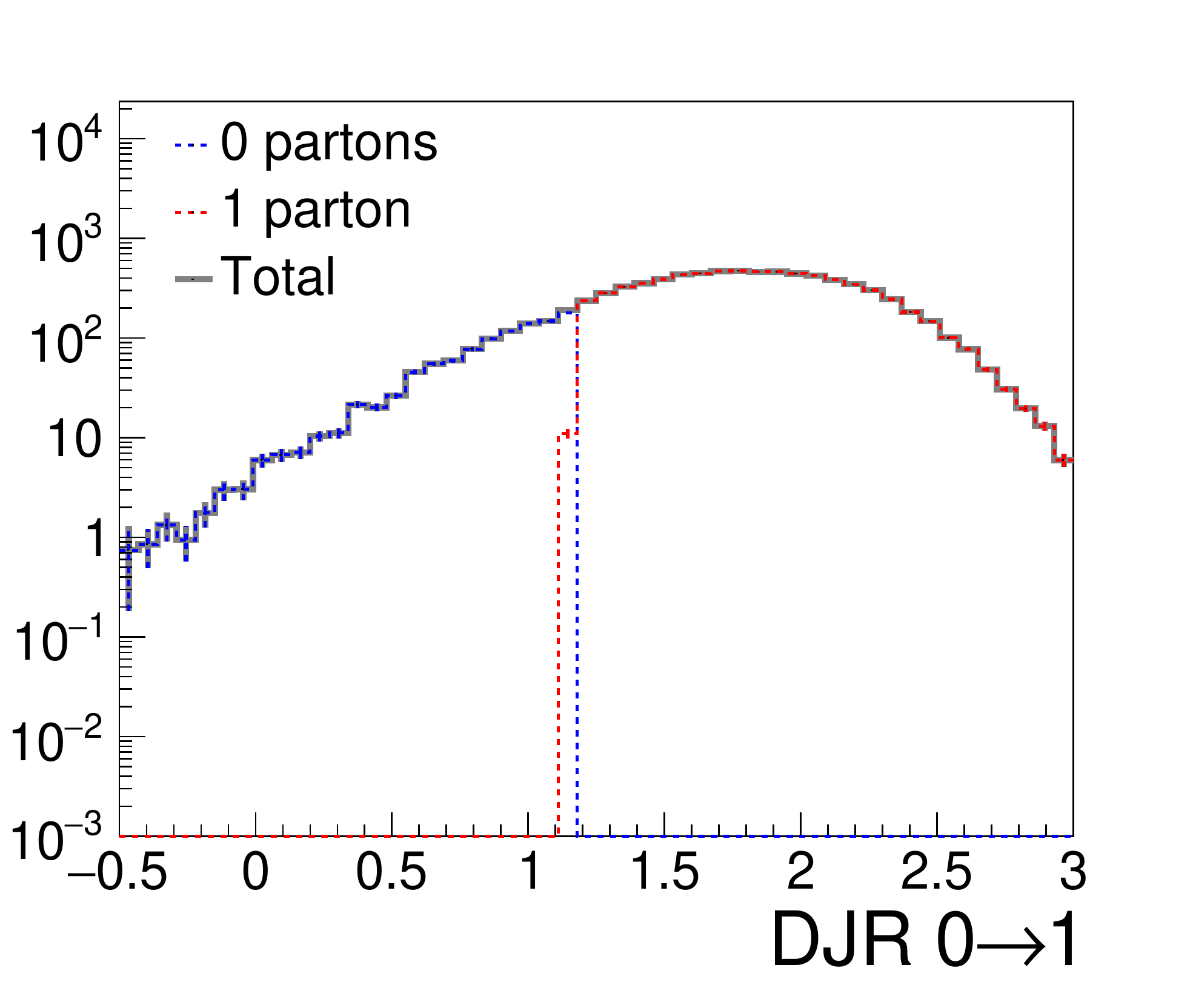}
\includegraphics[width=.32\textwidth]{plots/Final_plots/DJR/ttHJet_HanV4_LitPt_basePtSM.pdf}
\includegraphics[width=.32\textwidth]{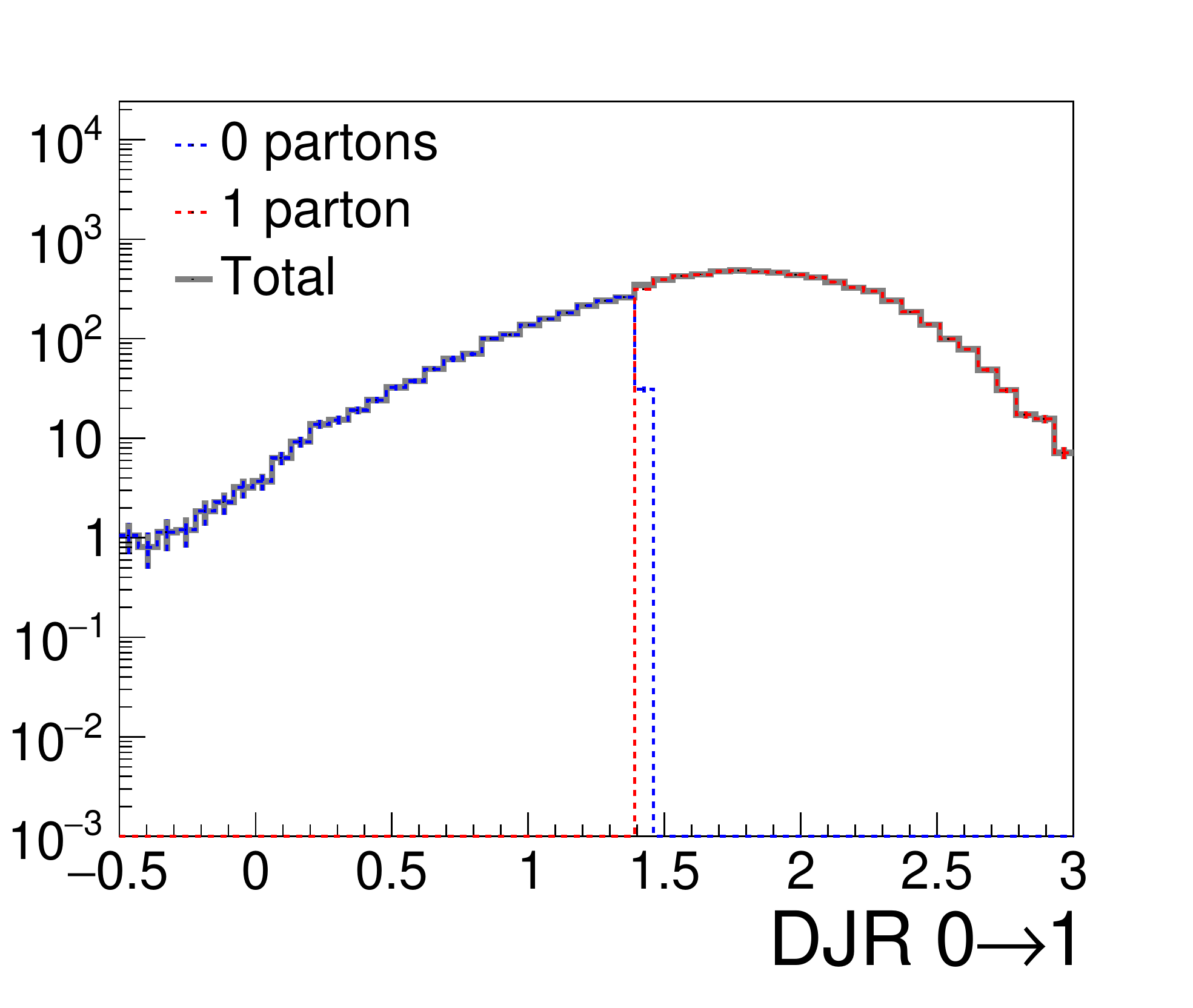}
\caption{\tth DJR plots for $\xqcut=10$, $\qcut=15$ (left), 19 (middle) and 25 (right). Here all Wilson coefficients are set to non-SM values. The x axis and the lines in the plots are the same as described in Figure~\ref{fig:djr_ttH-ttW-ttZ_NOT_ttll-ttlnu}.}
\label{fig:djr_tth_qCutScan}
\end{figure}

\begin{figure} [ht!]
\centering
\includegraphics[width=.32\textwidth]{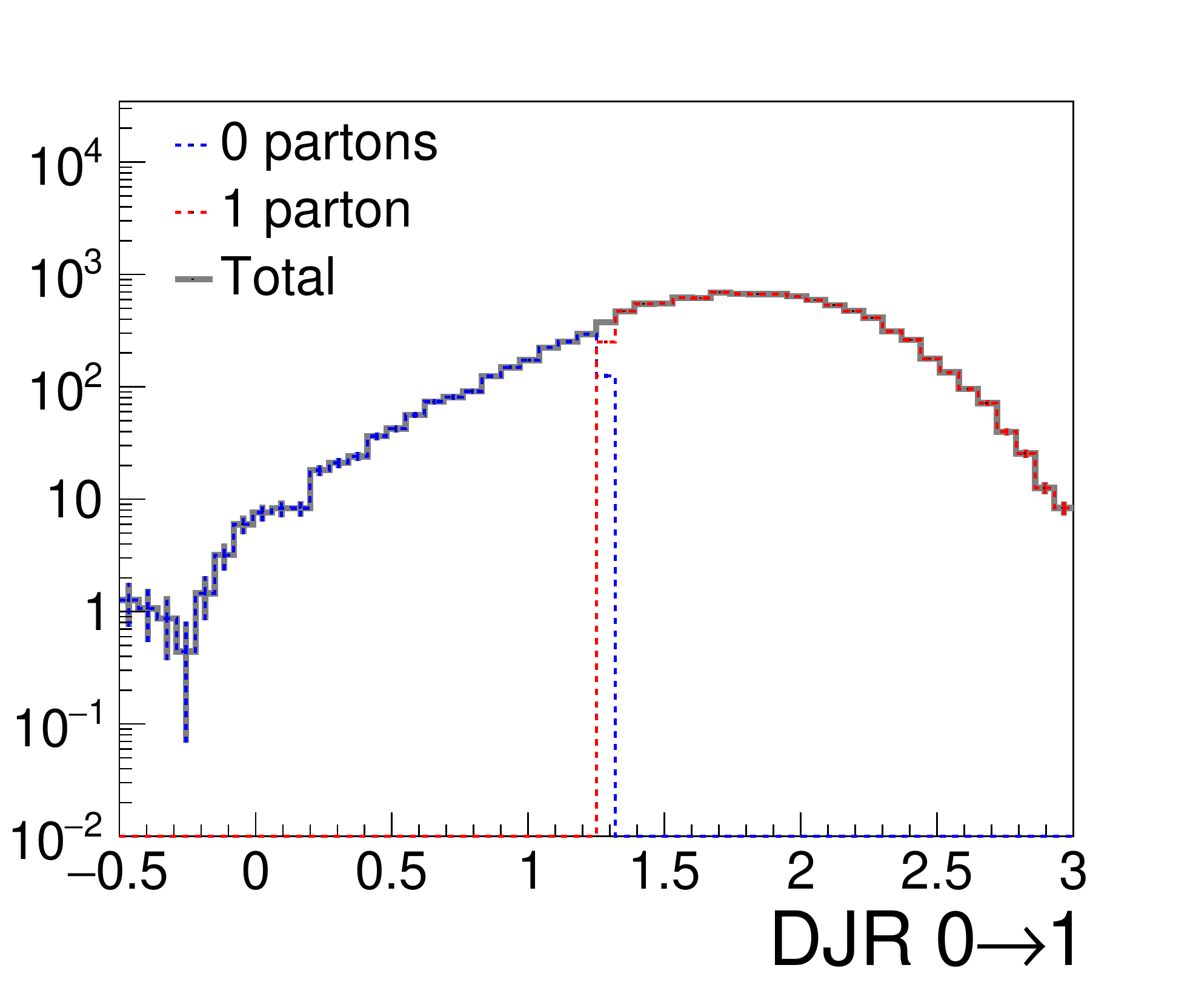}
\includegraphics[width=.32\textwidth]{plots/Final_plots/DJR/ttHJet_HanV4_LitPt_basePtSM.pdf}
\includegraphics[width=.32\textwidth]{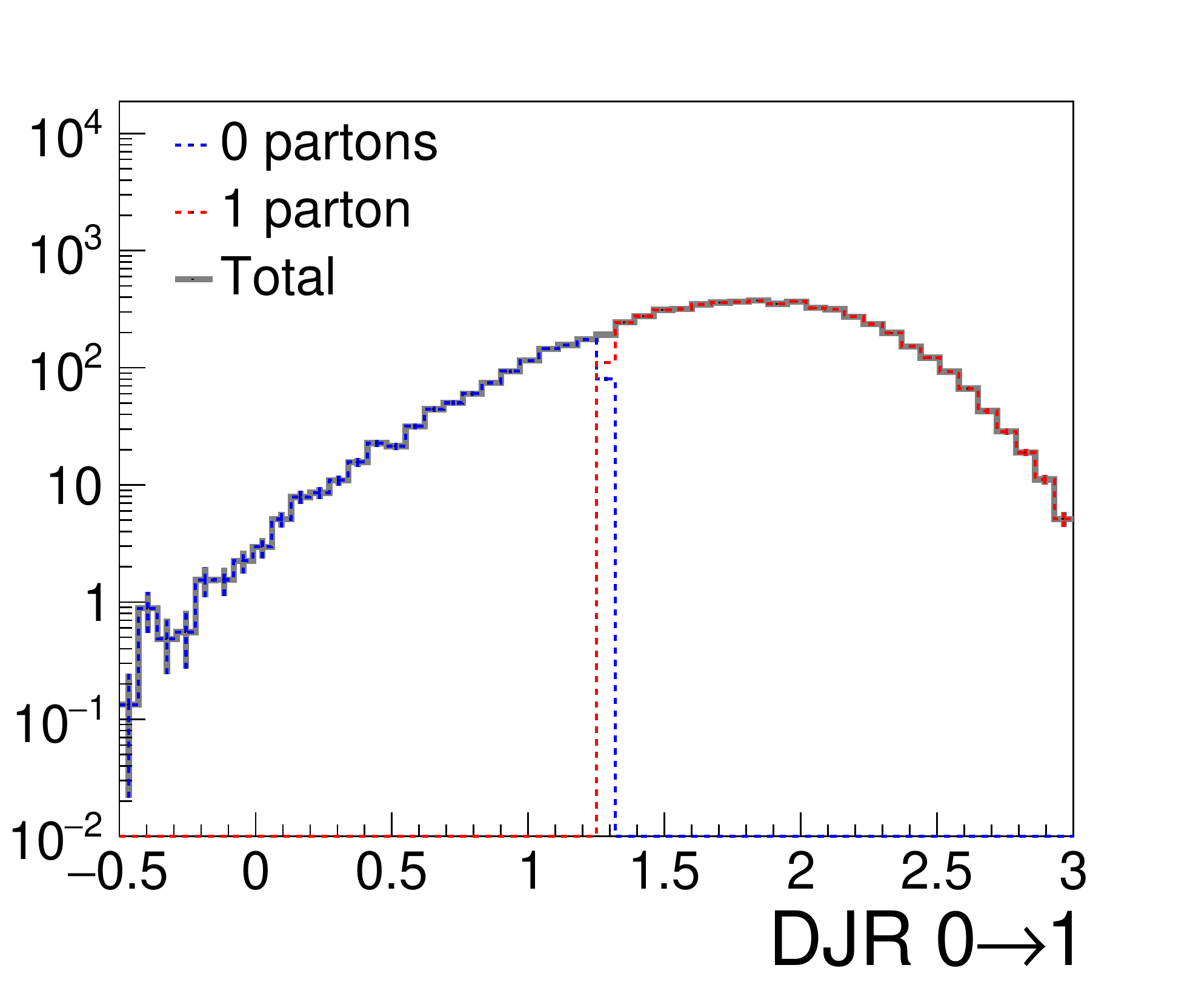}
\caption{\tth DJR plots for $\qcut=19$, $\xqcut=5$ (left), 10 (middle) and 15 (right). Here all Wilson coefficients are set to non-SM values. The x axis and the lines in the plots are the same as described in Figure~\ref{fig:djr_ttH-ttW-ttZ_NOT_ttll-ttlnu}.}
\label{fig:djr_tth_xqcutScan}
\end{figure}

Figure~\ref{fig:djr_ttH-ttW-ttZ_NOT_ttll-ttlnu} shows DJR plots for the three processes considered in this study. For each process, the transition from the 0 parton line to the 1 parton line is smooth, indicating that the matching is being handled appropriately. If the \xqcut and \qcut are varied around their nominal values of $\xqcut=10$, $\qcut=19$, the DJR plots should also remain smooth. Figure \ref{fig:djr_tth_qCutScan} and Figure \ref{fig:djr_tth_xqcutScan} show that the \tth DJR plots remain consistently smooth as we vary the \qcut and \xqcut, respectively.

\begin{figure} [htb!]
\centering
\includegraphics[width=.32\textwidth]{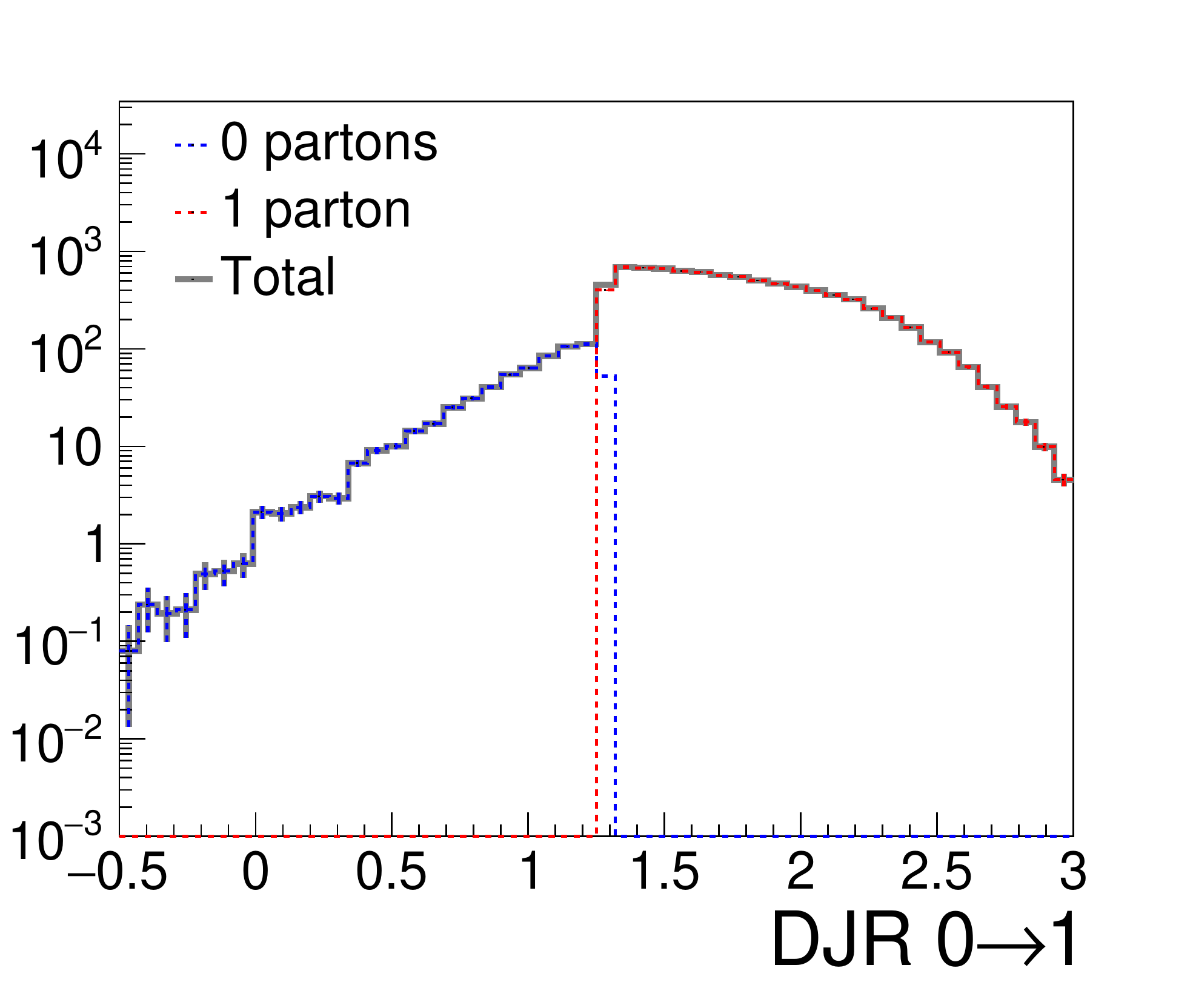}
\includegraphics[width=.32\textwidth]{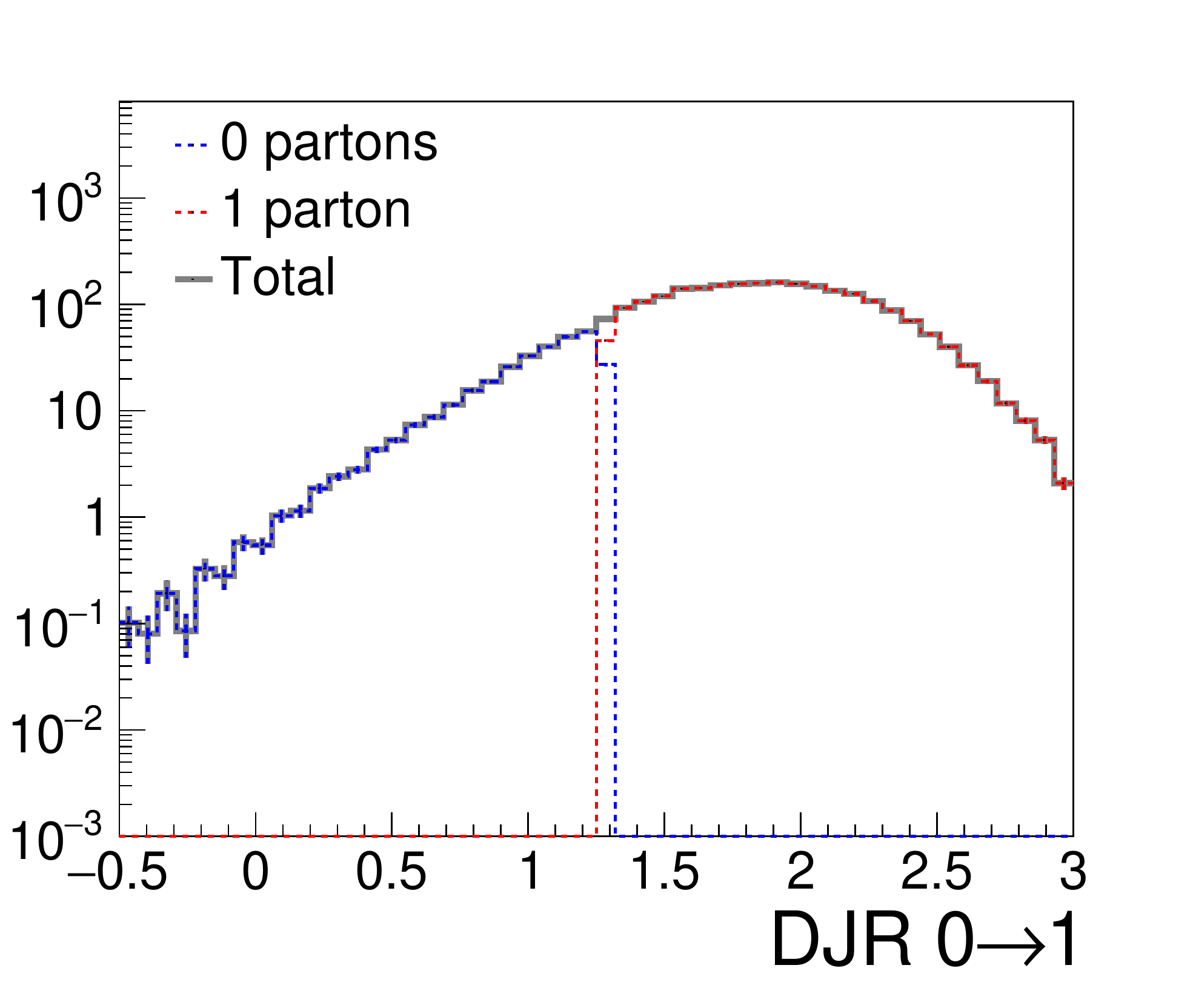}
\caption{DJR plots for \tth with the $g$-$g$-$t$-$t$-$h$ vertex missing (left) and with the vertex included (right). In both plots, the value of \ctG is set to 2, and all other Wilson coefficients are set to 0. The x axis and the lines in the plots are the same as described in Figure~\ref{fig:djr_ttH-ttW-ttZ_NOT_ttll-ttlnu}.}
\label{fig:djr_tth_ttggh}
\end{figure}

In addition to being useful for identifying problems associated with matching and validating matching parameters, DJR plots can also be useful for discovering problems with simulated samples that are unrelated to matching. For example, when we began this study, we were using an older version of the model described in Ref.~\cite{AguilarSaavedra:2018nen} that did not include five particle vertices. This is important for \tth, since the \OuG operator gives rise to a five particle vertex involving two top quarks, two gluons, and a Higgs boson. Before the inclusion of this missing vertex, the \tth DJR plot showed a significant discontinuity when \ctG was set to non-zero values. In fact, this discontinuity is what initially alerted us to the fact the five-particle \ctG vertex was missing from the model file. After the missing \ctG five-particle vertex was added to the model, the \tth DJR plots were found to be smooth even when \ctG was set to non-zero values. The DJR plots produced with and without the five-particle \ctG vertex included in the model are show in Figure~\ref{fig:djr_tth_ttggh}. 

\section{Results}
\label{sec:results}
Having established that including extra partons with matching in an LO calculation provides a valid approach to approximating the impact of higher-order QCD corrections for the operators considered here, we will now look at the results obtained with a matched LO calculation.  In particular, we are interested in evaluating whether there are any cases where a matched LO calculation produces different results from an LO calculation that does not include an additional parton.  After all, if the matched calculation is always essentially equivalent to the simple LO calculation, then it might not be advantageous to introduce the complexities of the matching procedure.  On the other hand, if including an additional parton results in significant differences in some of the EFT predictions, then it will be important to ensure that these contribution are always included.

The LO matched samples used in this section are the same as those described in section~\ref{sec:validation_matching}, and the samples produced without an additional parton are generated similarly, but with the matching turned off. We will also look at a few selected processes and operators at NLO.  The NLO samples are generated with the SMEFT NLO model \cite{Degrande:2020evl}; consistent with the LO samples, we use the PDF set NNPDF3.1 \cite{Ball:2014uwa}, the default dynamical scale choice for the renormalization and factorization scales, and we do not make any parton level cuts.

The range over which each Wilson coefficient is varied is taken from the marginalized limits presented in Ref.~\cite{Hartland:2019bjb}.  We make this choice to ensure that we focus on differences in the predictions in an experimentally relevant range for the Wilson coefficients.  We choose to consider the marginalized limits rather than the individual ones---which tend to be significantly more narrow---because extracting marginalized limits requires that the dependence of the cross section on the Wilson coefficients over that entire range is well modeled.

In our calculations, we consider diagrams involving a single EFT vertex only. As such, the amplitudes will depend linearly (at most) on the Wilson coefficients, and the cross section will depend quadratically:
\begin{eqnarray}
\sigma(c_i) = s_0 + s_1 c_i + s_2 c_i^2,
\label{eq:xsec}
\end{eqnarray}
where $c_i$ is a Wilson coefficient, $s_0$ is the standard model contribution, $s_1$ represents the interference between the EFT and the standard model, and $s_2$ is the pure EFT contribution. In principle, $s_2$ can contain the effects of interference between two different EFT operators. However, in this study we will turn on only one Wilson coefficient at a time, so the EFT cross section is a quadratic function of a single $c_i$.\footnote{Strictly adhering to EFT power counting, keeping the $c^2_i$ terms in the cross section requires we keep the $c^2_i$ terms in the amplitude and add the effects from dimension-8 operators interfering with the SM. As the number of dimension-8 operators is large~\cite{Lehman:2015coa,Henning:2015alf}, a consistent $\mathcal O(c^2_i)$ result is unwieldy, especially as our goal in this work is to show the effects of matching. We will therefore use Eq.~\eqref{eq:xsec} with the understanding that the EFT effects it models do not fully describe the complete SMEFT calculation.}

\subsection{Comparison methodology}
\label{sec:methodology}

As we examine these results, it is important to define carefully what quantities are interesting to consider.  For example, it is well known that the overall normalization obtained from LO calculations, even including the contributions from extra partons with matching, falls short of the NLO value, resulting in the standard practice of applying ``\kfactors'' to correct the normalization of LO samples, for example, when comparing to experimental data.  As we look at the impacts of including extra partons with matching compared to LO without matching or NLO calculations, we do not want these known inclusive normalization differences to obscure more relevant differences in how the various samples model the dependence of the cross section on the Wilson coefficients.  Therefore, in this paper, we will always look at the value of the inclusive cross section at a given Wilson coefficient relative to the SM cross section value, calculated with the same method; for example, the LO matched samples will be normalized to the SM value predicted by the LO matched calculation.  This ratio can be thought of as a ``relative \kfactor'' in that it can be converted into a physical cross section by multiplying by the SM cross section.  We will denote this ``relative \kfactor'' with the symbol $\mu$.  An added benefit to this approach is that sources of uncertainties that just impact the overall normalization, even in the SM case, cancel out, so we can directly visualize the impact of uncertainties on the EFT dependence. 

\begin{figure} [ht!]
\centering
\includegraphics[width=.34\textwidth]{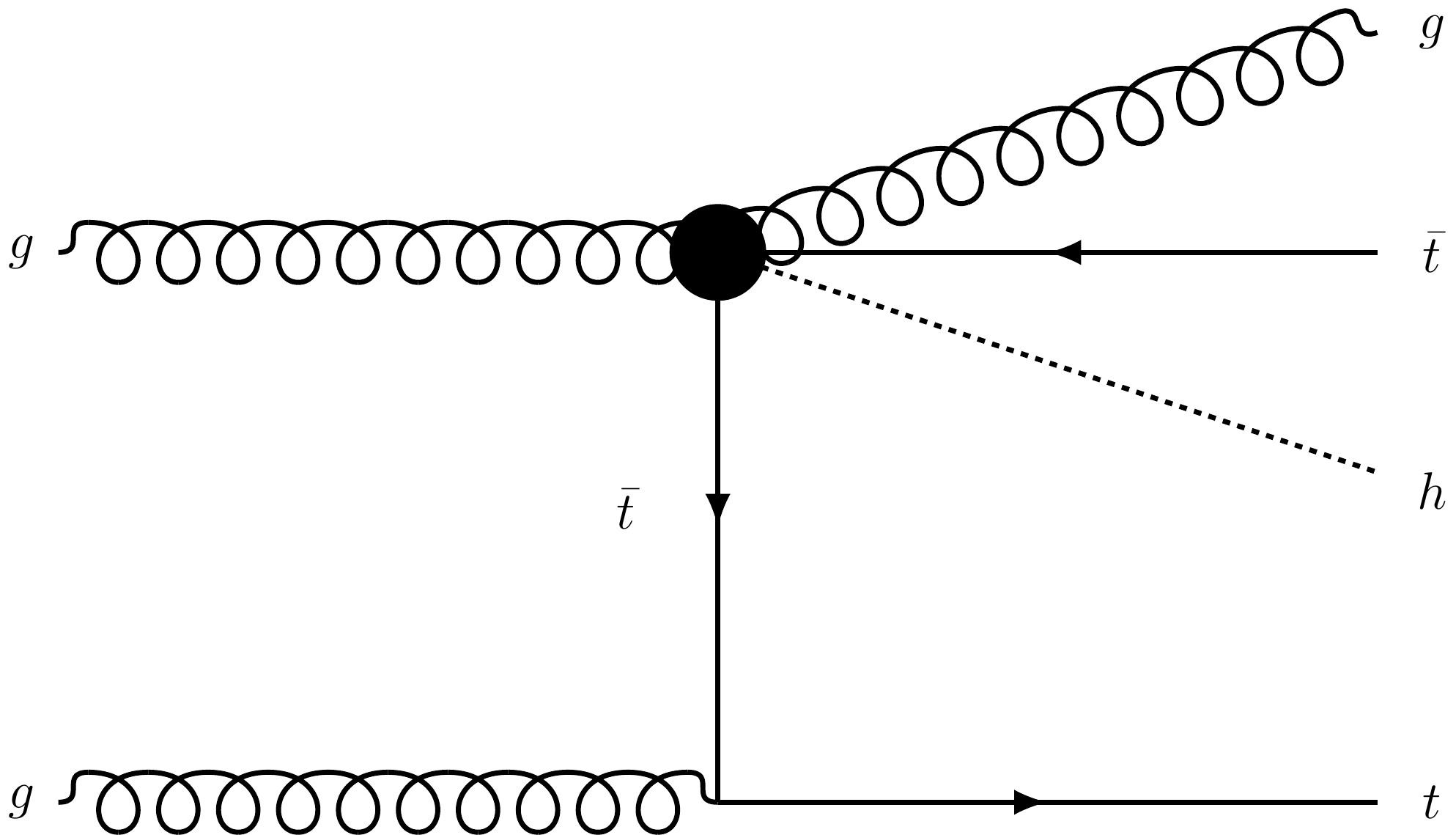} \hskip 8ex
\includegraphics[width=.34\textwidth]{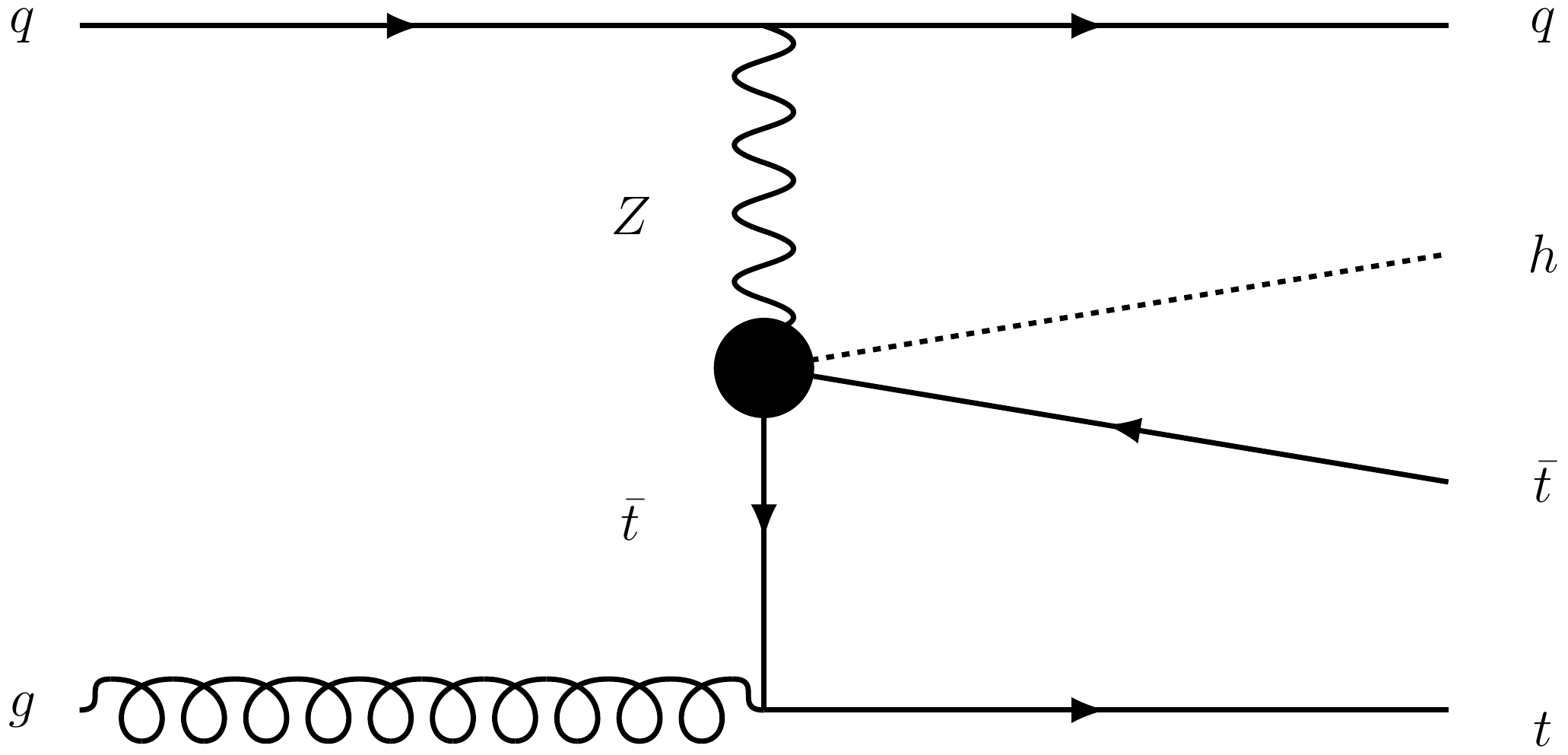}
\caption{Example diagrams that contribute to \tth. The EFT vertex in the diagram on the left is a \ctG vertex, and the QED order of the diagram is 1; we can therefore take this contribution into account at NLO. For comparison, the diagram on the right contains a \cpt vertex, and the QED order of the diagram is 3; we therefore cannot easily take this contribution into account at NLO.}
\label{fig:example_eft_diagrams}
\end{figure}

As a starting point, we use the methodology described above to compare a set of matched LO calculations to full NLO calculations.  The primary motivation for this is to highlight how our method of comparison is insensitive to the overall normalization, while making clear the comparison in inclusive cross section dependence on the Wilson coefficients between different calculations. Due to the challenges described in section~\ref{sec:motivation} associated with including higher-order QED effects in {\MADGRAPH} NLO calculations, we will focus on Wilson coefficients that can enter the processes we consider at the lowest QED order. For example, the effect of \ctG on \tth falls into this category, as \ctG vertices can enter into the \tth process when the QED order is constrained to be less than or equal to 1; an example of a \tth diagram where QED=1 that involves a \ctG vertex is shown in Figure~\ref{fig:example_eft_diagrams}. Combinations of operators and processes where the operator cannot contribute to the process at the lowest QED order are not the focus of this section due to the complications associated with modeling these effects at NLO.

\begin{figure} [ht!]
\centering
\includegraphics[width=.32\textwidth]{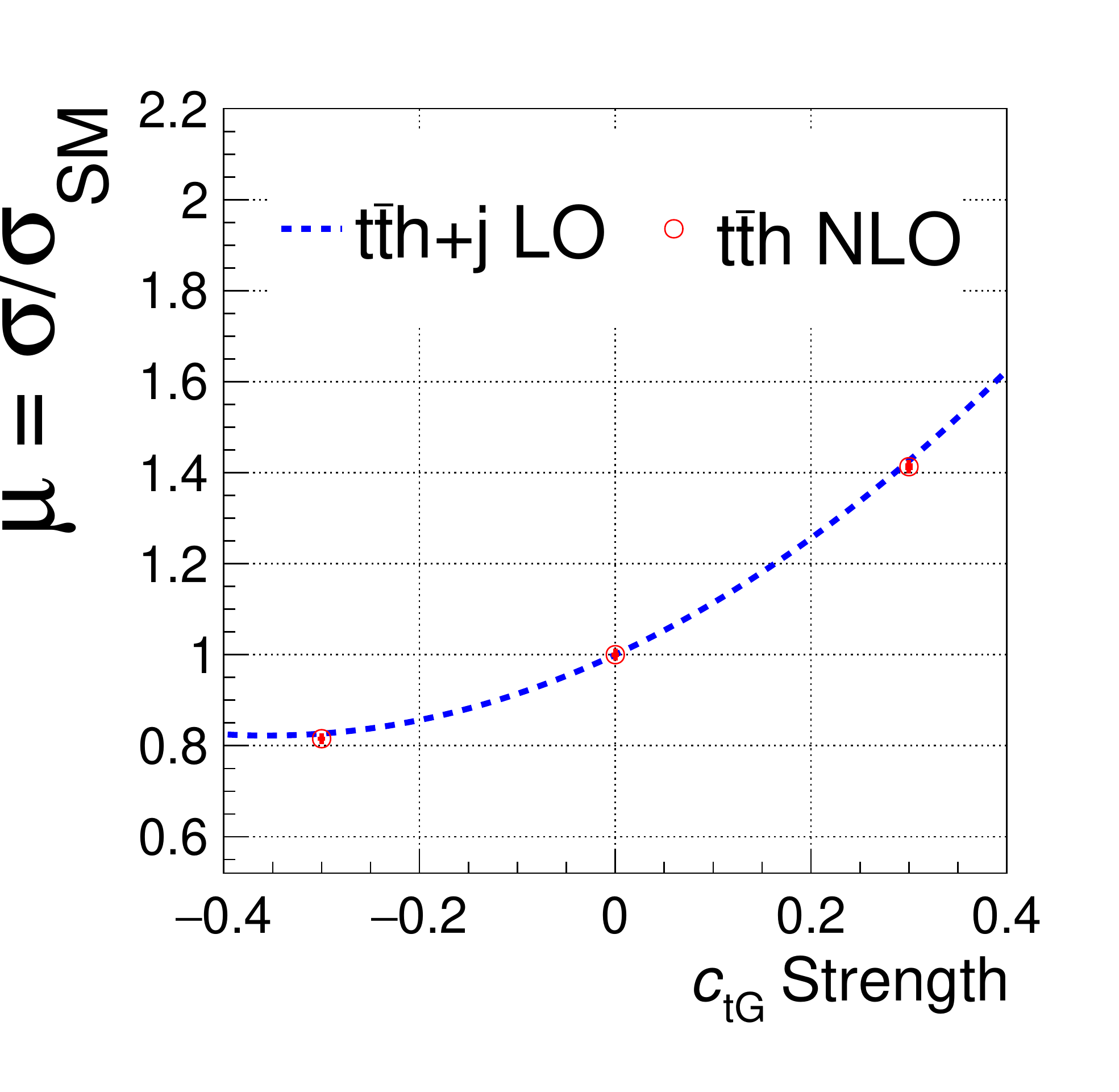}
\includegraphics[width=.32\textwidth]{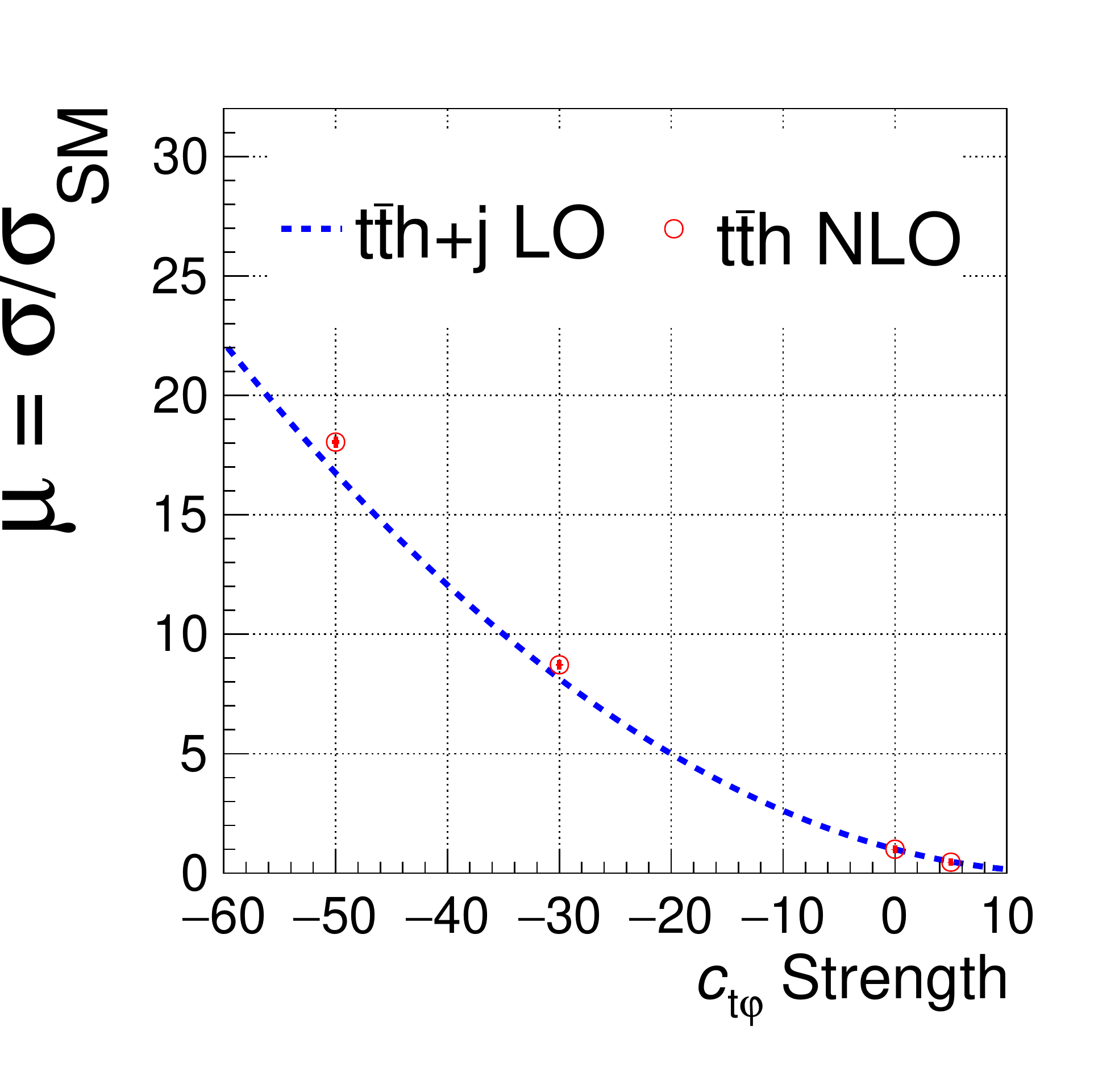}
\includegraphics[width=.32\textwidth]{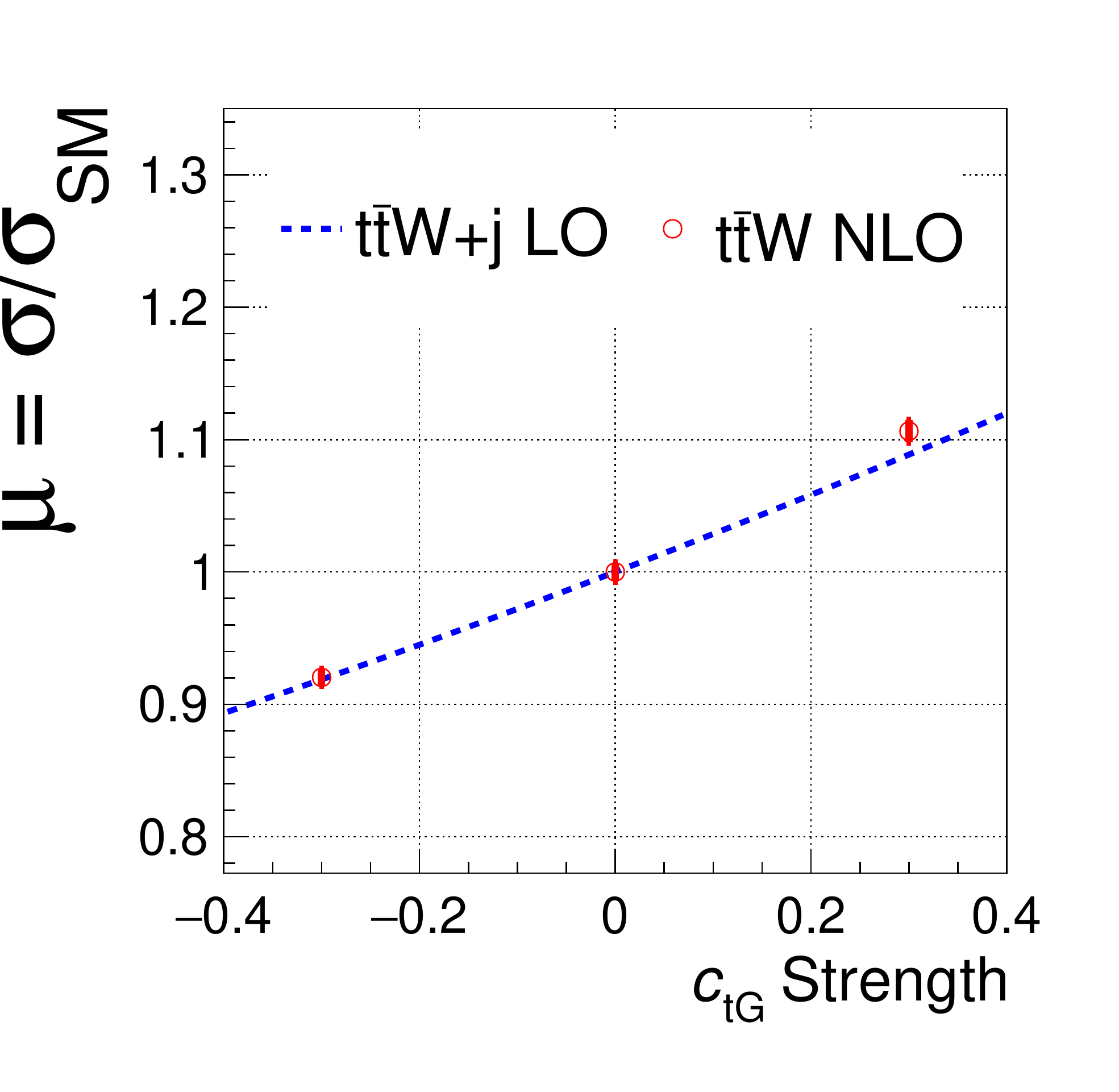}
\includegraphics[width=.32\textwidth]{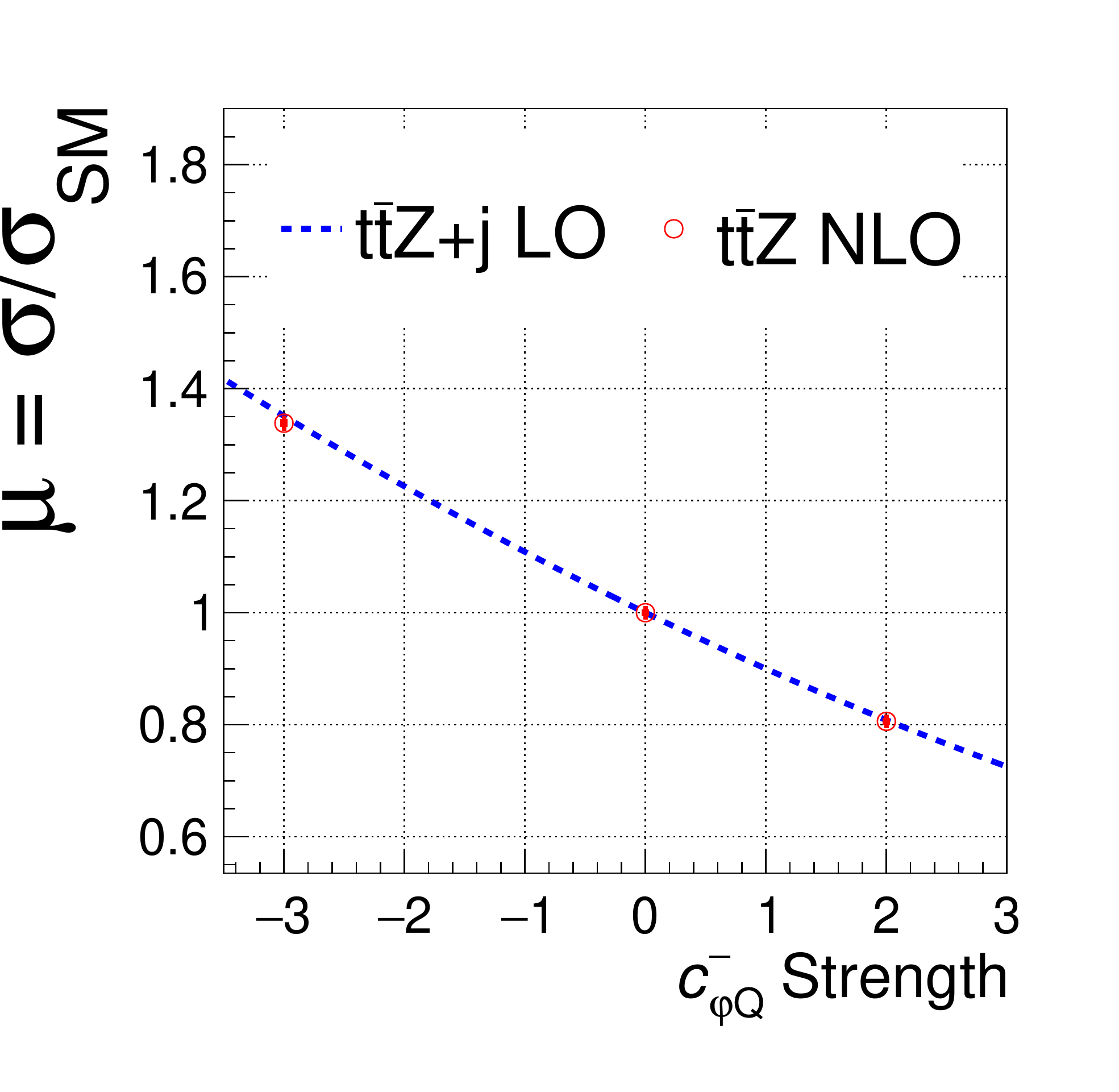}
\includegraphics[width=.32\textwidth]{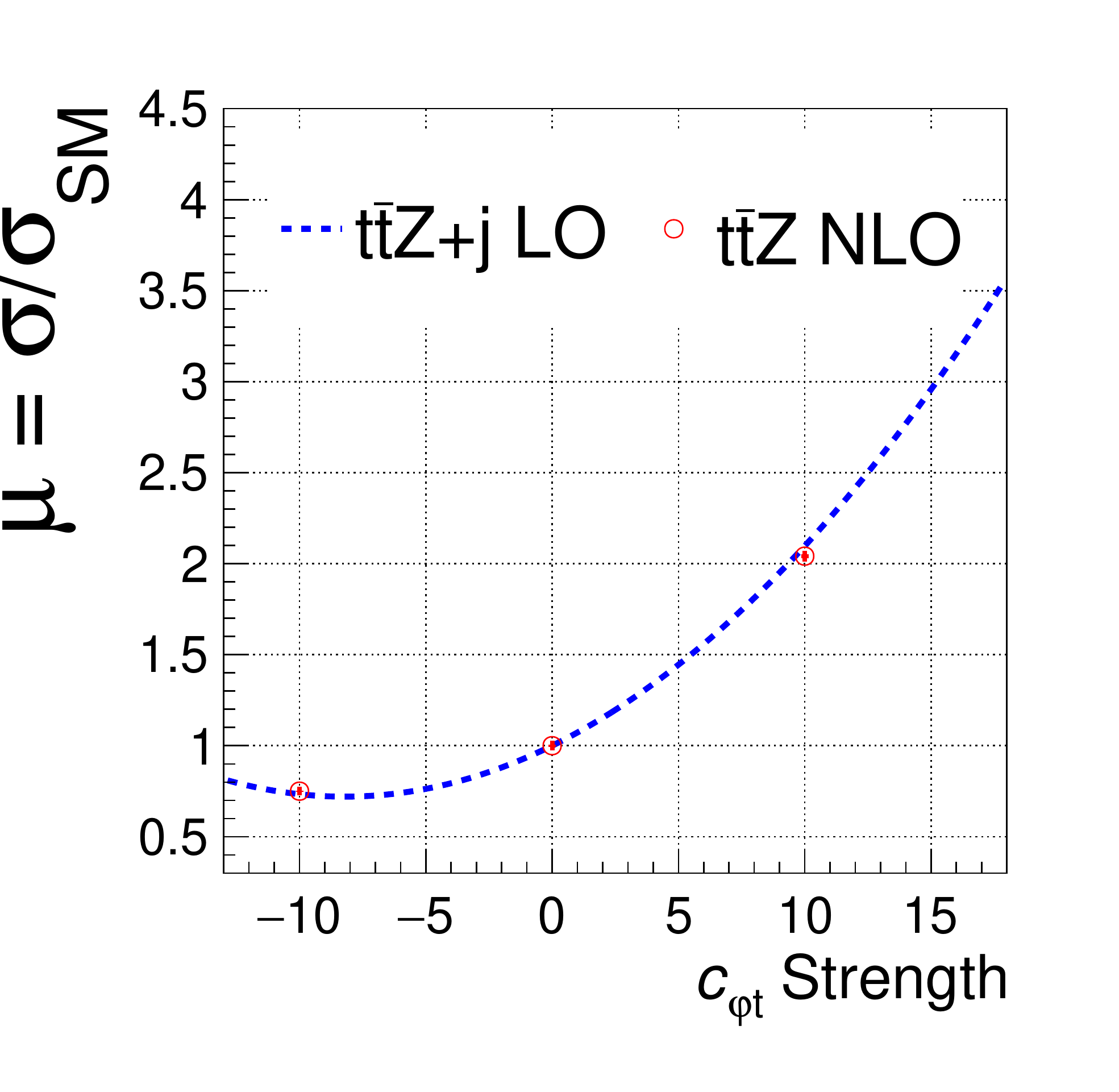}
\includegraphics[width=.32\textwidth]{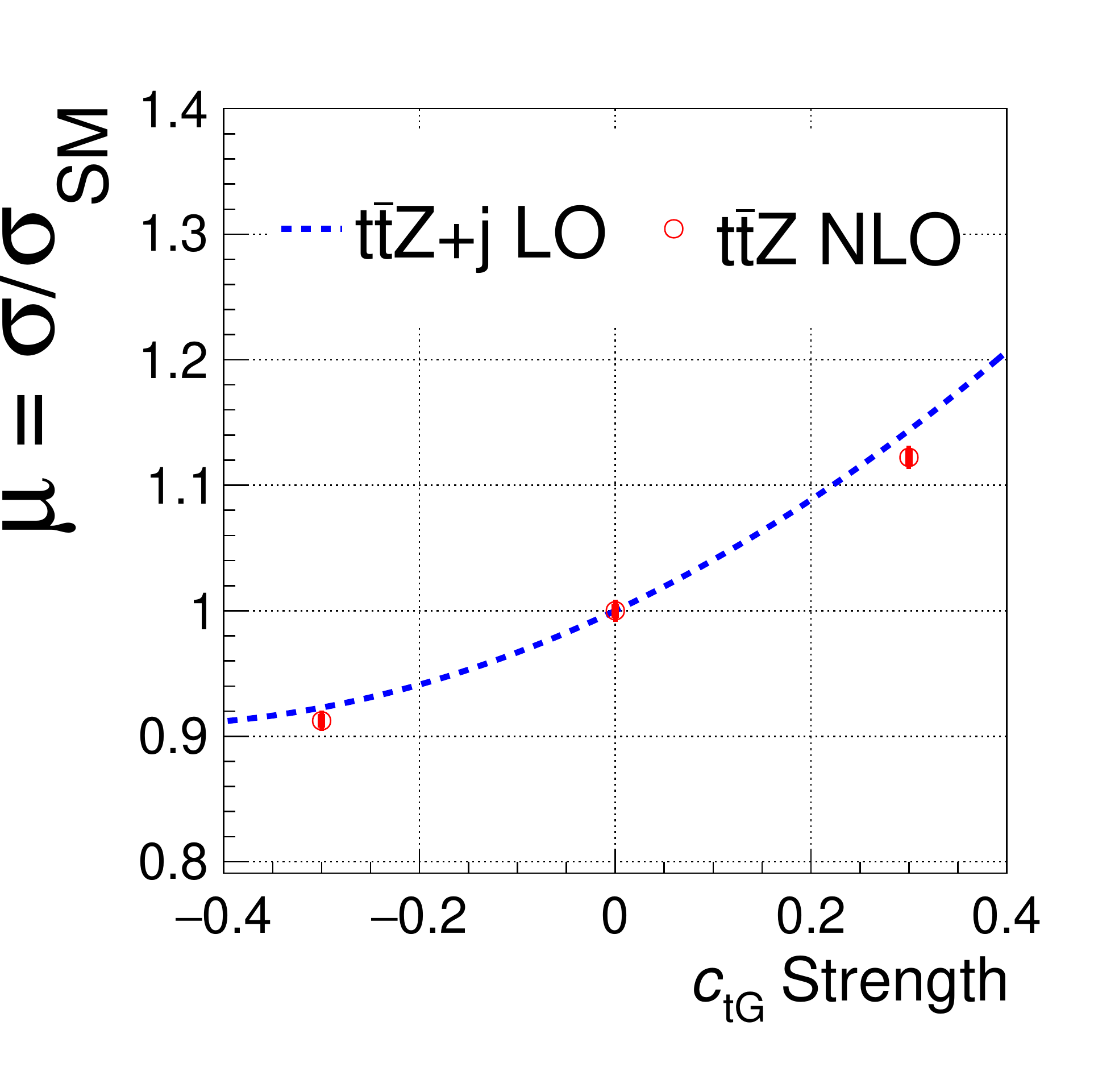}
\caption{Example quadratic cross section plots showing the LO samples (produced with an additional parton) compared with the NLO samples. The y axis shows the ratio of the cross section to the SM cross section; the LO fit is scaled by the LO SM prediction, and the NLO points are scaled by the value of the SM NLO point. As described in the text, the range for each Wilson coefficient is taken from the marginalized limits presented in Ref.~\cite{Hartland:2019bjb}.}
\label{fig:results_NLO_vs_LO_cross_section_comp}
\end{figure}

The results of matched LO predictions compared with NLO predictions are shown in Figure~\ref{fig:results_NLO_vs_LO_cross_section_comp}.  As mentioned above, we plot the ratio of the inclusive cross section at a given Wilson coefficient value ($\sigma$) to the cross section predicted by the same calculation for the SM point ($\sigma_\mathrm{SM}$), and refer to this ratio as $\mu$.  Therefore, the y axis  indicates the relative scaling of the cross section with respect to the SM as we vary the given Wilson coefficient.  For the matched LO calculation, the results are shown as curves generated by fitting a quadratic function to inclusive cross sections calculated at a number of different Wilson coefficient values.  The MC statistics are sufficiently large such that the statistical error on the fitted curves is negligible. Since generating NLO MC samples is significantly more computationally expensive (as discussed in Appendix~\ref{sec:appendix}), we calculate the NLO cross section at just three distinct points (always including the SM as one of the points) and plot those points (scaled to the NLO SM point) with the statistical uncertainty for each point indicated with an error bar.

In general, Figure~\ref{fig:results_NLO_vs_LO_cross_section_comp} shows that the dependence of the inclusive cross section on the Wilson coefficients, relative to the SM prediction, are in good agreement between the matched LO and NLO calculations for the processes and operators shown here.  Of course, one should remember that the overall normalization of the two calculations is different, but by comparing ratios, that difference drops out.  Therefore, any differences visible in the plot of $\mu$ represent a difference in the quadratic dependence of the cross section on the Wilson coefficient. The largest discrepancies between matched LO and NLO are associated with the dependence of \tth on \ctp, and are 7\% or less for the range of NLO points calculated. This amount of discrepancy is reasonable given that the matched LO calculation includes only additional radiation contribution to the corrections, but not the loop contributions. 

The fact that the NLO and matched LO calculations agree reasonably well in terms of the inclusive cross section as a function of the Wilson coefficient relative to the SM expectation is perhaps not surprising; it is also known~\cite{Maltoni:2016yxb,Bylund:2016phk} that the relative dependence of the cross section for $\tth$ and $\ttZ$ on these operators calculated at LO without matching is in reasonable agreement with the NLO calculation as well. As mentioned above, this is essentially equivalent to saying that the NLO \kfactor is roughly the same for the SM, EFT, and interference contributions.


\subsection{The EFT effects of including an extra parton in the LO calculation}
\label{sec:results_LO_extra_partons}

In this section, we will compare the LO samples produced with and without matching in order to determine if there are any combinations of operators and processes for which the inclusion of an additional jet with matching changes the observed dependence of the inclusive cross section on the Wilson coefficient.
This can be viewed as an interesting way to probe when an NLO \kfactor applied to a LO calculation might not be a suitable way to model the EFT dependence. Furthermore, because matched LO calculations are computationally easier to obtain, especially when higher-order QED corrections are involved, we can use these calculations to explore cases that are not currently feasible at NLO. 

To avoid preconceptions, we check each of the operators of interest for the \tth, \ttW, and \ttZ processes over the ranges from Ref.~\cite{Hartland:2019bjb}, as described above.  For each combination of operator and process considered, we compare the dependence on the inclusive cross section normalized to the SM expectation as a function of the operator's Wilson coefficient.  The cross section for the matched sample ($\sigma_{\ttXj}$) scaled by that sample's SM expectation is referred to as \muPlusOneP. The cross section for the sample calculated without an extra parton ($\sigma_{\ttX}$) normalized to that sample's SM prediction is referred to as \muZeroP. Since these samples are all generated at LO, we place no limits on the QED or QCD order of the diagrams included by \MADGRAPH. Using each processes' quadratic parameterizations, we find the normalized cross section at the positive and negative limits of the ranges from Ref.~\cite{Hartland:2019bjb}, and if \muPlusOneP is more than 10 percent different than \muZeroP, we classify the combination of process and coefficient as being significantly impacted by the inclusion on an additional parton.

The processes showing large differences for each operator are summarized in Table~\ref{tab:eft-ops-results}. Note that in the case of \OuB, while for \tth both \ctW and \ctZ show a significant impact, for \ttZ, only \ctZ shows a large difference according to the criteria defined above; additionally, we note that for \Opqa, only \cpQa has a large impact on ttW according to the criteria above. Section~\ref{sec:results_discussion} provides a discussion of these differences.

\begin{table}[hbt!]
\caption{List of operators and corresponding Wilson coefficients considered in this paper indicating for which of the processes we observe a large \muPlusOneP/\muZeroP ratio as described in the text. For operators that are associated with two coefficients, if the dependence on either of the coefficients is significantly impacted by the inclusion of an additional parton, the process is listed.}
\begin{center}
\begin{tabular}{cccc}
Operator \rule{0pt}{3.0ex} & Definition & Wilson Coefficient & Processes with large \muPlusOneP/\muZeroP \\ \hline
\Oup  & \OupDef  & $\ctp$          & \ttW \\
\Opqa & \OpqaDef & $\cpQM + \cpQa$ & \ttW \\
\Opqb & \OpqbDef & $\cpQa$         & \ttW \\
\Opu  & \OpuDef  & $\cpt$          & \tth, \ttW \\
\Opud & \OpudDef & $\cptb$         & - \\
\OuW  & \OuWDef  & $\ctW$          & \tth \\
\OdW  & \OdWDef  & $\cbW$          & - \\
\OuB  & \OuBDef  & $ (\cW \ctW - \ctZ)/\sW$ & \tth, \ttZ \\
\OuG  & \OuGDef  & $\ctG$  & - \\

\end{tabular}
\label{tab:eft-ops-results}
\end{center}
\end{table}

\begin{figure} [htb!]
\centering
\includegraphics[width=.32\textwidth]{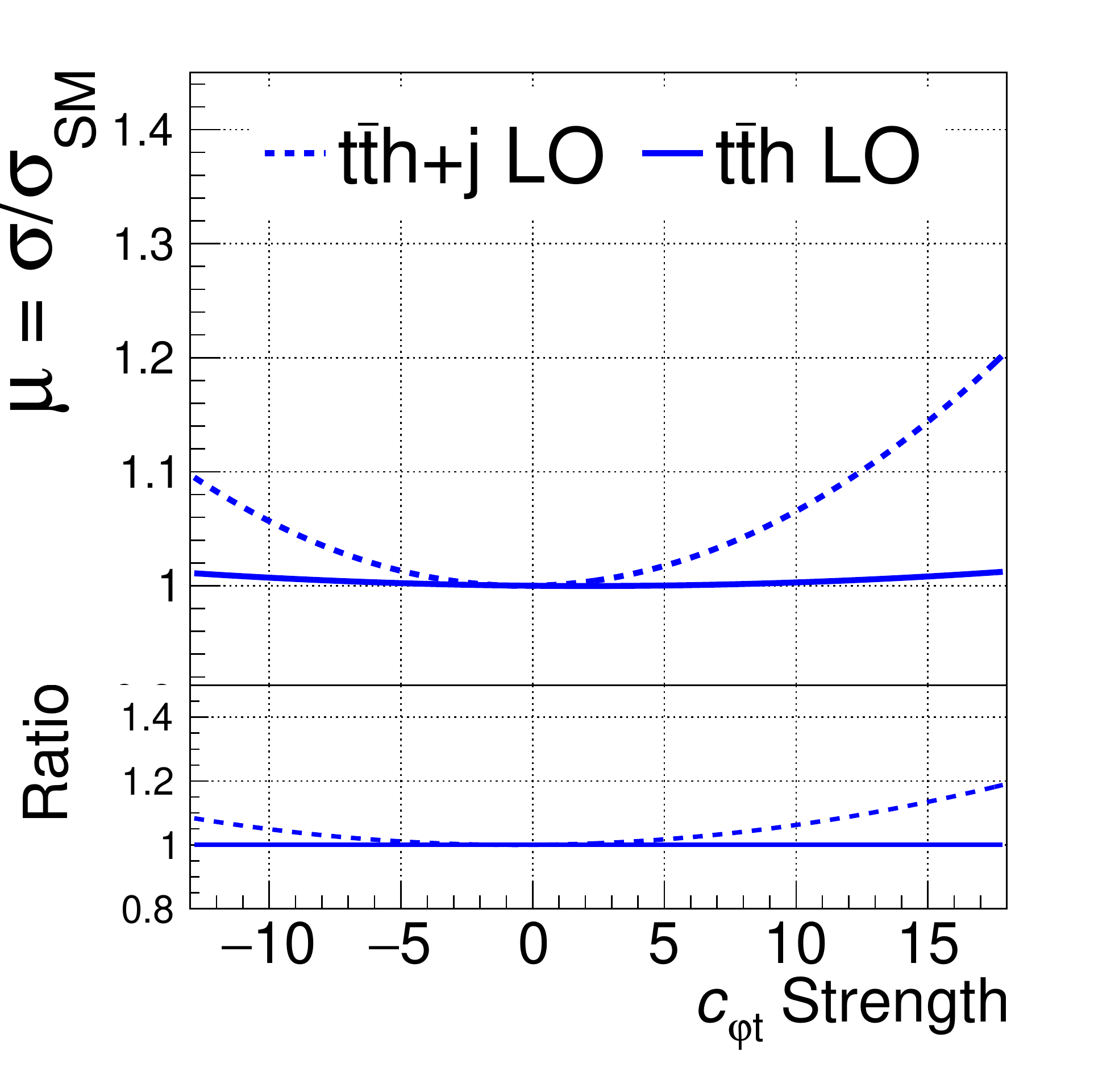}
\includegraphics[width=.32\textwidth]{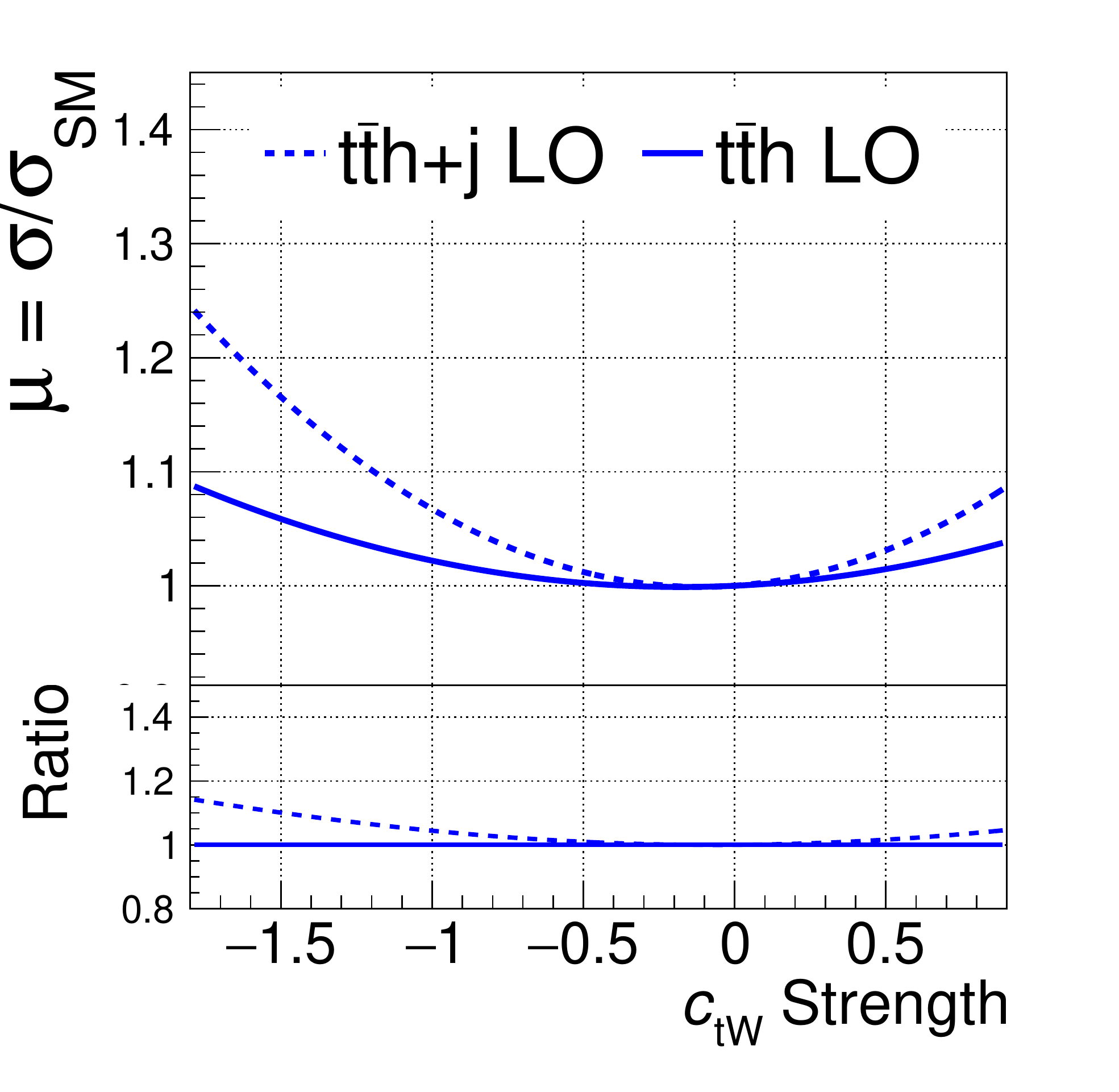}
\includegraphics[width=.32\textwidth]{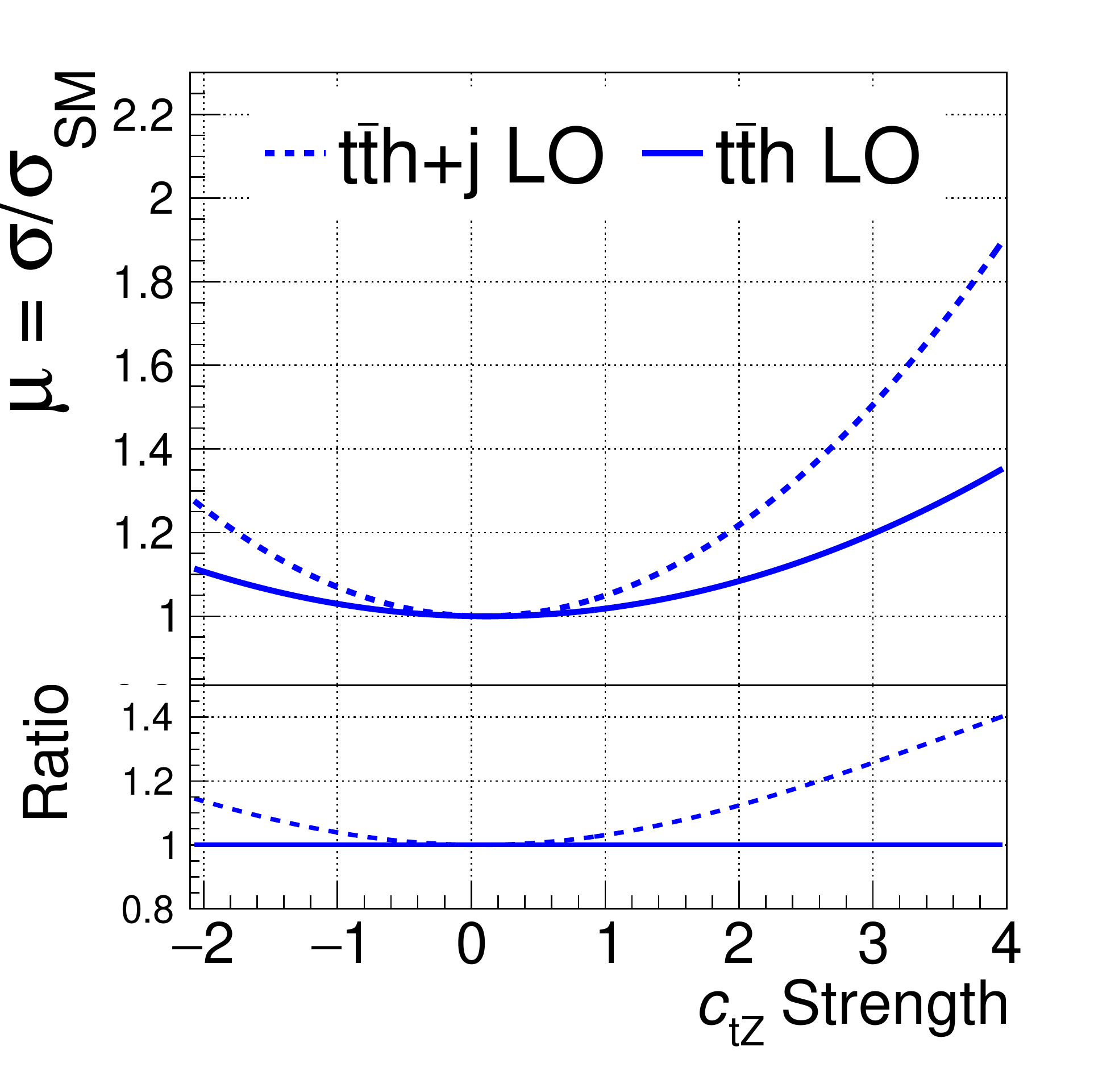}
\includegraphics[width=.32\textwidth]{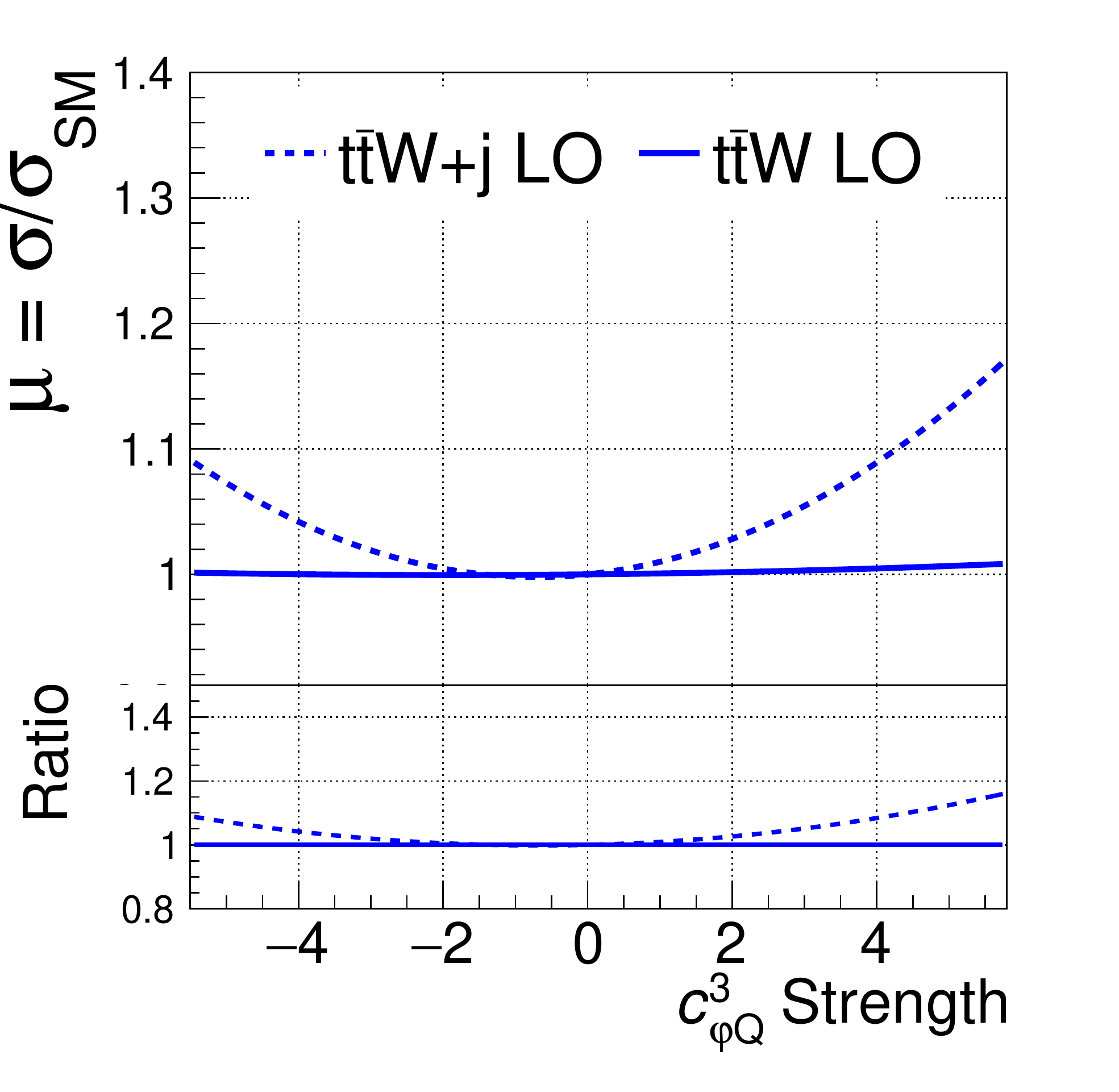}
\includegraphics[width=.32\textwidth]{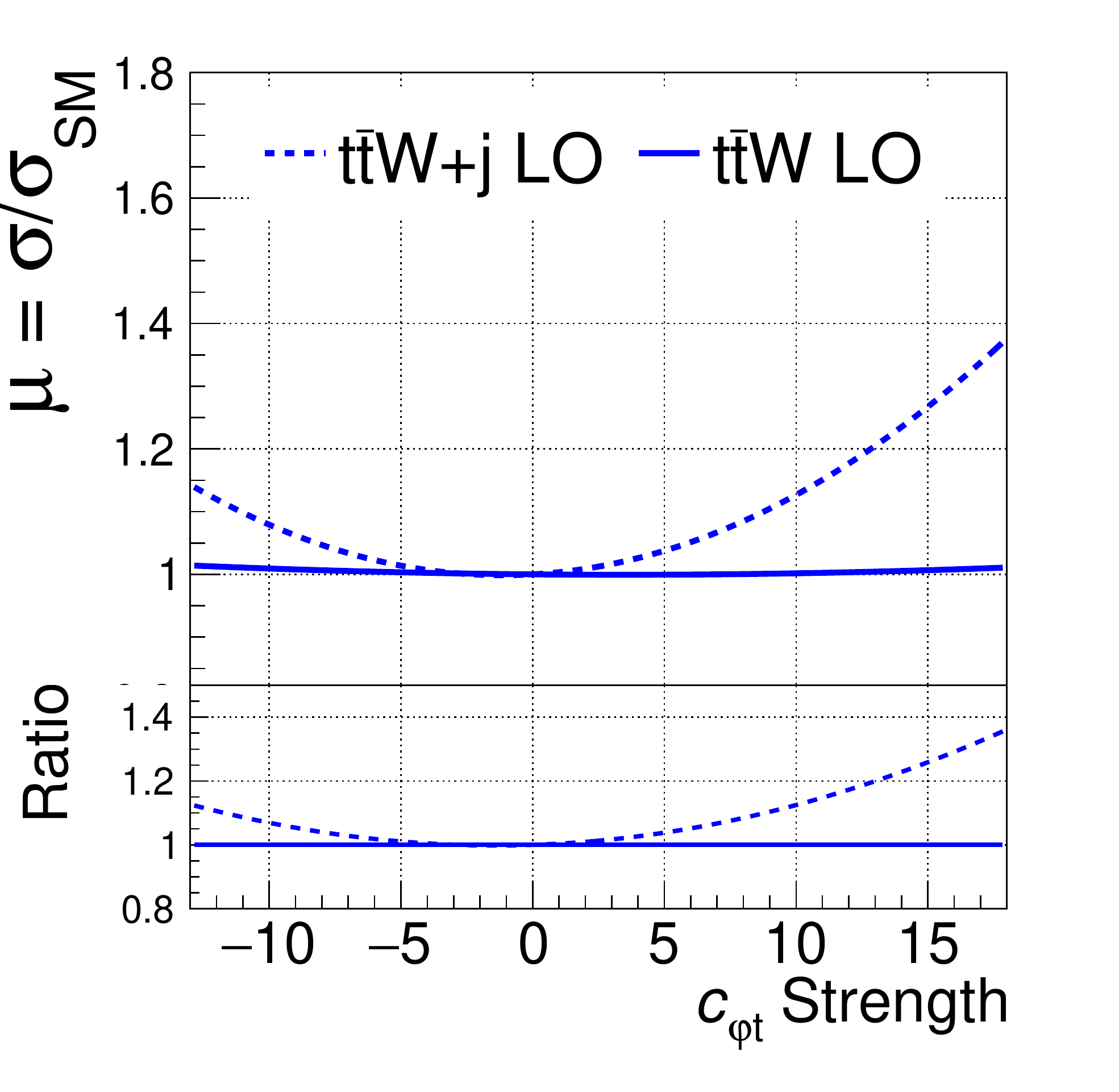}
\includegraphics[width=.32\textwidth]{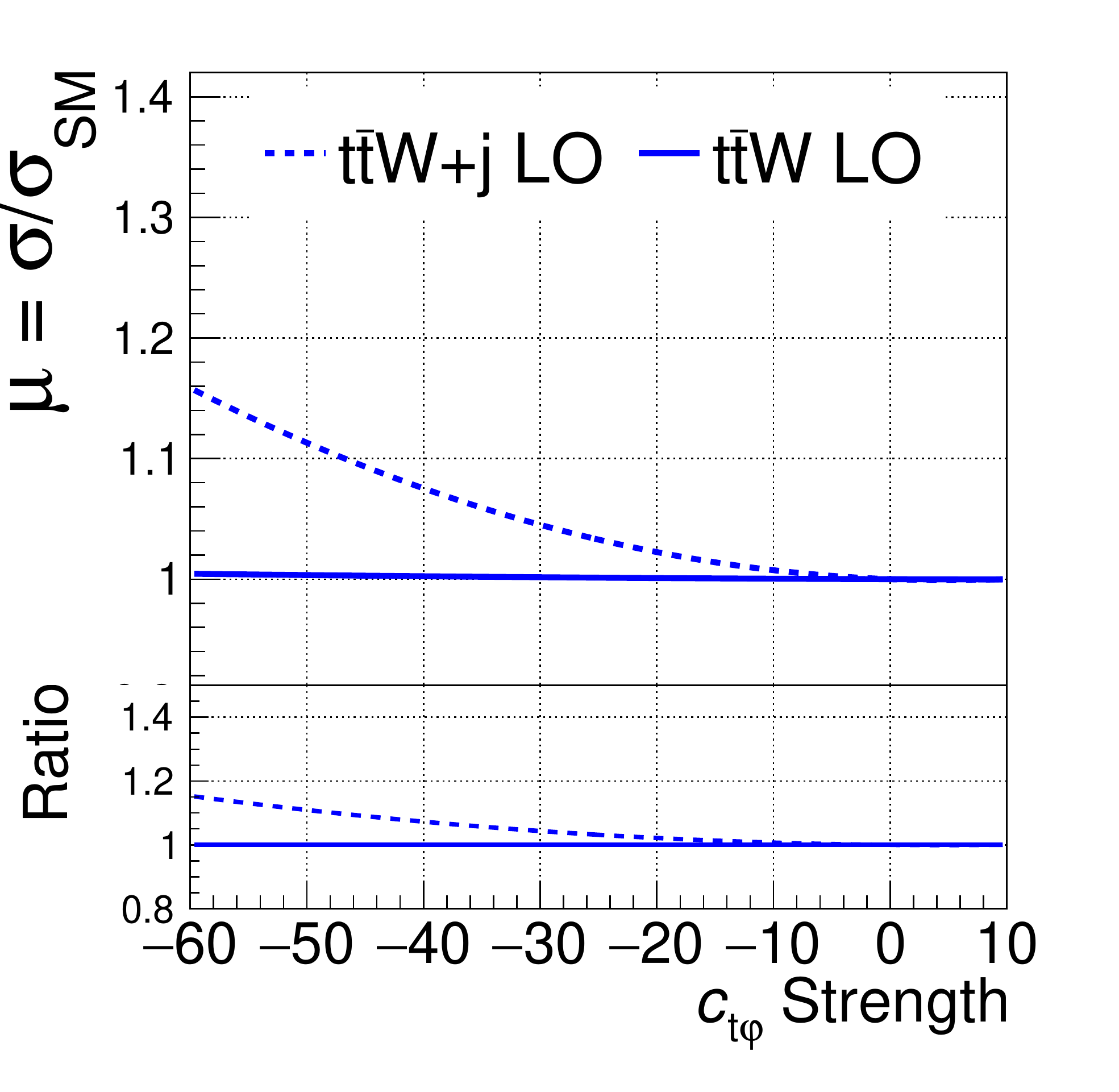}
\includegraphics[width=.32\textwidth]{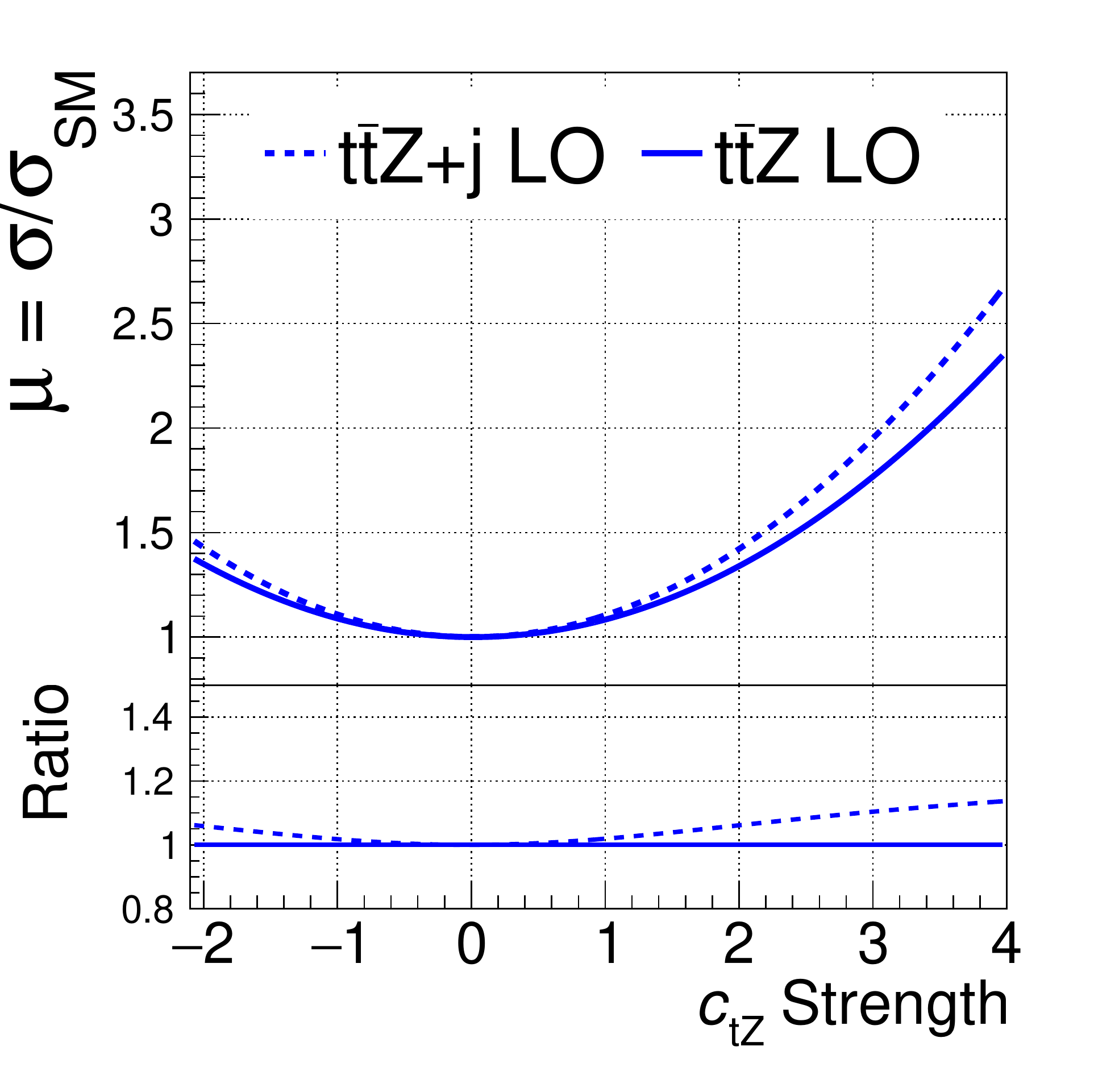}
\caption{Inclusive cross section for \tth, \ttW, and \ttZ (with and without an additional parton) as a function of the Wilson coefficient. The y axis shows the ratio of the EFT prediction to the SM. The quadratic parameterizations are generated by fitting a quadratic function to inclusive cross sections calculated at a number of different Wilson coefficient values; the MC statistics are sufficiently large such that the statistical error on the fitted curves is negligible. The ratio plots show the \muPlusOneP/\muZeroP ratios. As described in the text, the range for each Wilson coefficient is taken from the marginalized limits presented in Ref.~\cite{Hartland:2019bjb}.}
\label{fig:quad0p1p_ttX}
\end{figure}

Plots of the inclusive cross section relative to the SM as a function of the relevant Wilson coefficient for these seven combinations of operator and process are shown in Figure~\ref{fig:quad0p1p_ttX}.
Within the ranges of Wilson coefficient values considered, the largest \muPlusOneP/\muZeroP ratios are seen in the effect of \ctZ on \tth and the effect of \cpt on \ttW. Tables~\ref{tab:mu_ratios} lists the \muPlusOneP/\muZeroP ratios (evaluated at the upper and lower limits identified in Ref. \cite{Hartland:2019bjb}) for each of the seven combinations of processes and coefficients for which the inclusion of an additional parton significantly changes the dependence of the process on the given coefficient. 
The coefficients for the \ttX and \ttXj quadratic fits for all of the Wilson coefficients considered in this paper are shown in Appendix~\ref{sec:appendix_fit_coeffs}. 

\begin{table}[hbtp]
  \caption{Comparison of \muPlusOneP and \muZeroP at the limits identified in \cite{Hartland:2019bjb} for \tth, \ttW, and \ttZ. Wilson coefficients whose \muPlusOneP/\muZeroP ratio is greater than 1.10 at either the lower or upper limit are included in the table.}
  \centering
  \begin{tabular}{cccc}
   Process \rule{0pt}{3.0ex}  & WC &  Limits from \cite{Hartland:2019bjb}  &  \muPlusOneP/\muZeroP at limits \\ \hline
   \tth \rule{0pt}{3.0ex}& \cpt & [-13, 18] & 1.09, 1.19 \\
   \tth & \ctW  & [-1.8, 0.9] & 1.14, 1.05 \\ 
   \tth & \ctZ  & [-2.1, 4.0] & 1.15, 1.41 \\
   \ttW \rule{0pt}{3.0ex}& \cpQa  & [-5.5, 5.8] & 1.09, 1.16 \\
   \ttW & \cpt  & [-13, 18] & 1.13, 1.36 \\ 
   \ttW & \ctp  & [-60, 10] & 1.15, 1.00 \\
   \ttZ \rule{0pt}{3.0ex}& \ctZ & [-2.1, 4.0] & 1.06, 1.14 \\
  \end{tabular}
  \label{tab:mu_ratios}
\end{table}


\subsection{Discussion}
\label{sec:results_discussion}

In Section~\ref{sec:results_LO_extra_partons}, we performed a comparison between the LO samples produced with and without matching, finding that the ratio \muPlusOneP/\muZeroP is large for several of the Wilson coefficients and processes considered in this paper. In this section, we will discuss some of the factors that may contribute to how large of an impact the inclusion of an additional parton has on the dependence of a particular process' cross section for a given Wilson coefficient. After introducing these factors in general context, we will discuss how they may apply to the specific combinations of processes and operators identified in Section~\ref{sec:results_LO_extra_partons}.

The first factor we consider is how the parton distribution functions (PDFs) in the processes involving an additional parton compare to the PDFs in processes that do not involve an additional parton. Only quark-anti-quark (\qqbar) and gluon-gluon (\gluglu) and initial states are possible for \tth and \ttZ without an additional parton, and only \qqbarprime initial states are available for \ttW without an extra parton. However, the inclusion of an additional parton allows for new gluon-quark (\qglu) initiated diagrams to contribute in all cases. At the relevant LHC energy ranges, contributions from \gluglu and \qglu initiated processes are significantly larger than \qqbar initiated processes, so the new \qglu initiated diagrams that become available with an extra parton can have a large effect on the cross section, especially in cases where there were no \gluglu initiated diagrams available without an extra parton. This enhancement can affect both the SM and the EFT contributions, so the expected impact on the \muPlusOneP/\muZeroP ratio depends on the initial states available to each. Stepping through each case in turn, we will discuss the potential impacts on the \muPlusOneP/\muZeroP ratio: 
\begin{itemize}
    \item Case 1: Without an additional parton, there are already \gluglu initiated SM diagrams, but only \qqbar initiated diagrams are available for the given Wilson coefficient. In this situation, we would not expect the SM cross section to be strongly impacted by the inclusion of the additional parton, because the new \qglu initiated contributions represent only an incremental increase to the \gluglu initiated contribution already present without an extra parton. However, we would expect to see a significant impact on the EFT dependence, since in this case, the EFT contribution only arises for \qqbar initiated diagrams without an extra parton, and would thus expect an enhancement in \muPlusOneP, and a correspondingly large \muPlusOneP/\muZeroP ratio.  
    \item Case 2: Without an additional parton, there are only \qqbar initiated diagrams for both the SM and EFT contributions. In this situation, the new \qglu initiated diagrams should have a substantial impact on both the SM and EFT contributions, leading to a larger \ttXj cross section for both. However, we would not necessarily expect the increase in the SM cross section to fully ``cancel" the increase in the EFT's impact on the cross section in the \muPlusOneP/\muZeroP ratio. The contribution of the EFT diagrams can grow with energy, making their contribution more significant than the corresponding SM diagrams; furthermore, there will generally be a larger number of new EFT diagrams than new SM diagrams, since the EFT contribution can include not only diagrams that are topologically equivalent to SM diagrams, but also new diagrams (with e.g. four-particle vertices) that have no SM counterparts. For these reasons, we may still expect to see an increase in \muPlusOneP, though it might not be as large as in Case 1.
    \item Cases 3: Without an additional parton, there are already \gluglu initiated SM diagrams and \gluglu initiated EFT diagrams for the given Wilson coefficient. In this case, the impact of the new \qglu diagrams would not necessarily be large for either the SM for the EFT contributions. However, even if neither enhancement is large in absolute terms, the increase in the EFT contribution may still be larger than the increase in the SM contribution. Thus, similar to Case 2, the relative enhancement could still lead to a significant \muPlusOneP/\muZeroP ratio, even if it is not as large as would be expected in Case 1.
\end{itemize}

As mentioned in Case 2 above, another potential contribution to the differences between \muPlusOneP and \muZeroP is related to the energy scaling of the EFT operators. Since all dimension-six coefficients are suppressed by $1/\Lambda^2$, a quantity with units of energy squared is required in the numerator to keep the coupling dimensionless; these powers of energy can either come from the Higgs vacuum expectation value (\vev), or from the energy flowing through the vertex ($\sim$\sqrtshat). Numerators containing \sqrtshat will scale with the energy of the process, and numerators proportional to \shat will depend even more strongly on the energy of the process, so at high energies, the contributions from these types of vertices will be enhanced with respect to contributions proportional to $\vev^2$. The inclusion of an additional parton allows us to put more energy into the system, i.e. a $\ttX$ system can now recoil against an ISR jet, so we may expect to observe larger \muPlusOneP/\muZeroP ratios for contributions involving a stronger dependence on the energy scale of the process. To determine the energy scale dependence of a particular vertex, we should examine the structure of the operator that gives rise to the vertex of interest. For example, we note that both \ctZ and \cpt have a \ttZhvertex vertex, but the operator associated with \ctZ contains one Higgs field, while the operator associated with \cpt contains two. Because of this, the \ctZ \ttZhvertex vertex does not contain a \vev, and thus scales as $\shat/\Lambda^2$. On the other hand, for the \cpt \ttZhvertex vertex, one of the Higgs fields is set to its vev, and the vertex thus scales as $\vev\sqrtshat/\Lambda^2$. Since the \ctZ contribution scales like \shat, it will grow faster with increasing energy than the \cpt contribution, and we may potentially see a larger impact from the inclusion of an additional parton.

The energy scaling is also tied to the topology of the diagrams available for a given Wilson coefficient and process, which can  play an additional role in the relative impact of the coefficient. For example, if all of the diagrams for a particular Wilson coefficient involve off-shell s-channel propagators, we would not expect the effect of these diagrams to be large, since the contribution of each diagram should be suppressed by a factor of $1/\shat$. Including an additional parton can allow new t-channel diagrams to enter; though the propagators in these diagrams may still be off-shell, their contributions could be larger than the contributions from the s-channel diagrams due to the angular dependence in the denominator of the t-channel propagators. In cases like these, we may therefore expect that the given coefficient could have a significantly lager impact on the \ttXj process than on the \ttX process.

As noted in Equation~\ref{eq:xsec}, EFT effects can impact the cross section both linearly (i.e. through interfere with the SM) and quadratically. While the factors outlined above apply to both contributions, there are also several important factors that are only relevant to the interference effects. For example, let us first consider the chirality of the final state fermions. Although this factor would not lead to differences between \ttX and \ttXj, understanding the role it plays can be important in determining how an operator's linear term may or may not contribute to both \ttX and \ttXj, and is thus useful in understanding the full picture a given operator's impact on a given process. In SM \tth production, the top quarks are of opposite chirality (i.e. one top quark is a $t_R$ while the other is a $t_L$). For SM \ttW and \ttZ processes, the top quarks are of the same chirality. However, the processes involving EFT vertices contain both same-chirality and opposite-chirality contributions, so in the cases where the chirality does not agree with the SM chirality, interference with the SM will be suppressed. Since these contributions enter proportional to the mass of the given particle, the suppression will be less significant for interactions involving top quarks of the wrong chirality, and more significant for interactions involving bottom quarks of the wrong chirality. Another factor that can affect the SM interference is the color structure of the final state fermions. Except for \ctG, the operators considered in this paper contain EW bosons, so the \ttbar pairs produced by these operators are color singlets. These final states cannot interfere with color octet pairs, such as those produced by gluon splitting. However, when we account for additional radiation in the final state, we have the potential for more complex color structure, which may open up interference with some SM contributions and lead to a larger \muPlusOneP compared to \muZeroP.

In this section, we will step through each of the three processes and discuss how the factors outlined above may affect the cross section's dependence on the relevant Wilson coefficients when an additional parton is included in the process. We note that this discussion does not represent a definitive explanation for the \muPlusOneP/\muZeroP ratios; due to the competing influences of the multifarious effects described above, it is challenging to determine which combinations of processes and coefficients will be strongly impacted by the inclusion of an additional parton. However, since we have shown that the extra parton can affect the predicted EFT dependence in many cases, we argue that one should consider accounting for the additional parton whenever possible to avoid inadvertently neglecting significant EFT contributions.

\subsubsection{Discussion of \tth}
\label{sec:results_discussion_tth}

In section~\ref{sec:results_LO_extra_partons} we found that the three Wilson coefficients whose effects on \tth significantly change when an additional parton is included are \ctW, \ctZ, and \cpt. For all three of these instances, there are no \gluglu initiated EFT diagrams available without an extra parton (corresponding to Case 1 described above), so we may expect the new \qglu initial states to significantly impact on the EFT dependence. Furthermore, the new \qglu diagrams include t-channel diagrams that do not suffer from the 1/\shat suppression incurred by the off-shell s-channel propagators present in the \qqbar initiated diagrams. While we posit that the combination of these factors can explain the large \muPlusOneP/\muZeroP  ratio observed for these coefficients, there are subtleties to this argument that should be addressed. 

Since the operators associated with \cpQa and \cpQM are structurally very similar to the operator associated with \cpt, we might initially expect to see a corresponding increase in the cross section's dependence on these coefficients when considering an additional parton, but no such increase is reported in section~\ref{sec:results_LO_extra_partons}. For \cpQa, the apparent discrepancy can be explained by the fact that the quark in the new \qglu initiated diagrams must be a b quark, so the effect is suppressed in comparison to the case of \cpt, since the new \qglu diagrams for \cpt do not have this constraint. However, this is not the case for \cpQM, as the quarks in the \qglu initial states are not required to be b quarks; the apparent discrepancy is instead resolved by noting that the limits on \cpQM are much tighter than the limits on \cpt. When we consider a comparable range of values for the coefficients, we see a comparable difference between the \tth and \tthj curves, and there is thus no discrepancy between the effects we observe for \cpQM and \cpt.

\subsubsection{Discussion of \ttZ}
\label{sec:results_discussion_ttZ}

In section~\ref{sec:results_LO_extra_partons} we found that the only Wilson coefficient whose effect on \ttZ significantly changed with the inclusion of an additional parton is \ctZ. To understand why this is the case, we begin by considering the initial states that become available when an additional parton is included. For \ctZ, the additional parton allows new \qglu initiated diagrams (where the quark does not have to be a bottom quark). However, there are \gluglu initiated \ttZ diagrams involving \ctZ already available without an additional parton, so we may expect the impact of including an additional parton to be smaller than those reported for \tth. It should be noted that \ctZ is not the only coefficient for which the inclusion of an additional parton opens up new \qglu initiated diagrams where the quark does not have to be a bottom quark; \ctG, \ctW, \cpt, \cpQM, and \ctp all share this feature. Furthermore, for \ctW and \ctp, there are no \gluglu contributions involving these coefficients when an extra parton is not included, so we may expect the effect of the additional parton to be even more significant than for \ctZ. 

 To understand the cause of these apparent discrepancies, we will first consider the limits on the relevant coefficients. The limits on \ctG and \ctW are indeed tighter than the \ctZ limits, and when comparable ranges are considered, the \muPlusOneP/\muZeroP ratio are in fact comparable to (or even larger than) the \ctZ ratio as we would expect. However, the limits on \cpt, \cpQM, and \ctp are comparable to or looser than the \ctZ limits, so a different explanation is required. For \cpt and \cpQM, let us consider the energy dependence. As discussed previously, the operators associated with \cpt and \cpQM involve two Higgs fields, while the operator associated with \ctZ only contains one; though all three coefficients are associated with a \ttZvertex vertex, the \ctZ vertex will scale more strongly with energy, so we may expect to see larger impact for \ctZ when including an additional parton. 
 
 The final coefficient to discuss is \ctp. Contributing to \ttZ through a \tthvertex vertex, this coefficient has essentially no impact on \ttZ before the inclusion of an additional parton, and only a small effect after. Without the additional parton, all \ctp diagrams that are not initiated by a bottom quark pair require an s-channel Z propagator, and since the energy flowing through the Z propagator is much larger than the mass of the Z, we would not expect these diagrams to have a large contribution. Including an additional parton allows new t-channel diagrams involving \ctp to participate in the process, so their contributions may be larger than the contributions from the s-channel diagrams due to the angular dependence in the denominator of the t-channel propagators. This is consistent with the observation that \ctp has a larger impact on \ttZj than on \ttZ. However, the effect is still much smaller than that of \ctZ, which can be understood by considering the energy dependence of the relevant vertices. As shown in Table~\ref{tab:eft-ops-results}, the operator associated with the \ctp coefficient contains three Higgs fields, so the \ctp \tthvertex vertex will be proportional to $\vev^2/\Lambda^2$, while the \ctZ vertices scale as either $\vev\sqrtshat/\Lambda^2$ or $\shat/\Lambda^2$; it is therefore expected that the impact of \ctZ on \ttZ would be more significant than that of \ctp.

\subsubsection{Discussion of \ttW}
\label{sec:results_discussion_ttW}

For \ttW, we note that the inclusion of an additional parton leads to new \qglu initiated diagrams (where the quark does not have to be a bottom quark) for all nine of the coefficients considered in this paper. As \ttW does not involve any \gluglu initiated contributions, we might expect that the inclusion of an additional parton should have a non-negligible impact for all of the Wilson coefficients we consider. This scenario corresponds to Case 2 described above. 

In section~\ref{sec:results_LO_extra_partons}, we noted that within the limits identified in Ref.~\cite{Hartland:2019bjb}, the additional parton indeed leads to a significant change in the dependence of \ttW on \cpQa, \cpt, and \ctp. It should be noted that \ctW is also strongly affected, but the comparatively tight limits on the coefficient yield a \muPlusOneP/\muZeroP ratio that does not meet the 10\% cutoff used in section~\ref{sec:results_LO_extra_partons} when we evaluate the ratio at these tight upper and lower limits.\footnote{This upper limit, and all of \ttW, may deserve more attention given the observed discrepancy between the observed cross sections~\cite{ATLAS-CONF-2020-013, CMS-PAS-HIG-19-008} and the state-of-the-art SM calculations~\cite{Broggio_2019,frixione2015electroweak,frederix2020subleading,vonBuddenbrock:2020ter}.}

However, there are two coefficients that are notably not impacted by the additional parton; namely, both \cbW and \cptb have no effect on \ttW either before or after the inclusion of the additional parton. To understand why this is the case, we first note that the only vertices involving these coefficients that can enter \ttW or \ttWj are \tbWvertex vertices. As shown in the operator definitions listed in Table~\ref{tab:eft-ops-results}, the bottom quarks in these vertices are right handed, so the interference with the SM is  suppressed. The fact that the linear terms are zero is consistent with this observation. To understand why the quadratic contributions is also zero, we note that there are no \ttW or \ttWj diagrams involving \cbW or \cptb that may be initiated from bottom-quark PDFs. In all of the diagrams, the bottom quark in the EFT \tbWvertex vertex is connected to a SM \tbWvertex vertex. Since the SM vertex requires a bottom quark with the opposite chirality required by the EFT vertex, the effect of these diagrams is suppressed, and, in the case of a massless bottom quark, the diagrams are not possible. As mentioned in section~\ref{sec:validation_matching}, the five-flavor scheme is used for this study, so the mass of the bottom quark is set to zero, so these coefficients therefore cannot contribute to \ttW or \ttWj \footnote{In the four-flavor scheme, the dependence of on \cptb and \cbW is in principle non-zero; however, within the range of \cptb and \cbW values considered in this paper, the effect on the cross section is negligible.}.

\section{Matching systematics}
\label{sec:systemaics}
In the previous section, we examined the effect of including an extra parton in the \tth, \ttW, and \ttZ processes, finding relatively large differences from the LO without matching in several cases. However, it is fair to ask how the size of these differences compares with the uncertainties on the LO prediction.  If these effects are not significantly larger than the uncertainties, then it would still be safe to ignore them.  Conversely, if the differences observed when adding an extra parton with matching are large compared to the uncertainties, then these contributions should not be ignored.

In this section, we will explore the size of systematic uncertainties associated with the matched samples. As we did in prior sections, we will focus on $\mu = \sigma/\sigma_\mathrm{SM}$.  This will allow us to focus on uncertainties that change the quadratic shape of the inclusive cross section's dependence on the Wilson coefficients without becoming distracted by effects impacting only the overall normalization, even in the SM.  The following uncertainties are included:
\begin{itemize}
    \item We varied the nominal matching scale (i.e. \qcut = 19 GeV) between 15 GeV and 25 GeV.
    \item We varied the renormalization ($\mu_R$) and factorization scale ($\mu_F$) scales of the hard process.  Each scale was varied independently up or down by a factor of two and an envelope was constructed from the various combinations of up/down variations of the two scales.
    \item We varied the initial and final state radiation (ISR and FSR, respectively) scales in the parton shower up and down by a factor of $\sqrt{2}$.
    \item We use the PDF set NNPDF3.1, so the uncertainties are computed with eigenvector PDF members using the LHAPDF tools as described in Ref.~\cite{Buckley:2014ana}. 
\end{itemize}
We will consider how these uncertainties affect several example combinations of processes and Wilson coefficients, as a function of the Wilson coefficient. 

Three combinations of Wilson coefficients and operators are chosen for this study. For \tth, we select \cpt as an example of a Wilson coefficient whose effect on \tth is significantly different when an extra parton is included. As an example of a Wilson coefficient whose effect on the process does not significantly change when an additional parton is included in the process, we also study the effect of \ctG on \tth.  It is also interesting to consider \ctG because it is connected with the only operator among the subset under consideration whose vertices involve gluons. To broaden the study by including a process besides \tth, we also examine the effect of \cpQa on \ttW as an example of an operator that affects \ttW significantly differently when an extra parton is included in the calculation.

\begin{figure} [ht!]
\centering
\includegraphics[width=.32\textwidth]{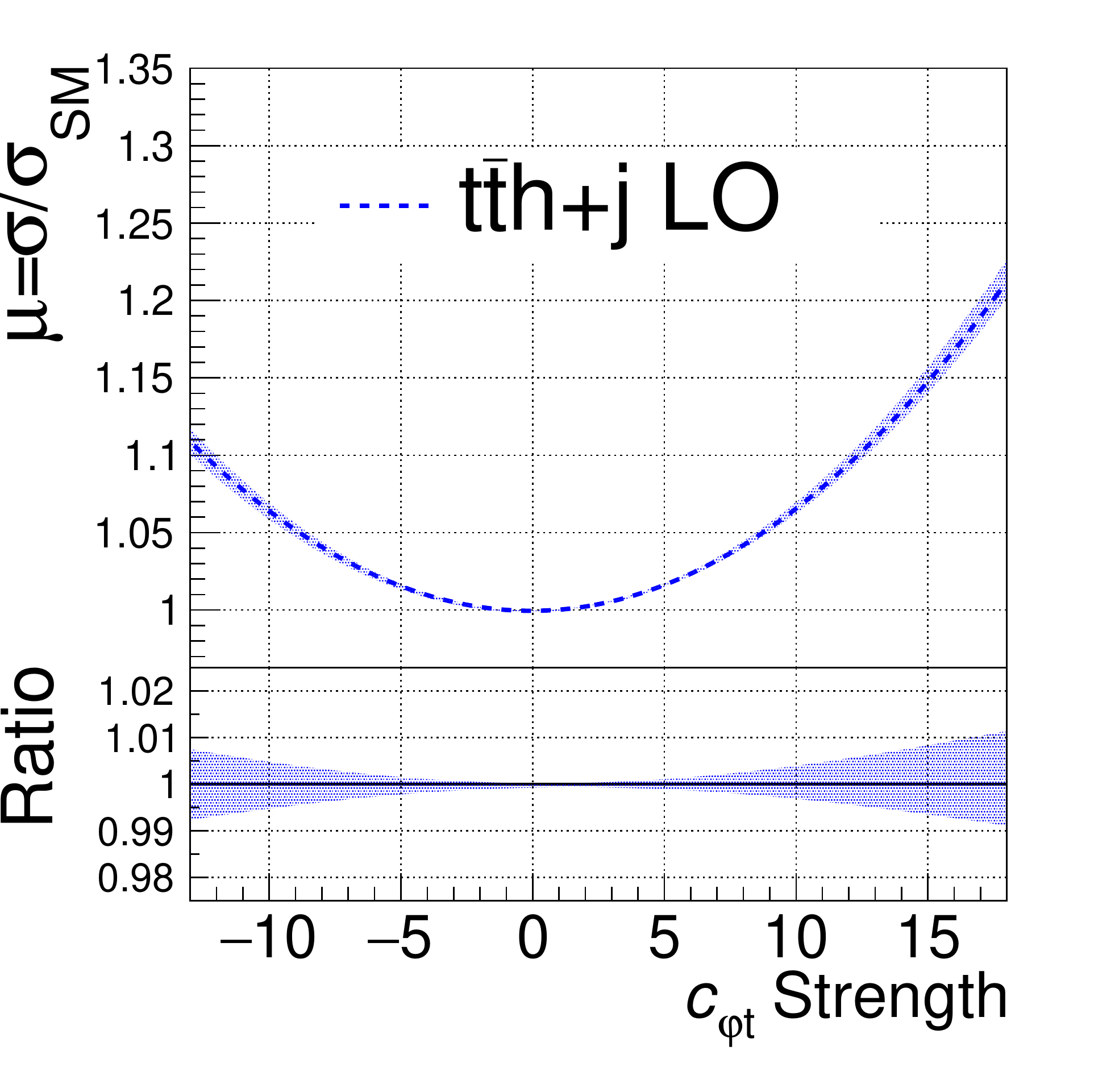}
\includegraphics[width=.32\textwidth]{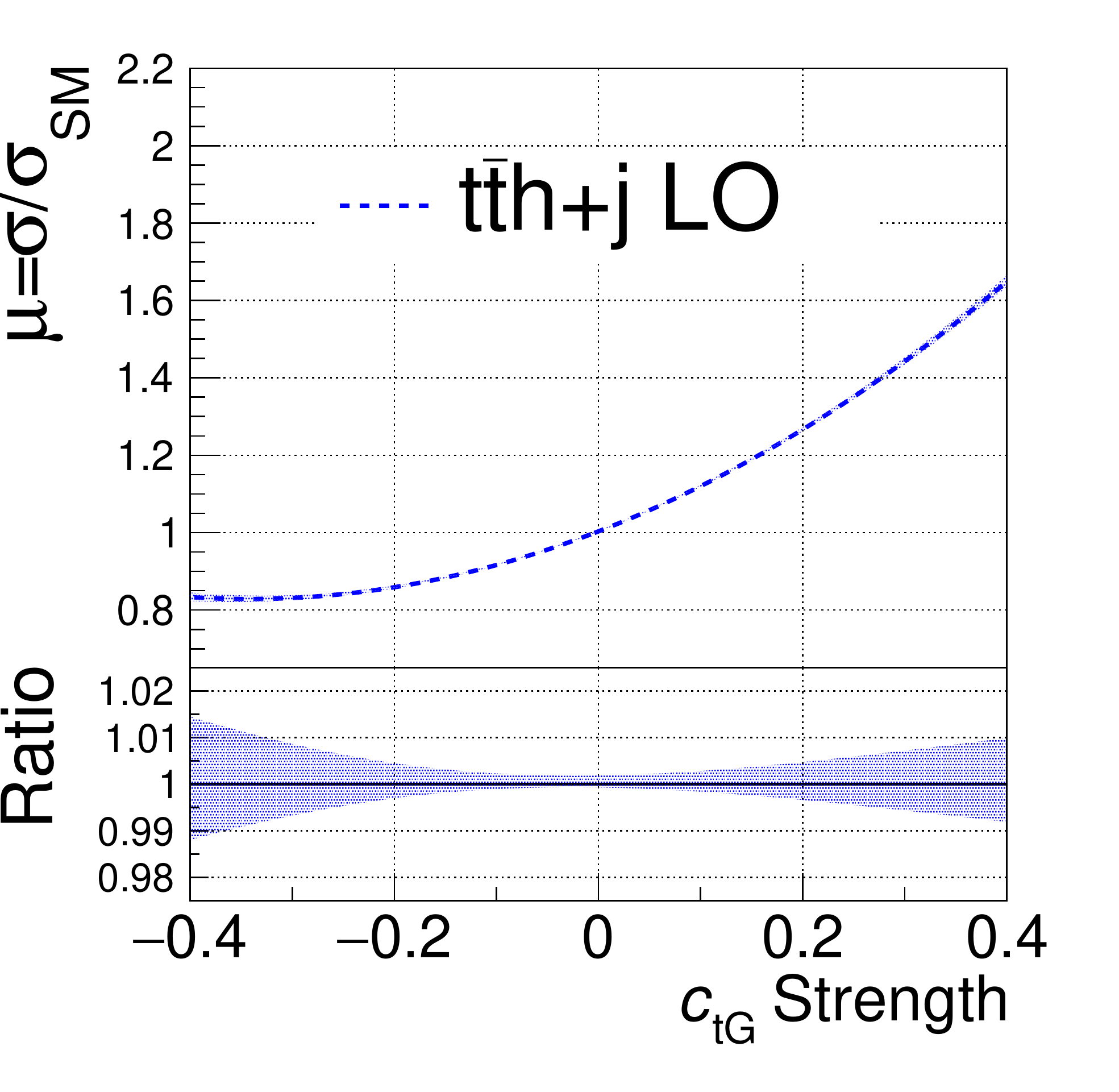}
\includegraphics[width=.32\textwidth]{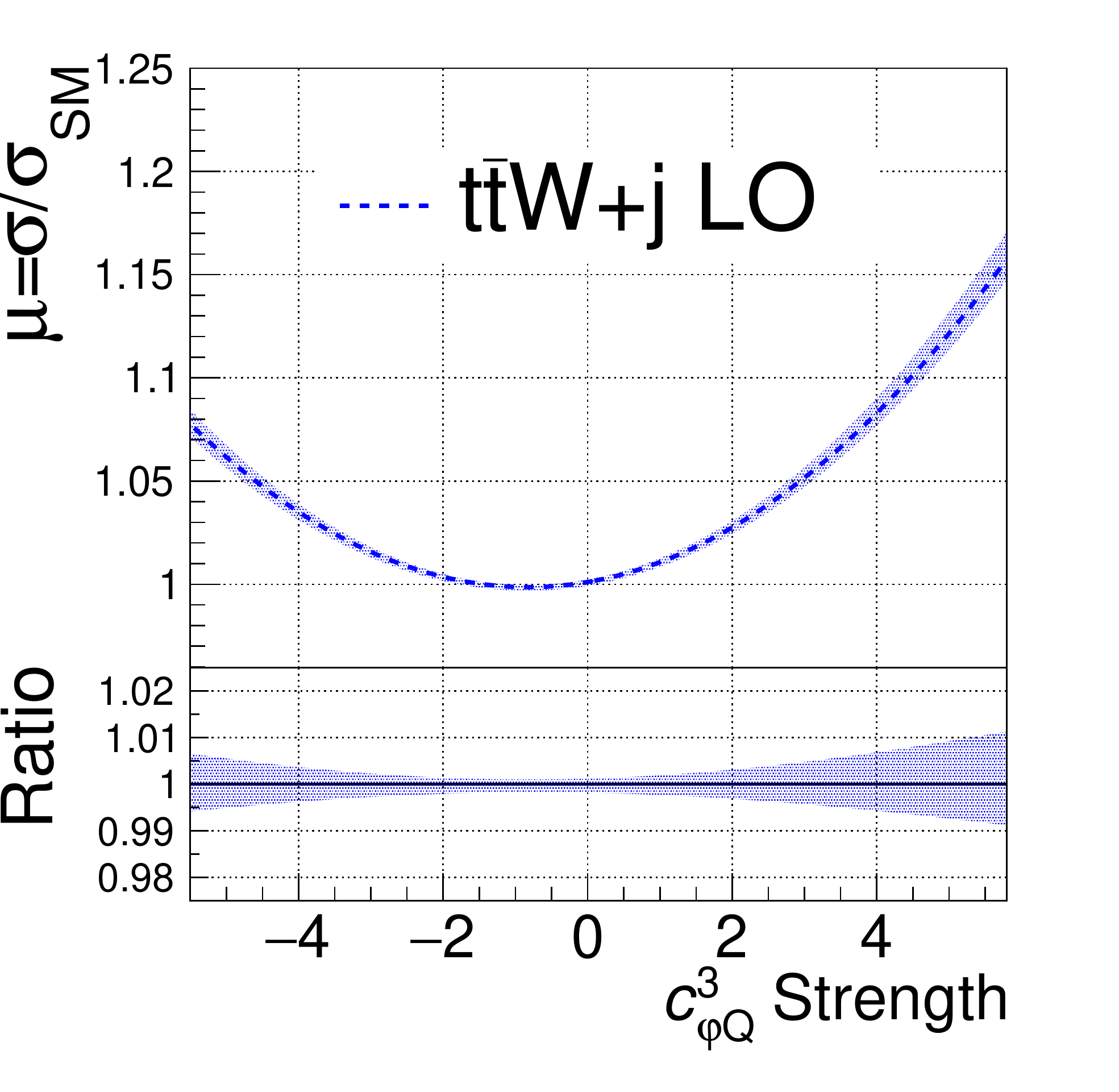}
\caption{Systematic uncertainties on the inclusive \tth cross sections for \cpt and \ctG, and on the inclusive \ttW cross section for \cpQa. The cross sections are normalized to the SM, as discussed in the text. Additionally, the range for each Wilson coefficient is taken from the marginalized limits presented in Ref.~\cite{Hartland:2019bjb}.}
\label{fig:systematics_quad_sum}
\end{figure}

For each of these three representative combinations of processes and Wilson coefficients, we vary the systematic uncertainties listed above for a range of Wilson coefficient values. Fitting a quadratic to the points, we can determine how the uncertainties scale with the Wilson coefficient; the effects of the systematics are combined in quadrature and shown as error bands on the quadratic curves in Figure~\ref{fig:systematics_quad_sum}.  Normalized to the SM, these plots show how the effects of the systematic uncertainties evolve as we move away from the SM; sources of uncertainties that just impact the overall normalization cancel out. 

Within the range of values considered for each Wilson coefficient (taken from the marginalized limits presented in Ref.~\cite{Hartland:2019bjb}) the combined effects of the systematic uncertainties are small, on the order of one percent. The relative size of each systematic is shown in Table~\ref{tab:systematics}.

\begin{table}[ht!]
  \caption{Examples of the relative sizes of the systematic uncertainties considered in this section. For each Wilson coefficient shown, the symmetrized fractional uncertainty for each systematic at the limits from \cite{Hartland:2019bjb} is listed.}
  \centering
  \begin{tabular}{c c c c c c c c}
   Process \rule{0pt}{4.0ex} & WC value & $\mu_R$ $\mu_F$ & PDF & ISR & FSR & qCut & Total \\ \hline
   \tth & \cpt = 18  &  0.5\% & 0.6\% & 0.1\% & 0.2\% & 0.5\% & 1.0\%  \\
   \tth & \cpt = -13  & 0.4\% & 0.4\% & 0.1\% & 0.1\% & 0.2\% & 0.8\%  \\
   \tth & \ctG = 0.4 &  0.8\% & 0.1\% & $<$0.1\% & $<$0.1\% & 0.3\% & 0.9\%  \\
   \tth & \ctG = -0.4 & 1.1\% & 0.2\% & 0.1\% & 0.3\% & 0.5\% & 1.3\%  \\
   \ttW & \cpQa = 5.8 &  1.0\% & 0.2\% & 0.1\% & 0.2\% & 0.1\% & 1.0\%  \\
   \ttW & \cpQa = -5.5 & 0.7\% & $<$0.1\% & 0.1\% & 0.2\% & $<$0.1\% & 0.6\%  \\
  \end{tabular}
  \label{tab:systematics}
\end{table}

\begin{figure} [htbp!]
\centering
\includegraphics[width=.35\textwidth]{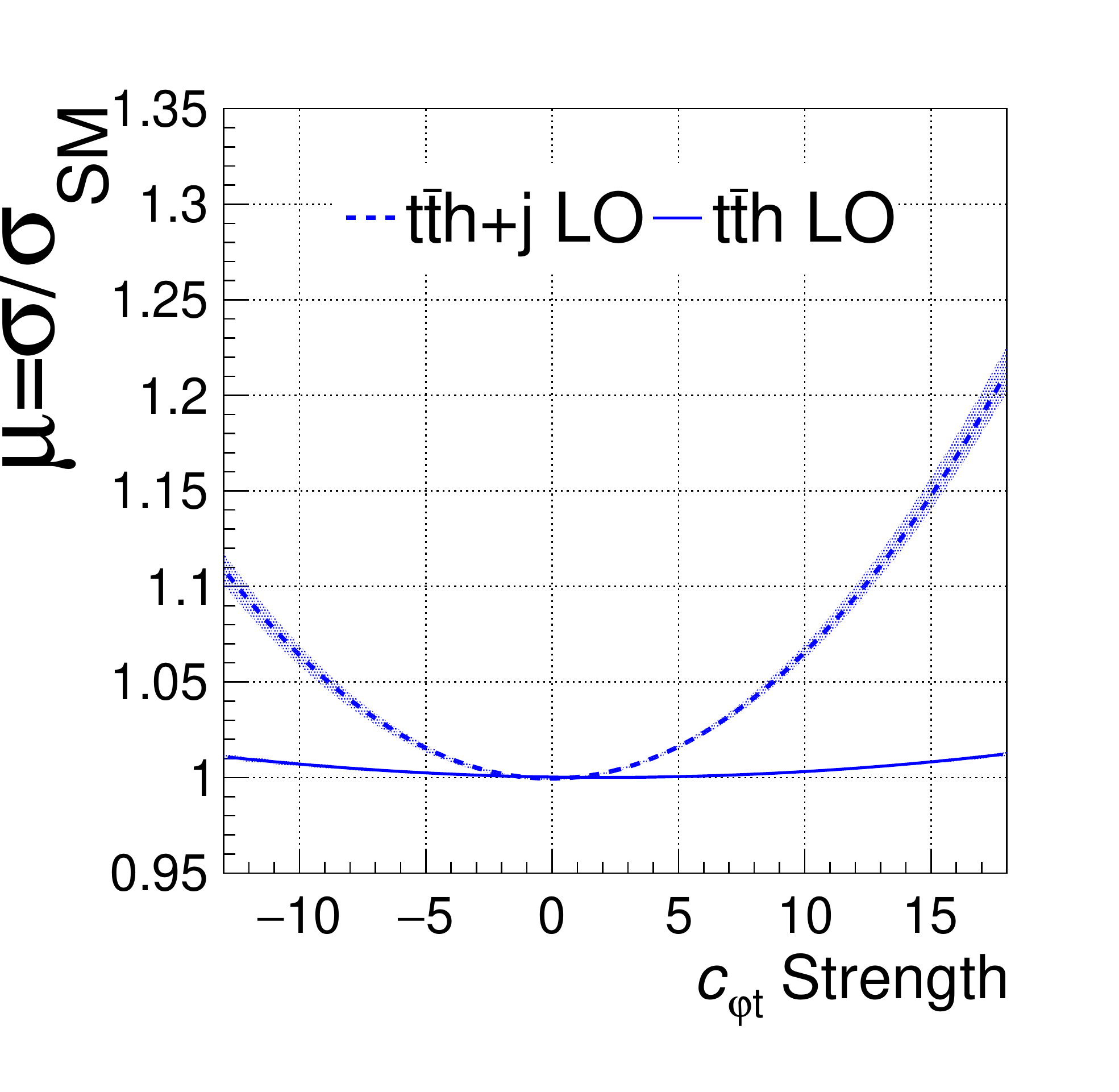}
\includegraphics[width=.35\textwidth]{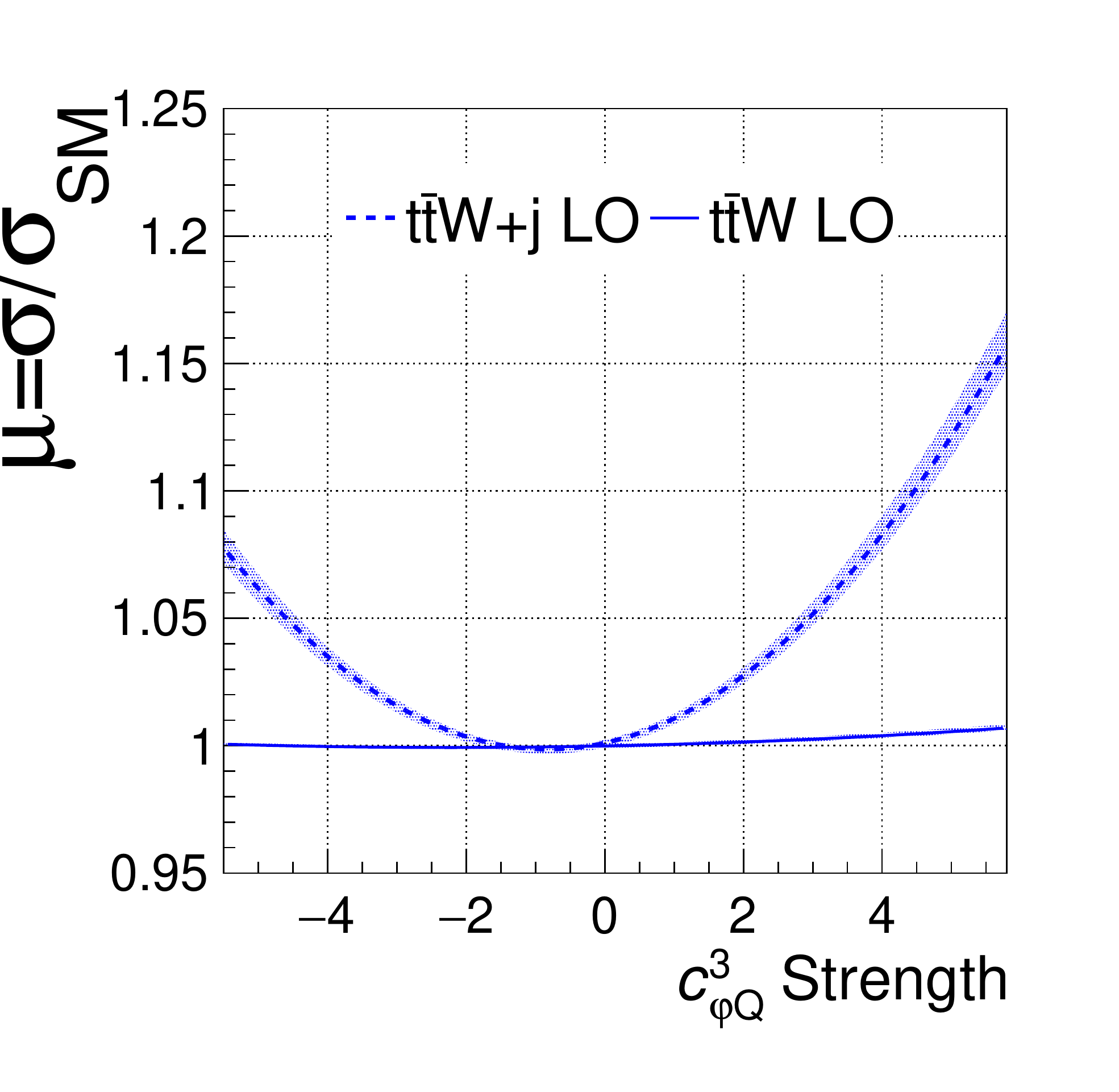}
\caption{Systematic uncertainties on the inclusive \tth and \tthj cross sections for \cpt and the inclusive \ttW and \ttWj cross section for \cpQa. Both \ttX and \ttXj curves are normalized to the SM as discussed in the text, and both curves include the uncertainty band, but the uncertainties on the \ttX curve are too small to be seen on this plot. We note that the size of the uncertainties are much smaller than the size of the effect of including an extra parton. As described in the text, the range for each Wilson coefficient is taken from the marginalized limits presented in Ref.~\cite{Hartland:2019bjb}.}
\label{fig:systematics_0p_1p_comp}
\end{figure}

Recall that in section~\ref{sec:results_LO_extra_partons}, we found that including an additional parton can have a sizeable effect, with a \muPlusOneP/\muZeroP ratio of 10\% or larger for several combinations of processes and Wilson coefficients.  As seen in Table~\ref{tab:systematics}, the size of the systematic uncertainties for the \ttXj processes (relative to the SM) are less than 2\%. This is much smaller than the effects described section~\ref{sec:results_LO_extra_partons}. In Figure~\ref{fig:systematics_0p_1p_comp}, we plot both the \ttX and \ttXj curves together with the uncertainty bands on both predictions.  It is easy to see that the size of effect of including an extra parton is much larger than the size of the associated systematic uncertainties. 

\section{Conclusion}
\label{sec:conclusions}
In this paper we have motivated the use of leading order calculations plus matching/merging for SMEFT \ttXj, $X = h/W/Z$ inclusive cross section studies. Being leading order, the numerical calculations are fast, have no issue with negative event weights, and can be done without restriction on the coupling order, thereby avoiding any model-dependence or subtlety of the QED or QCD order associated to SMEFT couplings. As LO plus matching is only an approximation of higher order effects, we do not expect it to correctly determine the overall normalization. However, it does capture the relative SMEFT vs. SM cross section. Matching/merging introduces new uncertainties in the form of the phase space boundary where the calculation is split between the matrix element and parton shower; however, we find that the uncertainty introduced by varying this boundary is less than 2\% in the cases we have considered. The impact of including an additional parton can be much larger than this. For example, the systematic uncertainties related to matching for \tth \cpt are approximately 1\%, while the inclusion of an additional parton has a 19\% effect at the upper limit for \cpt identified in Ref.~\cite{Hartland:2019bjb}.

As SMEFT vertices cannot be added easily to the parton shower, one may worry that certain regions of phase space are being inappropriately treated. While a valid concern, we find that, if we focus on operators that are most strongly constrained by \ttX processes (where X is on-shell), the only operator missing from the parton shower is $\OtG$. Studying the impact of this operator in soft and collinear regions of phase space, we find that its impact is minimal due to the chirality and dimensionality of the operator. This conclusion is similar to what Ref.~\cite{Englert:2018byk} found regarding $\mathcal O_G$. Monte Carlo studies, in the form of DJR plots, back up our analytical results, provided we make sure to include all relevant five-point vertices in the SMEFT UFO.

To characterize the benefits of adding one additional parton with matching, we compared LO with no extra partons (\ttX) to the matched LO with an extra parton (\ttXj).  For many of the operators considered for these \ttX processes, NLO calculations with the current \MADGRAPH SMEFT implementation are difficult to obtain due to complications involving the QED order assigned to the EFT vertices, so the matched LO result provides a particularly interesting estimate of the impact of higher-order corrections on the cross section dependence (relative to the SM) on the Wilson coefficients.  We find several cases in which the dependence on Wilson coefficient is significantly modified by the addition of an extra parton.  The dominant factor accounting for this change is that when including diagrams containing an additional parton, diagrams with quark-gluon initial states are able to contribute, while the diagrams without extra partons are limited only to quark-quark initiated diagrams.  Factors such as the range of Wilson coefficient considered, as well as the derivative, chiral and color structure of the operator under consideration also play an important role.

In closing, while there is no doubt that when practical, NLO calculations provide more accurate predictions, matched LO calculations do provide a computationally affordable alternative.  Matched LO calculations are preferable to strictly LO calculations without extra partons, because contributions from diagrams with an extra parton can provide important modifications to the dependence of the cross section on the Wilson coefficient.  Given that matched LO calculations evade complications connected to the QED order assigned to various EFT vertices in current SMEFT MC, they can even be seen to provide a useful cross check on the full NLO predictions.

\section*{Acknowledgments}

We thank Steve Mrenna for numerous helpful discussions.
The work of AM was supported in part by
the National Science Foundation
under Grant Number PHY-1820860.
The work of RG, KL, KM, and AW was supported in part by the National Science Foundation under Grant Number PHY-1914059.
The work of JHK was supported by Chungbuk National University Korea National University Development Project (2020). JHK is grateful to
Haider Alhazmi for valuable help and discussions.

\appendix


\section{Comparison of CPU and negative weights for NLO vs LO calculations}
\label{sec:appendix}

While it is clear that NLO MC are superior in terms of precision and accuracy of modeling QCD effects, there are reasons why matched tree-level calculations might be more practical. For example, the matched tree-level calculations typically require far less CPU-time for event generation.  Another consideration is the fact that the NLO calculation produces negatively weighted events, which can affect the statistical power of the samples generated at NLO.  Below we quantify some of these differences for a representative example calculation, \MADGRAPH which can produce both tree-level matched and NLO results.

\subsection{CPU comparison}
\label{sec:NLO_CPU_comp}

In this section, we will investigate the differences in CPU required to generate events with both the NLO and LO models. For this comparison, we used \MADGRAPH to produce \ttX gridpacks with both the SMEFT NLO model and our LO model. The LO gridpacks were produced with QED=1, QCD=3 order constraints for consistency with the NLO gridpacks. For each process (\tth, \ttW, and \ttZ), we generated a set number of events from both the NLO and LO gridpacks and recorded the CPU time required. Repeating this procedure for several set numbers of events, we can plot the CPU time required versus the number of events generated to understand how the CPU time scales with the number of events for the NLO and LO models. The same random seed is used for each running of each gridpack. 
These tests were performed using a single core from an AMD Opteron 6276 2.3 GHz processor.

\begin{figure} [htbp!]
\centering
\includegraphics[width=.32\textwidth]{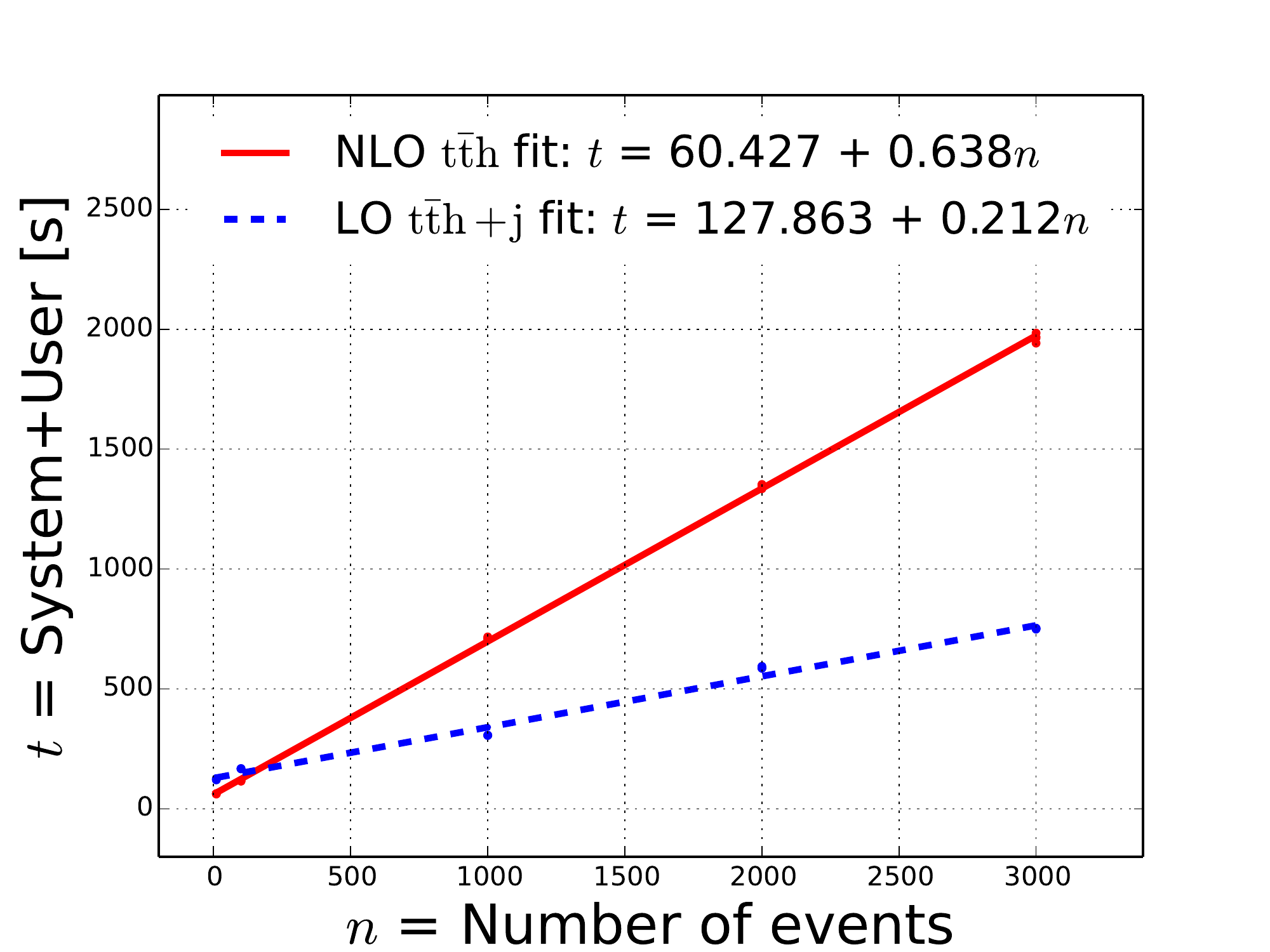}
\includegraphics[width=.32\textwidth]{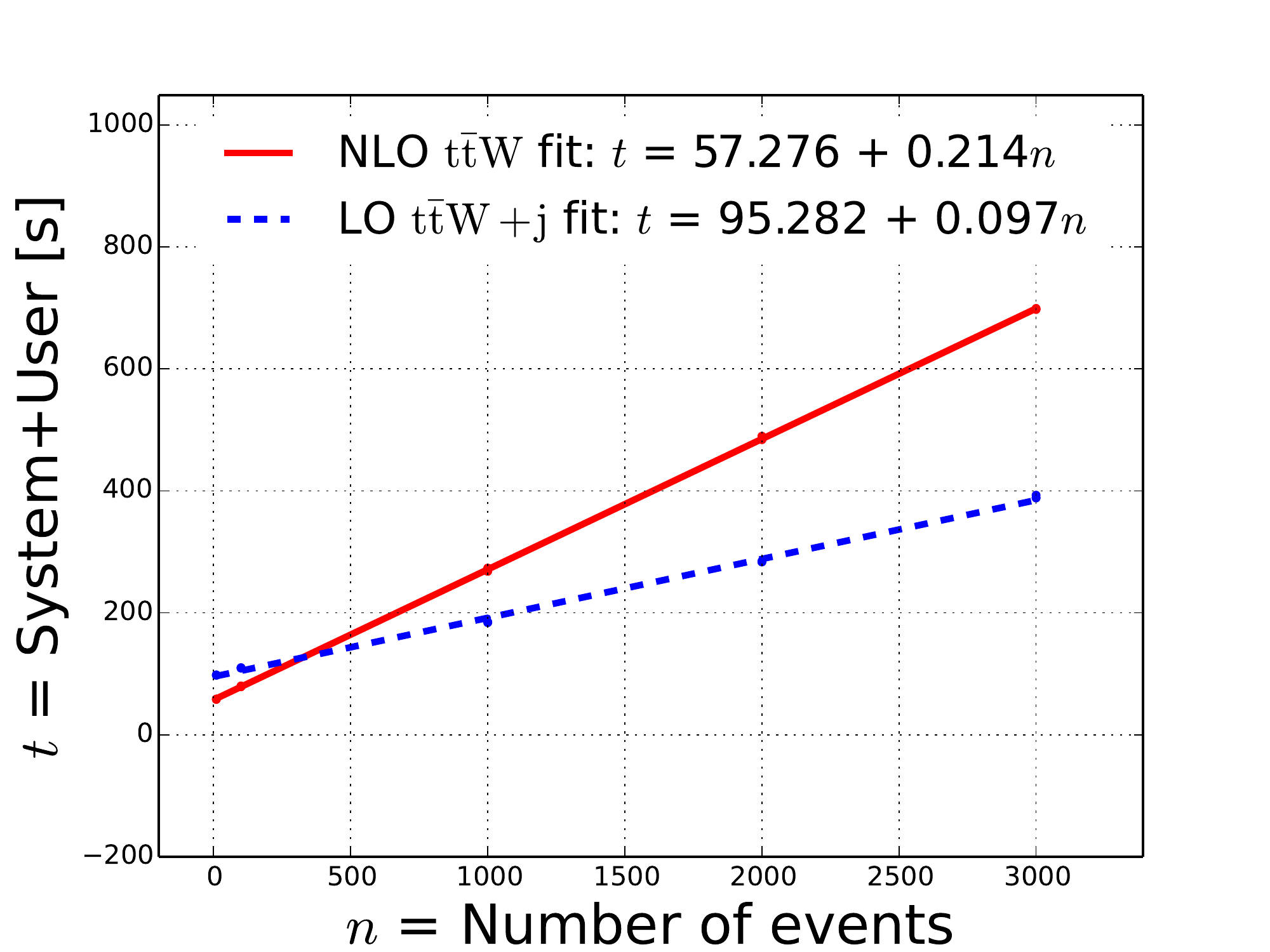}
\includegraphics[width=.32\textwidth]{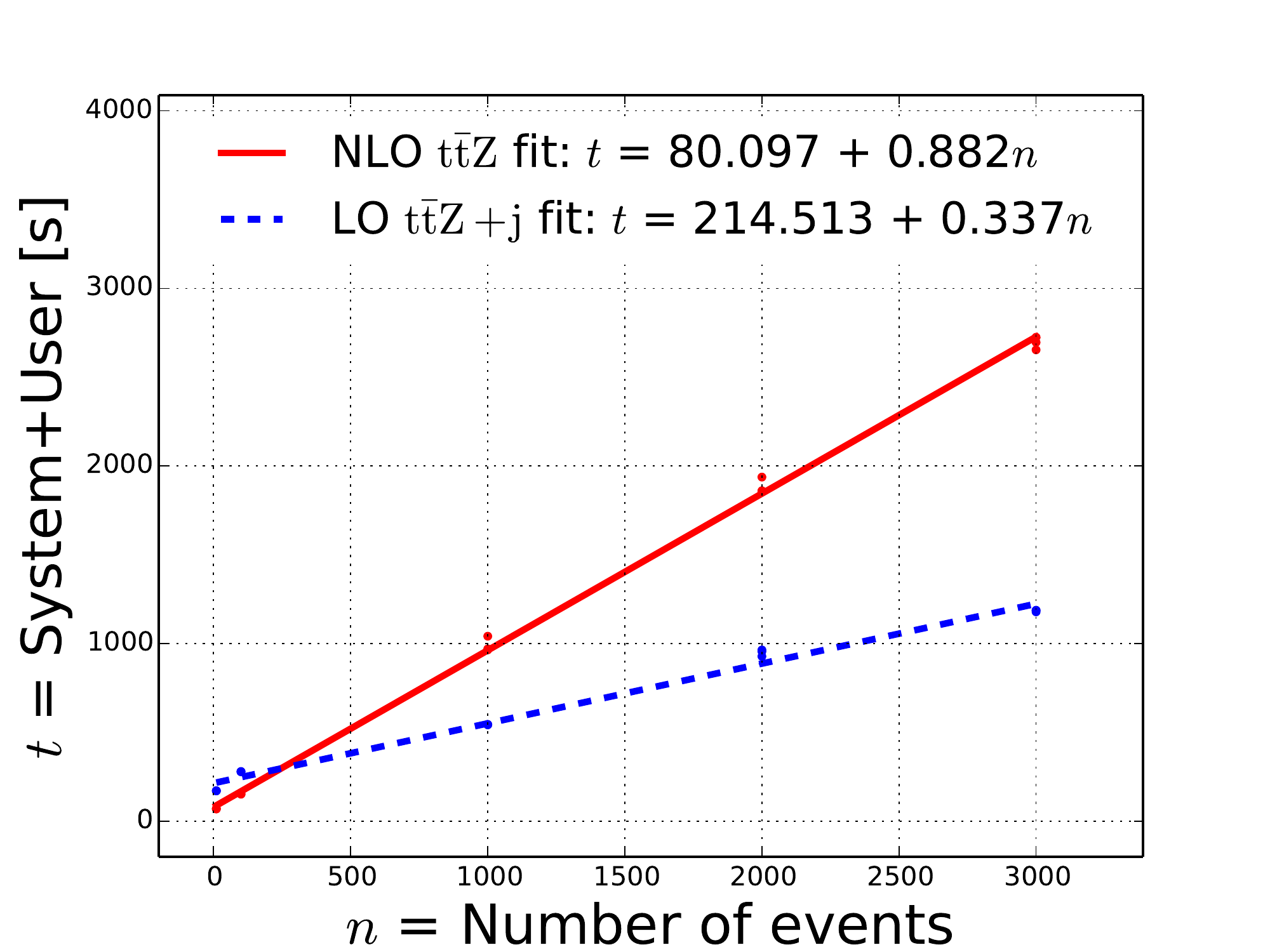} \\

\caption{Plots showing how the CPU time scales with the number of events generated for the NLO and LO models. For \tth (left), the slope of the fit to the NLO points is about 3 times larger than the slope of the line fit to the LO points. For \ttW (center), the slope of the fit the NLO points is about 2 times larger than the slope of the line fit to the LO points. For \ttZ (right), the slope of the line fit to the NLO points is about 3 times larger than the slope of the line fit to the LO points. For consistency with the NLO samples, the LO samples were generated with a QED=1 constraint.
}
\label{fig:NLO_vs_LO_CPU_comp}
\end{figure}

Figure \ref{fig:NLO_vs_LO_CPU_comp} shows the NLO and LO CPU comparisons \tth, \ttW, and \ttZ. Though the overhead time (indicated by the y-intercept of the fit) for the LO samples is somewhat larger than the NLO overhead, the slopes of the NLO fits are larger than the LO slopes by a factor of about two to three, depending on the process. 

\subsection{Negative weights}
\label{sec:NLO_negative_weights}
It is known that NLO cross sections are not always positive locally in the phase space which would imply events with negative weights in simulations. NLO calculations involve cancellation of soft/collinear infrared effects between virtual corrections and radiation, and negative weights can arise as a Monte Carlo artifact of this cancellation. The negative weighted events reduce the statistical accuracy of the simulated samples at the level of physical observable. For example, the statistical power of a sample with a fraction of 20-30\% negative weighted events is 3-5 times less than a sample with the same number of events but all weighted positively \cite{Frederix:2020trv}. Lowest order events, which require no subtle cancellations, have positive definite weight.

Currently the vast majority of studies about the SMEFT are based on the predictions by the \MADGRAPH. The NLO simulated samples, generated by the \MADGRAPH, have negative-weight events which their fractions vary for different processes. In Simulated \ttW, \ttZ and \tth samples by the \MADGRAPH for an arbitrary reference point in WC phase space, far from the SM point, we find 17\%, 27\%, and 28\% negative-weight events, respectively.
This means we need to simulate 3-5 times larger sample at NLO for finding similar statistical precision as LO sample which will cause more time and CPU cost for a global EFT search.


\section{Feynman rules for \texorpdfstring{$t$-$t$-$g$}{t-t-g} couplings}
\label{sec:appendix2}

\begin{figure}[ht]
\begin{center}
\includegraphics[width=0.37\textwidth,clip]{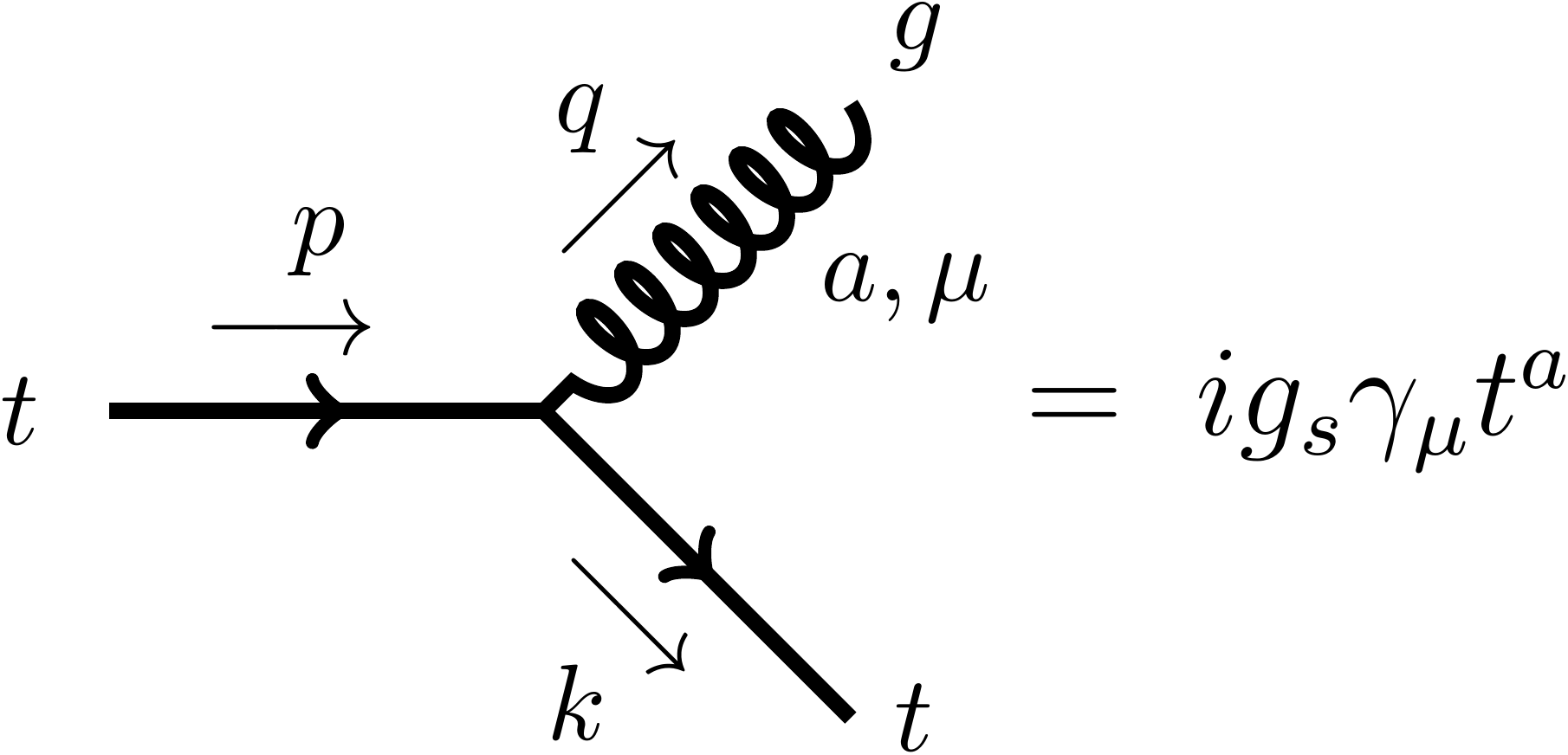} \hspace{2.0em}
\includegraphics[width=0.57\textwidth,clip]{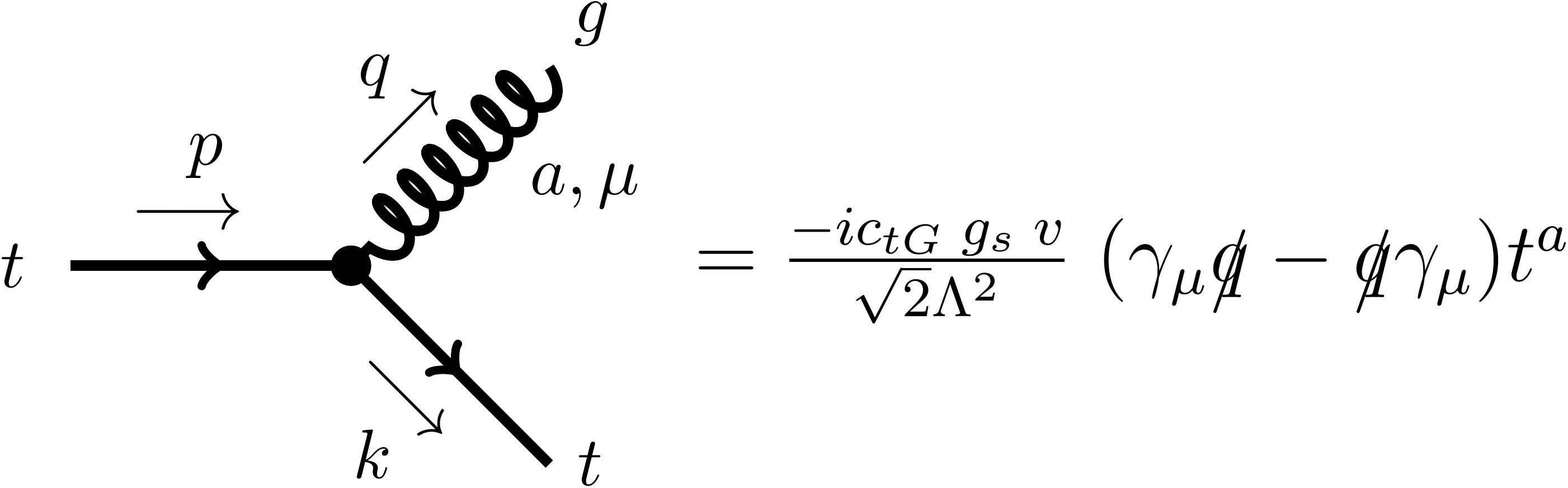} 
\caption{Feynman rules for (left) the SM $t$-$t$-$g$ vertex and (right) the anomalous $t$-$t$-$g$ coupling originated from the $O_{tG}$ operator in Eq.(\ref{eq:Otg}). }\label{ttg_FR}
\end{center}
\end{figure}

Feynman rules for the $t$-$t$-$g$ couplings are shown in Figure \ref{ttg_FR}, where the index $a$ runs over the eight color degrees of freedom of the gluon field.

 \section{Comparison of fit parameters for \ttX and \ttXj processes}
\label{sec:appendix_fit_coeffs}
 
In tables \ref{tab:tthSummary}, \ref{tab:ttWSummary}, and \ref{tab:ttZSummary} we report the coefficients of the quadratic \muZeroP and \muPlusOneP fits. Since the fits have been normalized to the SM, the constant term in each parameterization is 1, and we only report the coefficients for the linear terms (corresponding to the interference between the given Wilson coefficient and the SM) and quadratic terms (corresponding to the purely EFT term). The statistical uncertainties on the fit coefficients are smaller than the precision with which we report the coefficients.

\begin{table}[hbt!]
    \caption{Summary of quadratic fit coefficients for the \tth parameterization for \mutth and \mutthj. As discussed in the text, the cross sections are normalized to the SM.}
    \label{tab:tthSummary}
    \centering
    \begin{tabular}{|c|cc|cc|}
    \hline
    \multirow{2}{*}{WC} & \multicolumn{2}{c|}{\mutth} &  \multicolumn{2}{c|}{\mutthj}   \\ \cline{2-5} & Linear coefficient  & Quad coefficient & Linear coefficient & Quad coefficient \\
    \hline
    \ctW & 0.01215 & 0.03415 & 0.01903 & 0.08627 \\
    \ctp & -0.12269 & 0.00382 & -0.12268 & 0.00384 \\
    \cpQM & -0.00044 & 0.00012 & -0.00130 & 0.00069 \\
    \ctZ & -0.00568 & 0.02382 & -0.01009 & 0.05951 \\
    \ctG & 1.00811 & 1.23562 & 0.99949 & 1.40486 \\
    \cbW & 0.00000 & 0.00178 & 0.00000 & 0.00153 \\
    \cpQa & 0.00202 & 0.00248 & 0.00163 & 0.00215 \\
    \cptb & 0.00000 & 0.00036 & 0.00000 & 0.00029 \\
    \cpt & -0.00021 & 0.00005 & 0.00044 & 0.00061 \\

    \hline
    \end{tabular}
\end{table}
 
 \begin{table}[hbt!]
    \caption{Summary of quadratic fit coefficients for the \ttW parameterization for \muttW and \muttWj. As discussed in the text, the cross sections are normalized to the SM.}
    \label{tab:ttWSummary}
    \centering
    \begin{tabular}{|c|cc|cc|}
    \hline
    \multirow{2}{*}{WC} & \multicolumn{2}{c|}{\muttW} &  \multicolumn{2}{c|}{\muttWj}   \\ \cline{2-5} & Linear coefficient  & Quad coefficient & Linear coefficient & Quad coefficient \\
    \hline
    \ctW & 0.01384 & 0.02293 & 0.02332 & 0.05624 \\
    \ctp & -0.00003 & 0.00000 & -0.00036 & 0.00004 \\
    \cpQM & -0.00085 & 0.00006 & -0.00553 & 0.00106 \\
    \ctZ & 0.00041 & 0.00139 & -0.00043 & 0.00196 \\
    \ctG & 0.28568 & 0.02875 & 0.28313 & 0.04146 \\
    \cbW & 0.00000 & 0.00000 & 0.00000 & 0.00000 \\
    \cpQa & 0.00058 & 0.00015 & 0.00590 & 0.00409 \\
    \cptb & 0.00000 & 0.00000 & 0.00000 & 0.00000 \\
    \cpt & -0.00038 & 0.00006 & 0.00237 & 0.00103 \\
    \hline
    \end{tabular}
\end{table}
 
  \begin{table}[hbt!]
    \caption{Summary of quadratic fit coefficients for the \ttZ parameterization for \muttZ and \muttZj. As discussed in the text, the cross sections are normalized to the SM.}
    \label{tab:ttZSummary}
    \centering
    \begin{tabular}{|c|cc|cc|}
    \hline
    \multirow{2}{*}{WC} & \multicolumn{2}{c|}{\muttZ} &  \multicolumn{2}{c|}{\muttZj}   \\ \cline{2-5} & Linear coefficient  & Quad coefficient & Linear coefficient & Quad coefficient \\
    \hline
    \ctW & 0.00521 & 0.00946 & 0.00780 & 0.02654 \\
    \ctp & -0.00001 & 0.00000 & -0.00014 & 0.00001 \\
    \cpQM & -0.10464 & 0.00415 & -0.10434 & 0.00424 \\
    \ctZ & -0.00262 & 0.08617 & -0.00188 & 0.10632 \\
    \ctG & 0.35923 & 0.26202 & 0.36826 & 0.37035 \\
    \cbW & 0.00000 & 0.00154 & 0.00000 & 0.00167 \\
    \cpQa & 0.00368 & 0.00109 & 0.00292 & 0.00106 \\
    \cptb & 0.00000 & 0.00017 & 0.00000 & 0.00016 \\
    \cpt & 0.06876 & 0.00410 & 0.06820 & 0.00415 \\
    \hline
    \end{tabular}
\end{table}
 
 \clearpage
 
\bibliographystyle{JHEP}
\bibliography{refs}

\end{document}